\documentclass{ar-1col}
\pdfoutput=1

\usepackage{graphicx}
\usepackage{rotating}
\usepackage[T1]{sansmath}
\usepackage{bm,accents}
\SetMathAlphabet{\mathsfbf}{sans}{\sansmathencoding}{\sfdefault}{bx}{sl}
\usepackage{etoolbox}
\AtBeginEnvironment{sansmath}{\let\bm\mathsfbf}{}{}

\usepackage[comma]{natbib}
\usepackage{url}

\def\spose#1{\hbox to 0pt{#1\hss}}
\def\la{\mathrel{\spose{\lower 3pt\hbox{$\mathchar"218$}}
     \raise 2.0pt\hbox{$\mathchar"13C$}}}
\def\ga{\mathrel{\spose{\lower 3pt\hbox{$\mathchar"218$}}
     \raise 2.0pt\hbox{$\mathchar"13E$}}}

\newcommand{\etal}{et al.}
\newcommand{\apj}{\textit{Ap. J.}}
\newcommand{\apjl}{\textit{Ap. J. Lett.}}
\newcommand{\apjs}{\textit{Ap. J. Suppl.}}
\newcommand{\aj}{\textit{Astron. J.}}
\newcommand{\aap}{\textit{Astron.\,Astrophys.}}
\newcommand{\mnras}{\textit{MNRAS}}
\newcommand{\nat}{\textit{Nature}}
\newcommand{\sci}{\textit{Science}}
\newcommand{\araa}{\textit{Annu. Rev. Astron. Astrophys.}}
\newcommand{\aarev}{\textit{Astron. Astrophys. Rev,}}
\newcommand{\pasa}{\textit{Publ. Astron. Soc. of Australia}}
\newcommand{\pasj}{\textit{Publ. Astron. Soc. of Japan}}
\newcommand{\pasp}{\textit{Publ. Astron. Soc. of the Pacific}}
\newcommand{\assl}{\textit{Astrophysics and Space Science Library}}
\newcommand{\RMxAA}{\textit{Revista Mexicana de Astronom\'{i}a y Astrof\'{i}sica}}
\newcommand{\arxiv}{\textit{arXiv:}}

\newcommand{\arcsec}{\hbox{$^{\prime\prime}$}}

\newcommand{\farcs}{\mbox{$.\!\!^{\prime\prime}$}}

\newcommand{\micron}{$\rm \mu m$}

\newcommand{\kmostd}{$\rm KMOS^{3D}$}

\newcommand{\zph}{$z_{\rm phot}$}
\newcommand{\zgr}{$z_{\rm grism}$}
\newcommand{\zsp}{$z_{\rm spec}$}
\newcommand{\hii}{H{\footnotesize II}}
\newcommand{\nii}{[N{\footnotesize II}]}
\newcommand{\oii}{[O{\footnotesize II}]}
\newcommand{\oiii}{[O{\footnotesize III}]}
\newcommand{\sii}{[S{\footnotesize II}]}
\newcommand{\niiha}{[N{\footnotesize II}]/H$\alpha$}

\newcommand{\vrot}{$v_{\rm rot}$}
\newcommand{\sigo}{$\sigma_{0}$}
\newcommand{\vsigo}{$v_{\rm rot}/\sigma_{0}$}
\newcommand{\sigt}{$\sigma_{\rm tot}$}

\newcommand{\vcirc}{$v_{\rm c}$}
\newcommand{\fbar}{$f_{\rm bar}$}

\newcommand{\fgas}{$f_{\rm gas}$}
\newcommand{\mdyn}{$M_{\rm dyn}$}
\newcommand{\mbar}{$M_{\rm bar}$}

\jname{Annu.\,Rev.\,Astron.\,Astrophys.}
\jvol{58}
\jyear{2020}
\doi{10.1146/annurev-astro-032620-021910}
\firstpagenote{Posted with permission from the
  Annual Review of Astronomy and Astrophysics, Volume 58 \textcopyright\
  2020 by Annual Reviews, http://www.annualreviews.org/.}

\begin{document}

\markboth{F\"orster Schreiber \& Wuyts}{Star-Forming Galaxies at Cosmic Noon}

\title{Star-Forming Galaxies at Cosmic Noon}

\author{Natascha M. F\"orster Schreiber$^1$ and Stijn Wuyts$^2$
\affil{$^1$Max-Planck-Insitut f\"ur extraterrestrische Physik,
       85748 Garching, Germany; email: forster@mpe.mpg.de}
\affil{$^2$Department of Physics, University of Bath,
       Bath, BA2 7AY, United Kingdom; email: S.Wuyts@bath.ac.uk}}

\begin{abstract}
Ever deeper and wider lookback surveys have led to a fairly robust outline
of the cosmic star formation history, which culminated around $z \sim 2$ --
a period often nicknamed ``cosmic noon.''
Our knowledge about star-forming galaxies at these epochs has dramatically
advanced from increasingly complete population censuses and
detailed views of individual galaxies.  
We highlight some of the key observational insights that influenced
our current understanding of galaxy evolution in the equilibrium
growth picture: \\
$\bullet$ scaling relations between galaxy properties are fairly well established
among massive galaxies at least out to $z \sim 2$, pointing to regulating
mechanisms already acting on galaxy growth; \\
$\bullet$ resolved views reveal that gravitational instabilities and efficient
secular processes within the gas- and baryon-rich galaxies at $z \sim 2$
play an important role in the early build-up of galactic structure; \\
$\bullet$ ever more sensitive observations of kinematics at $z \sim 2$ are
probing the baryon and dark matter budget on galactic scales and the links
between star-forming galaxies and their likely descendants; \\
$\bullet$ towards higher masses, massive bulges, dense cores, and powerful
AGN and AGN-driven outflows are more prevalent and likely play a role in
quenching star formation. \\
We outline emerging questions and exciting prospects for the next decade with
upcoming instrumentation, including the {\it James Webb Space Telescope\/}
and the next generation of Extremely Large Telescopes.
\end{abstract}

\maketitle

\tableofcontents

\section{INTRODUCTION}
\label{intro.sec}

\subsection{Background}
 \label{intro_bkgd.sec}

Star-forming galaxies at redshift $z \sim 2$, 10 billion years ago, trace
the prime formation epoch of today's massive disk and elliptical galaxies.
Our knowledge about their properties, and their place in the global context
of galaxy evolution, has undergone spectacular advances in the past two
decades from both increasingly complete population censuses at ever earlier
cosmic times and increasingly detailed descriptions of individual systems.
The identification and characterization of galaxies according to their global
colors, stellar populations, structure and morphologies, and environment is now
routinely done out to $z \sim 3$, encompassing 85\% of the Universe's history.
Comprehensive surveys of the kinematics and interstellar medium (ISM) properties
have been obtained from spatially- and spectrally-resolved observations of ionized
gas line emission out to $z \sim 3-4$.  The cold gas content has been measured,
and is being resolved on subgalactic scales for rapidly rising numbers.
Growing samples at $z \sim 4 - 8$ are being assembled and the first candidates
have been identified at $z \sim 9-11$ within 500~Myr of the Big Bang, yielding
insights into the progenitor populations of $z \sim 2$ star-forming galaxies.

\begin{marginnote}[30pt]
 \entry{ISM}{Interstellar medium.}
 \entry{SMBH}{Supermassive black hole.}
 \entry{SFR}{Star formation rate.}
 \entry{SFG}{Star-forming galaxy.}
 \entry{MS}{\looseness=-2 main sequence of SFGs,
            referring to the observed tight relationship between their
            stellar mass and star formation rate.}
\end{marginnote}

This body of observational work has led to a fairly robust outline of
the evolution of the stellar mass build-up and star formation activity of
galaxies and the growth of supermassive black holes (SMBHs) over most of
the Universe's history \citep{Mad14}.
As much as half of the stellar mass observed in galaxies today was formed
in just about 3.5~Gyr between $z \sim 3$ and $z \sim 1$.
After a rapid rise $\propto (1+z)^{-2.9}$, the cosmic star formation rate
(SFR) volume density peaked around $z \sim 2$ and subsequently declined as
$\propto (1+z)^{2.7}$ to $z = 0$.
The comoving rate of SMBH accretion follows a similar evolution, in support
of co-evolution of central black holes and their host galaxies.
This evolution in cosmically averaged rates finds its counterpart
in the observed properties of individual star-forming galaxies (SFGs),
which around $z \sim 2$ were forming stars and feeding their central
black holes $\sim 10$ times faster than today's SFGs.
At least up to $z \sim 3$, the vast majority
of SFGs follows tightly a roughly linear ``main sequence'' (MS) between SFR
and stellar mass \citep[e.g.,][]{Rod11, Spe14}, whose zero-point evolution
reflects the decline in cosmic SFR density to the present time.
Other scaling relations involving size, kinematics,
metal, and gas content are also observed as early as $z \sim 2 - 3$
\citep[e.g.,][]{vdWel14a, Ueb17, Mai19, Tac20}.
Detailed mapping of the distribution and motions of stars and gas within
galaxies have begun to probe the internal workings of galaxy evolution, and
the spatial and temporal progression of the build-up of galactic components.
Despite increasingly clumpy and irregular appearances at higher redshift,
more so in the rest-frame ultraviolet \citep[e.g.,][]{Con14}, there is now
compelling evidence that the bulk of (massive) SFGs have global disk-like
stellar light distributions and kinematics
\citep[e.g.,][]{Wuy11b, Wis19}.

The existence of scaling relations and the prevalence of disk structure at
$z \sim 2$ implies that regulating mechanisms already controlled the growth
and lifecycle of SFGs at early times.  Specifically, these observations have
highlighted the importance of internal processes in shaping galaxies and of
smoother modes of accretion, with a lesser role of (gas-rich) major merger
events able to dramatically alter the structure and drive large short-term
fluctuations in SFRs.
Taken together, these findings have laid out the empirical foundations for
the ``equilibrium growth model'' \citep[e.g..][]{Dek09a, Bou10, Lil13}, in
which the stellar mass growth of galaxies is governed by the balance between
accretion, star formation, and outflows, until they reach a stellar mass of
$M_{\star} \sim 10^{11}\,{\rm M_{\odot}}$ where their star formation is
quenched and they rapidly transition to the sequence of
quiescent galaxies \citep[e.g.,][]{Pen10}.


\subsection{Setting the Stage}
 \label{intro_motiv.sec}

Once dubbed the ``redshift desert'' because of the relative inaccessibility
of key spectral features for source identification with then available
instrumentation, our matured view driven by rapid observational progress
now shows that $z \sim 1 - 3$ is a pivotal epoch in galaxy evolution ---
it is ``cosmic noon''.

\begin{marginnote}[215pt]
 \entry{\begin{sansmath}$\bm{JWST}$\end{sansmath}}
       {{\it James Webb Space Telescope.\/}}
 \entry{ELT}{Extremely Large Telescope.}
 \entry{TMT}{Thirty Meter Telescope.}
 \entry{GMT}{Giant Magellan Telescope.}
 \entry{\begin{sansmath}$\bm{HST}$\end{sansmath}}
       {{\it Hubble Space Telescope.\/}}
\end{marginnote}

\looseness=-2
Lookback studies are at a turning point, with major leaps forward anticipated
in the next decade from cutting-edge instrumentation at existing observatories,
the imminent launch of the {\it James Webb Space Telescope} ({\it JWST\/}), and
the coming of next-generation large aperture ground-based telescopes such as
the 39\,m Extremely Large Telescope (ELT), the Thirty Meter Telescope (TMT),
and the 25\,m Giant Magellan Telescope (GMT).
Recent and future capabilities at current facilities will allow us to establish
the missing links between the distributions and kinematics of stars, gas,
and metals in and around galaxies, unraveling vital phases of the baryon cycle
and the interplay between baryons and dark matter.
With transformative boosts in sensitivity and angular resolution afforded by
{\it JWST\/} and the extremely large telescopes,
galaxy evolution at $z > 1$ will be charted with
unprecedented completeness well into the epoch of reionization and with
unrivaled sharpness down to the 100-pc scale of individual giant star-forming
complexes --- a landscape revolution akin to the advent of the
{\it Hubble Space Telescope\/} ({\it HST\/}) and the
first 8\,m-class telescopes in the 1990s.

Of the remarkably rich observational harvest of the past 5-10 years, we can
here only highlight select aspects that have been among the most influential
in advancing our knowledge about $z \sim 2$ SFGs.
We focus on the internal properties of galaxies as revealed by diagnostics
in emission and typical environments found in deep extragalactic fields,
which comprise the bulk of the galaxy population.
Section\ \ref{obsbkgd.sec} presents the observational landscape.
Section\ \ref{global.sec} discusses global properties providing the
population context and enabling evolutionary links, and
Section\ \ref{resolved.sec} zooms on resolved properties providing
insights into the physics shaping galaxies.
Section\ \ref{otherpops.sec} discusses subpopulations of SFGs
with extreme properties.
Section\ \ref{disc.sec} briefly comments on the theoretical landscape.
In closing, Section\ \ref{outlook.sec} summarizes the article, and
outlines open issues and future observational opportunities.

For simplicity, we refer throughout to the $1 \la z \la 3$ epochs as
``$z \sim 2$'' or ``high $z$'' unless explicitly stated otherwise.
We adopt a $\Lambda$-dominated cosmology with
$H_{0} = 70~{\rm km\,s^{-1}\,Mpc^{-1}}$, $\Omega_{\rm m} = 0.3$,
and $\Omega_{\Lambda} = 0.7$.
For this cosmology, 1\arcsec\ corresponds to 8.4~kpc at $z = 2$.
Magnitudes are given in the AB photometric system.
Where relevant, galaxy masses and star formation properties are adjusted
to a common \citet{Cha03} stellar initial mass function (IMF).
\begin{marginnote}[0pt]
 \entry{IMF}{Initial mass function of stars.}
\end{marginnote}

\section{OBSERVATIONAL LANDSCAPE}
\label{obsbkgd.sec}

The dramatic advances in our knowledge about galaxies at cosmic noon have
been driven by the confluence of novel observational techniques and sensitive
high-multiplex ground- and space-based instrumentation across the electromagnetic
spectrum.
The concentration of multi-wavelength campaigns in select fields targeted
as part of the Great Observatories Origins Deep Survey (GOODS), the Cosmic
Evolution Survey (COSMOS), the All-wavelength Extended Groth strip International
Survey (AEGIS), and the UKIDSS Ultra-Deep Survey (UDS) have yielded rich data
sets and have seen their legacy value fully realized by providing samples of
choice for many detailed follow-up studies.
Several reviews have covered various aspects of $z>1$ galaxy surveys
in the past decade \citep[notably][]{Sha11,Glaze13,Mad14,Lut14,Con14,Tac20}.
This Section gives an update incorporating recent programs with the goal
of highlighting the observational underpinnings of our current physical
understanding of cosmic noon galaxies.

\begin{figure}[!t]
$\begin{array}{rr}
 \includegraphics[scale=0.46,trim={25.0cm -0.2cm 5.0cm 3.0cm},clip=0,angle=0]{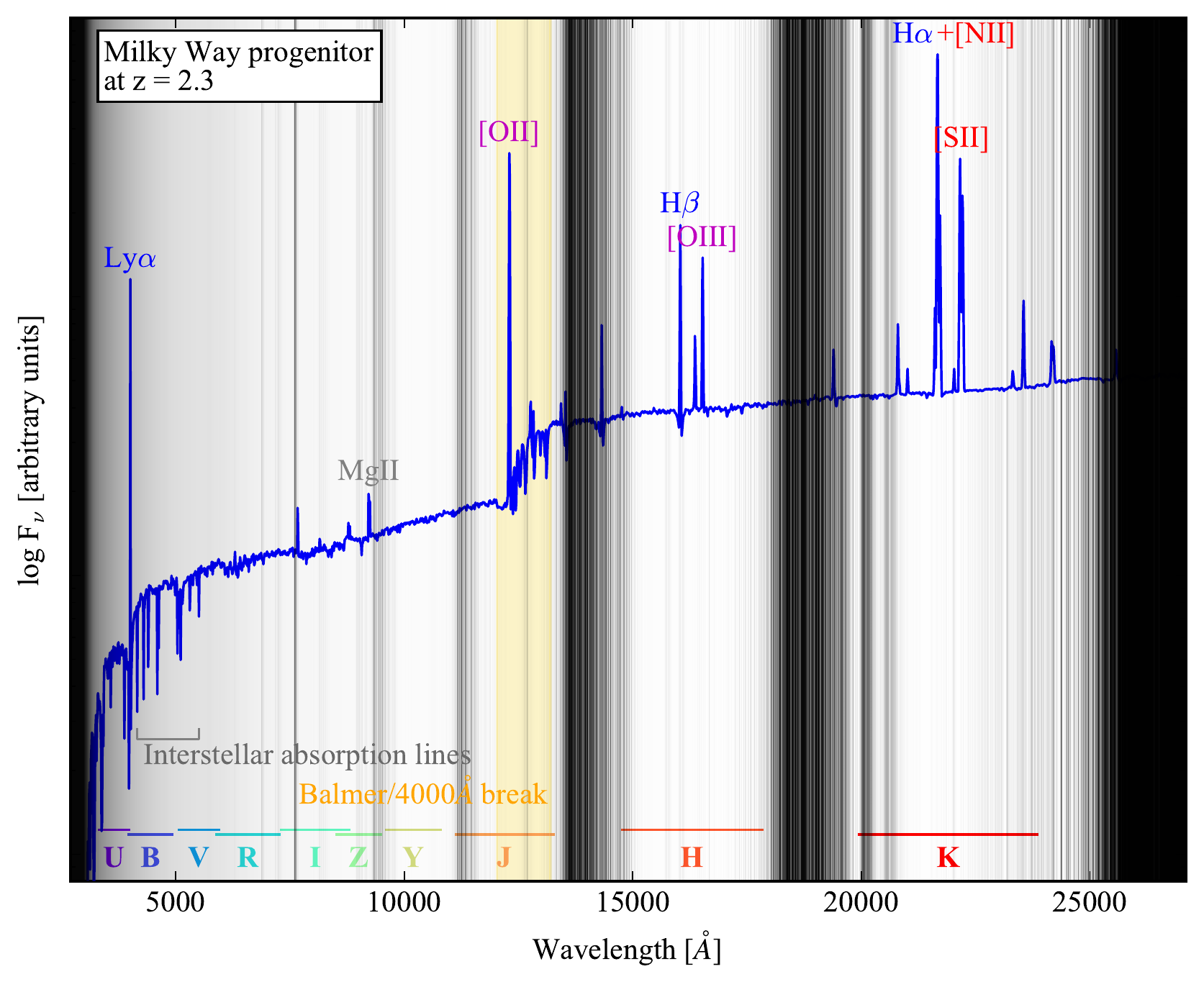}  &
 \includegraphics[scale=0.46,trim={45.5cm -0.2cm 4.0cm 3.0cm},clip=0,angle=0]{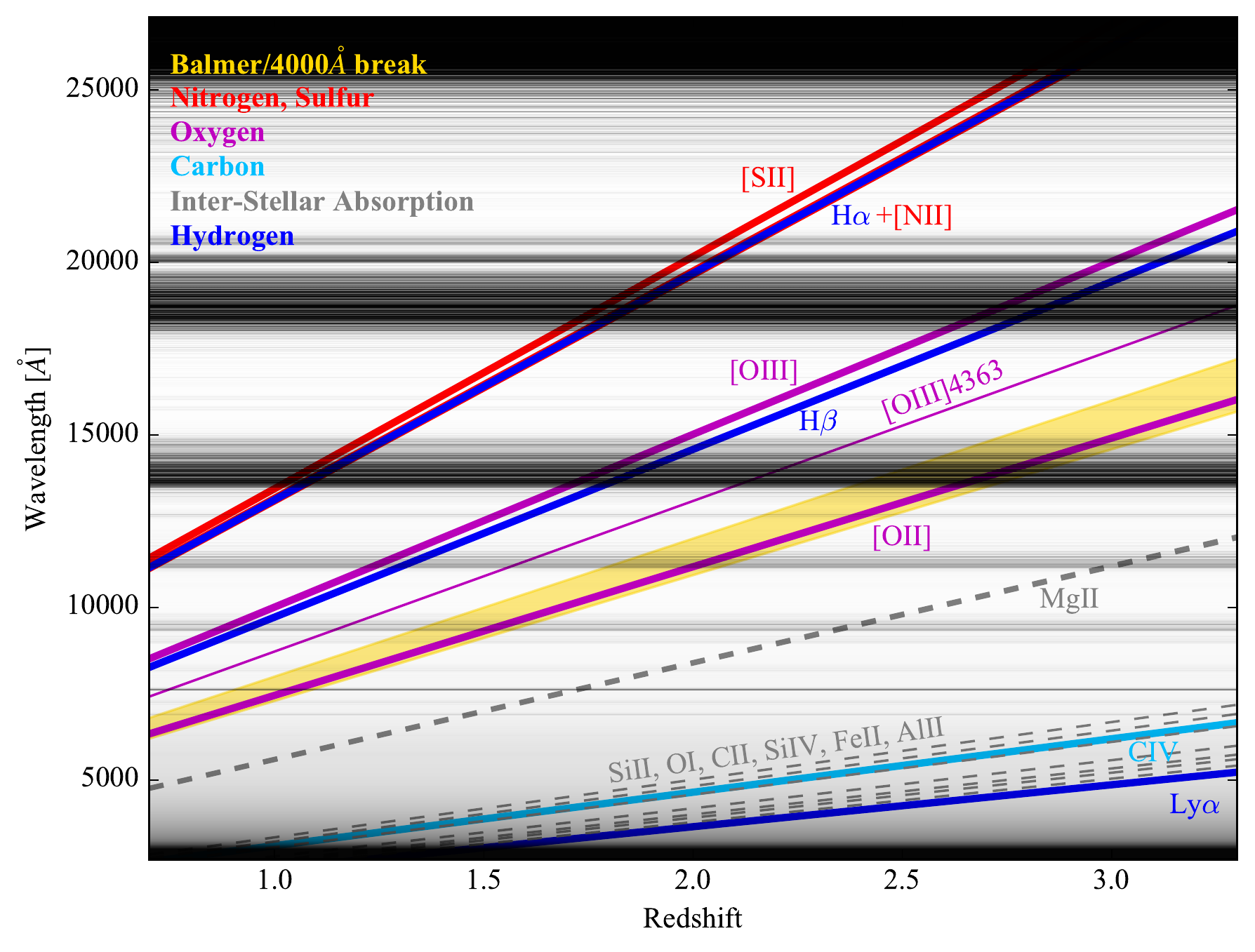}  \\
\end{array}$
\renewcommand\baselinestretch{0.85}
\caption{
{\it Left:\/}
Synthetic spectrum of an illustrative SFG at $z = 2.3$ plotted in observed
wavelength vs.\ logarithmic flux density units.
The spectrum was produced using the code BAGPIPES \citep{Carn18} for
the properties of a Milky Way-mass progenitor galaxy
based on the scaling relations and evolution thereof discussed in 
Section\ \ref{global.sec}.
Rest-frame far-UV absorption lines were incorporated with relative strengths
based on \citet{Ste16}.
{\it Right\/:}
Observed wavelengths of salient emission and absorption features (identified
on the spectrum and described in the text) as a function of redshift from
$z \sim 1$ to $\sim 3$.
In both plots, the dark-to-light grey shading scales with increasing
atmospheric transparency computed 
with the European Southern Observatory (ESO) SkyCalc tool
(http://www.eso.org/observing/etc/skycalc/skycalc.htm),
and the main photometric bandpasses are indicated at the bottom of
the left panel.
}
\label{atmosline.fig}
\end{figure}

\begin{marginnote}[210pt]
 \entry{IR}{Infrared.}
 \entry{UV}{Ultraviolet.}
\end{marginnote}

\begin{figure}[!t]
\centering
\includegraphics[scale=0.60,trim={6.0cm 3.5cm 4.0cm 6.5cm},clip=1,angle=0]
                {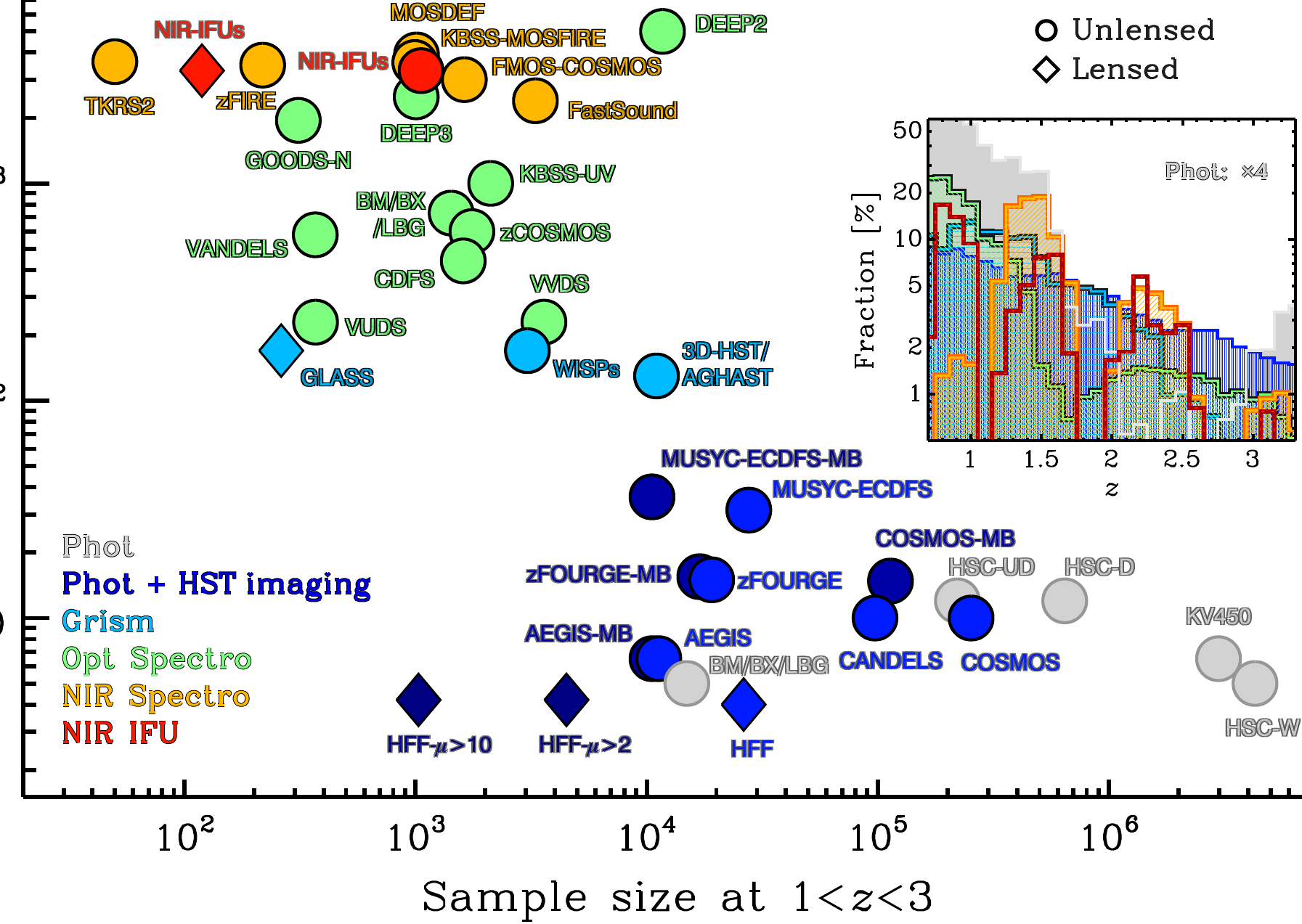}
\renewcommand\baselinestretch{0.85}
\caption{
Overview of selected optical/near-IR surveys covering $1 < z < 3$
as a function of number of sources in this interval and spectral resolution
(see list in Supplemental Tables~1 and 2).
The color-coding indicates the primary type of observations:
photometric imaging (light grey);
photometric imaging including high-resolution {\it HST\/} data (blue) and subsets
thereof with useful (i.e., $\rm S/N > 3$) medium-/narrow-band data (dark blue);
slitless grism data from {\it HST\/} (cyan);
optical spectroscopy (green); near-IR spectroscopy (yellow);
near-IR IFU data (red; detailed in Figure\ \ref{IFUsurveys.fig}).
Different symbols distinguish surveys of gravitationally lensed
targets and/or areas (diamonds) from unlensed ones (circles);
for the HFF, the full survey (including parallels)
and subsets magnified by $\mu > 2$ and $\mu > 10$ are plotted.
The inset shows the combined redshift distributions, grouped and color-coded
by observations type, normalized by the total number of $1 < z < 3$ galaxies,
and with fractions on a logarithmic scale.
Given the very heterogeneous nature of the samples (depth, detection/selection
function, etc.), the histograms merely
serve to illustrate the typical relative distributions.
The overall drop with increasing $z$, smoothest for photometric and grism surveys,
largely reflects the flux limits; the turn-up at $z \ga 3$ for photometric-only
surveys is driven by efficient Lyman-break dropout identification in optical
surveys.
Key spectral features falling between atmospheric windows cause the $z$ gaps
for ground-based spectroscopic and IFU surveys.
}
\label{surveys.fig}
\end{figure}

Our empirical knowledge rests on a ladder going from the identification of
galaxies from large photometric samples and their spectroscopic confirmation
enabling statistical descriptions of the population, to increasingly detailed
studies of subsets from spectrally/spatially resolved data.
Observations at optical to near-IR wavelengths form a major part of each step,
probing the redshifted, rest-frame UV to optical emission from $z \sim 2$ galaxies.
Figure~\ref{atmosline.fig} identifies salient spectral features on a model
spectrum created for an example SFG at $z = 2.3$ and shows how they
shift across the various atmospheric bandpasses from $z = 1$ to 3.
These features include (with rest wavelengths given in \AA):
\begin{itemize} \vspace{-1.5ex}
\item hydrogen recombination and atomic forbidden emission lines from warm
 ionized gas excited by star formation, AGN, and shock activity, which provide
 diagnostics of nebular conditions, dust attenuation, galaxy dynamics, and gas
 outflows (Ly$\alpha$\,$\lambda$1216, H$\beta$\,$\lambda$4861,
 H$\alpha$\,$\lambda$6563, [O{\footnotesize II}]$\lambda\lambda$3726,3729,
 [O{\footnotesize III}]$\lambda\lambda$4959,5007,
 [S{\footnotesize II}]$\lambda\lambda$6716,6730);
\begin{marginnote}[0pt]
 \entry{AGN}{Active galactic nucleus.}
\end{marginnote}
\item stellar continuum emission, encompassing the Balmer discontinuity at
 3646\AA\ and the 4000\AA\ break caused by hydrogen and multiple metallic species
 and molecules in the atmospheres of intermediate- to low-mass evolved stars, 
 and on which estimates of the stellar age, stellar mass, and dust reddening
 are based;
\item a rich suite of far-UV ($\sim 1200 - 2000$\AA) interstellar low- and
 high-ionization atomic absorption lines useful to trace gas outflows/inflows,
 alongside various other absorption and emission features from stellar
 photospheres and winds, and gas photoionized by hot stars and AGN
 (including Si{\footnotesize II}$\lambda$1260,
 the blend O{\footnotesize I}$+$Si{\footnotesize II}$\lambda$1303,
 C{\footnotesize II}$\lambda$1334,
 Si{\footnotesize IV}$\lambda\lambda$1393,1402,
 C{\footnotesize IV}$\lambda\lambda$1548,1550,
 Fe{\footnotesize II}$\lambda$1608,
 Al{\footnotesize II}$\lambda$1670);
\item weaker but important interstellar
 Mg{\footnotesize II}$\lambda\lambda$2796,2803
 absorption (another common ISM and outflow diagnostic) and the faint auroral
 [O{\footnotesize III}]$\lambda$4363 line (a temperature-sensitive indicator in
 direct-method gas metallicity estimates).
\end{itemize} \vspace{-1.5ex}
Figure~\ref{surveys.fig} illustrates the ladder of surveys in terms of spectral
resolution vs.\ the number of galaxies within the $1 < z < 3$ interval of
interest for this article.  The full list of surveys and main references
are compiled in the Supplemental Tables 1 and 2.

\subsection{Photometric Surveys in the Optical to Near-/Mid-infrared}
 \label{surveys_phot.sec}

Imaging in multiple photometric bandpasses is the most efficient way to
identify and characterize large numbers of galaxies over a wide redshift
range.  Imaging campaigns at optical to mid-IR wavelengths
($\rm \lambda_{obs} \sim 0.3 - 8\,\mu m$) with sensitive cameras at
ground-based telescopes and from space with {\it HST\/} and
the {\it Spitzer Space Telescope\/} (hereafter {\it Spitzer\/})
have provided the most extensive censuses of distant galaxies.
At $z \sim 2$, the multi-color information is primarily sensitive
to the shape of the stellar continuum modulated by interstellar dust.
The spectral energy distribution (SED) of galaxies is used to derive
photometric redshifts (\zph) and basic properties such as stellar mass
and SFR (for techniques, see \citealt{Salv19} and \citealt{Conroy13},
respectively, and Supplemental Text).
\begin{marginnote}[2pt]
 \entry{SED}{Spectral energy distribution.}
 \entry{\begin{sansmath}$\bm{z}_{\scriptsize\textbf{phot}}$\end{sansmath}}
       {Photometric redshift,~based~on the broad/medium/
        narrow-band~SED.}
\entry{\begin{sansmath}$\bm{R}$\end{sansmath}~$\boldsymbol{=\lambda/\Delta\lambda}$}
       {\looseness=-2 Spectral resolution given as the~ratio of wavelength~to~the
        full-width at half-maximum of a filter bandpass or spectral line spread
        function.}
\end{marginnote}

The inclusion of near/mid-IR wavelengths has been crucial to the inventory
of the full population by detecting red optically-faint galaxies, probing
wavelengths where outshining by young hot stars and attenuation by dust are
reduced, and allowing to better trace the light from cooler stars that dominate
the stellar mass.
At $z \sim 2$, near-IR data are particularly important to gain leverage
from the fairly sharp Balmer/4000\,\AA\ continuum breaks.
Photometry in broad bandpasses is most sensitive but delivers coarse spectral
resolution with typically $R = \lambda/\Delta\lambda \sim 5 - 10$.
The addition of medium-band ($R \sim 10 - 20$) and narrow-band ($R$ up to
$\sim 100$) information has proven vital to improve the accuracy and reliability
of photometric redshifts and galaxy parameters \citep[e.g.,][]{Ilb09, Whi11}.
In the GOODS-S and COSMOS fields, with most extensive photometry in $\sim 40$
optical to mid-IR bands, \zph\ estimates are as good as
$\sim 0.01-0.05 \times (1+z)$, with $\la 5\%$ of catastrophic
outliers \citep[e.g.,][]{Ske14, Lai16}.
Because of the wide variety of galaxy SEDs, the accuracy depends on galaxy type,
redshift range, specific set of filters, observational depth, treatment of line
emission contributions, and availability of spectroscopic redshifts to calibrate
the \zph.
Nonetheless, the wider wavelength coverage and finer SED sampling in many survey
fields has brought decisive improvements.  The tracking of similar rest-frame
wavelengths across a broad range of redshifts allows more consistent comparisons
of galaxy properties at different cosmic times.  By better encompassing the full
diversity of galaxy SEDs, more complete samples can be selected on the basis of
photometric redshifts rather than color criteria involving a few bandpasses
devised to isolate specific populations, or of more fundamental galaxy
parameters such as stellar mass rather than brightness in a given filter with
important $k$-corrections.  As a result, more robust distribution functions in
terms of intrinsic galaxy properties and the evolution thereof have been derived,
such as rest-frame luminosity functions and stellar mass functions.

\looseness=-2
Multi-band $0\farcs 1 - 0\farcs 2$ resolution imaging with {\it HST\/} has been
increasingly exploited to not only detect distant galaxies and characterize their
sizes and morphologies on $\rm \sim 1~kpc$ scales, but also to derive maps of
stellar properties from resolved color information. 
Here, the CANDELS survey \citep{Gro11,Koe11} played a prominent role, bringing new
sensitive near-IR and optical imaging over $\rm \sim 800~arcmin^{2}$ distributed
in five premier sky regions within the GOODS-S and N, COSMOS, AEGIS, and UDS
footprints.
Together with imaging from other {\it HST\/} programs, this created a
multi-tiered data set from ultra-deep ($5\sigma$ depths of $\rm \sim 29 - 30~mag$)
full 9-band imaging over $\rm \sim 5~arcmin^{2}$ \citep{Illing13}, through deep
($\rm \sim 125~arcmin^{2}$) and wide ($\rm \sim 800~arcmin^{2}$) 4$-$7-band
imaging to typical $5\sigma$ depths of $\rm \ga 27~mag$, to the widest areas
from the $I$-band $\rm 1.7~deg^{2}$ mosaic as part of COSMOS
\citep[$\rm \sim 28~mag$, $5\sigma$;][]{Sco07} and $H$-band imaging of a
$\rm 0.66~deg^{2}$ sub-area ($\rm \sim 25~mag$, $5\sigma$) largely from the
COSMOS-DASH program \citep{Mowla19b}.
The deepest pencil-beam surveys, reaching $\rm 29 - 30~mag$ or fainter in
areas magnified through gravitational lensing by massive foreground galaxy
clusters \citep[e.g., the Hubble Frontier Fields (HFF); ][]{Lotz17} probe
$z \sim 2$ galaxies down to $\la 0.01~L^{\star}$ and masses well into the
dwarf regime.
At the other end, some recently undertaken very wide-area surveys such as the
optical$+$near-IR KiDS$+$VIKING \citep{Wright19} and optical Hyper Suprime-Cam
Subaru Strategic Program \citep[HSC-SSP;][]{Aih18} are deep enough to already yield
$\sim 10^{6}$ sources at $z \sim 2$ and $\ga 0.1 - 1~L^{\star}$ in the
first few hundreds of square degrees mapped.
\begin{marginnote}[]
\entry{\begin{sansmath}$\bm{L^{\boldsymbol{\star}}}$\end{sansmath}}
    {characteristic value of the galaxy luminosity distribu- tion
    described by a Schechter function:
    $\Phi(L) = 
    (\phi^{\star}/L^{\star}) \times (L/L^{\star})^{\alpha}\,e^{-L/L^{\star}}$.
    See \citet{March12} and \citet{Parsa16} for rest- optical and UV
    luminosity functions out to cosmic noon.}
\end{marginnote}


\subsection{Spectroscopic Surveys in the Optical to Near-infrared}
 \label{surveys_spec.sec}

\begin{marginnote}[30pt]
 \entry{\begin{sansmath}$\bm{z}_{\scriptsize\textbf{spec}}$\end{sansmath}}
       {Spectroscopic redshift,~based~on a spectrum (typically
        at $R > 200$).}
 \entry{S/N}{Signal-to-noise ratio.}
\end{marginnote}

Spectroscopic redshifts (\zsp) are essential to validate and optimize \zph\
techniques, construct the most precise galaxy distribution functions from
confirmed samples, and provide secure targets for detailed and time-consuming
follow-ups.  Spectroscopy at $R > 200$ is adequate to measure redshifts to
within $\rm \sim 300~km\,s^{-1}$ or better from ISM emission lines and/or
from stellar absorption features.
To be secure, \zsp's rely on the identification of at least two spectral
features\footnote{
 A distinct profile such as the characteristic asymmetry of Ly$\alpha$ or
 doublets such as \oii\,$\lambda\lambda$3726,3729 can be sufficient if these
 lines are observed.
},
and the success also depends on the signal-to-noise ratio (S/N) of the data,
the wavelength range probed, and the galaxy type.  For instance, it is easier
to measure the redshift of a source with higher emission line or continuum
surface brightness, introducing a notorious bias towards bluer, more compact,
more star-forming galaxies at $z \ga 1.5$ in optical spectroscopic surveys.
The challenges of confirming large samples at $z \sim 2$ are manifold.
The galaxies are faint.  At $z = 2$, $L_{\rm UV}^{\star}$ corresponds to
$R \sim 24.5~{\rm mag}$ and $L_{V}^{\star}$ to $H \sim 22.3~{\rm mag}$,
often necessitating long integrations to reach a sufficient S/N for \zsp\
measurements even with 8\,m-class telescopes.
Absorption and emission features observable in the optical are
typically weak.  The stronger nebular emission lines are shifted into the
near-IR regime that is plagued by a dense forest of $> 1000$ bright and
variable emission lines mostly from OH radicals excited in the upper
atmosphere, broad intervals of low atmospheric transmission around 1.4
and 1.9\,\micron, and thermal background from instrument to infrastructure
and atmosphere at $\rm \lambda_{\rm obs} > 2~\mu m$.  

\begin{marginnote}[150pt]
 \entry{MOS}{Multi-object spectroscopy, or spectrograph.}
 \entry{VLT}{Very Large Telescope.}
 \entry{\begin{sansmath}$\bm{z}_{\scriptsize\textbf{grism}}$\end{sansmath}}
       {Redshift from grism spectroscopy, here~specifically~from {\it HST\/}
        $R \sim 130$ grism data supplemented with photometric SEDs.}
\end{marginnote}

In the optical, great progress has come from high throughput multi-object
spectrographs (MOS) such as Keck/LRIS and DEIMOS and VLT/VIMOS and FORS2,
optimized to extend bluewards to the atmospheric cutoff near 3000\,\AA\
or redwards to $\rm \sim 1~\mu m$ to overcome the ``redshift desert.''
The more recent arrival of sensitive cryogenic near-IR MOS, including Keck/MOSFIRE
and Subaru/FMOS and MOIRCS, further expanded confirmed $z \sim 2$
samples mainly through rest-optical emission lines.
Near-IR observations from space have an obvious advantage and use of the
{\it HST\/}/WFC3 grism G141 with $R \sim 130$ has been very productive at
yielding redshifts.  The lack of atmosphere ensures continuous coverage of the
full $\rm \lambda_{obs} = 1.1 - 1.7~\mu m$ grism window and greatly enhances
continuum sensitivity, reducing biases towards line-emitting sources.  The
slitless aperture maximizes multiplexing and avoids target pre-selection
biases, with the added ability to map spectral features at {\it HST\/}'s
angular resolution.
Reliable grism redshifts (\zgr) from the 3D-HST and AGHAST programs \citep{Mom16},
for instance, have nearly tripled the number of $1 < z < 3$ galaxies with secure
spectroscopic redshifts in the five CANDELS fields, with typical \zgr\ accuracy
of $0.003\,\times (1+z)$ ($\rm \sim 1000~km\,s^{-1}$ at $z \sim 2$) at
$JH \leq 24~{\rm mag}$, and only $2 - 3\times$ worse for the subset
of quiescent galaxies.

\looseness=-2
Besides redshift, spectra also provide a wealth of information on the
stellar, gas, dust, and AGN content of galaxies.  Detailed information is
more demanding in terms of S/N and spectral resolution, to measure accurate
emission and absorption line strengths and profiles for a range of fluxes and
equivalent widths, and to deblend spectral features (e.g., \oii\ and \sii\
doublets, or kinematic components such as host disk and gas outflow).
Among many results from MOS surveys at $z \sim 2$, scaling relations have
been constructed such as the MS using SFR estimates from Balmer lines or UV
luminosities, and the mass-metallicity(-SFR) relationship from strong line
diagnostics of the gas-phase oxygen abundance.
Excitation sequences in nebular line ratio diagrams have been examined to
characterize the evolving ISM conditions at high $z$.
Galaxy kinematics have been investigated from integrated line widths and, with
data subsets of sufficient spatial resolution and suitable slit alignment, from
velocity gradients.
The demographics and energetics of galactic outflows have been investigated
from the strength and velocity profile of rest-UV interstellar absorption and
rest-optical nebular line emission.
In addition, stellar and dynamical properties of smaller but important samples
of massive quiescent galaxies have been constrained from absorption (and in
some cases weak emission) features -- valuable to establish the fate of
massive SFGs from their likely immediate descendants.
These results are discussed throughout Sections\ \ref{global.sec},
\ref{resolved.sec}, and \ref{otherpops.sec}.

\subsection{Integral Field Spectroscopic Surveys}
   \label{surveys_ifu.sec}

Imaging spectroscopy at $R \ga 2000$ arguably provides the richest
datasets of individual sources --- a large multiplexing of its own.
Integral field unit (IFU) spectroscopy is the most efficient technique ---
collecting simultaneously the full three-dimensional (3D) spatial and spectral
information --- and became possible for $z \sim 2$ SFGs (with typical H$\alpha$
fluxes of $\rm \sim 10^{-16}~erg\,s^{-1}\,cm^{-2}$
or fainter) with sensitive near-IR
IFU instruments at 8\,m-class telescopes.  IFU studies have so far mainly used
H$\alpha$$+$\nii\ line emission (or \oiii $+$H$\beta$ at $z \geq 2.7$) to map
the internal gas motions of galaxies, the distribution of star formation, gas
excitation, and ISM metallicities within them, and the extent and properties
of the gaseous winds they expel.  Key results on these topics are covered in
Sections\ \ref{global.sec} and \ref{resolved.sec}.

\begin{marginnote}[90pt]
 \entry{IFU}{Integral field unit.}
 \entry{AO}{Adaptive optics.}
\end{marginnote}

\looseness=-2
First samples were obtained with single-IFUs including
VLT/SINFONI, Keck/OSIRIS, and Gemini/NIFS, all with resolving powers of
$R \sim 2000 - 5000$ and designed to be fed by an adaptive optics (AO)
system improving the angular resolution from typical near-IR seeing of
$\sim 0\farcs 5 - 0\farcs 7$ at their sites to the diffraction limit of their
host telescopes ($\rm \sim 50 - 60~mas$ at $\rm \lambda_{\rm obs} = 2~\mu m$).
To date, near-IR single-IFU samples amount to $\sim 400$ targets altogether,
with roughly half of these sources observed in AO mode.
These samples, all drawn from spectroscopically-confirmed subsets of parent
photometric samples with diverse primary selection criteria (magnitudes,
colors, narrow-band identification, strong lensing) form a heterogeneous
collection probing different parts in $z - M_{\star} - {\rm SFR}$ space
\citep[e.g.,][and references therein]{Glaze13,FS18}.
Larger and more complete surveys have been enabled with the advent in 2013 of
KMOS at the VLT, with 24 IFUs deployable over a $7^{\prime}$-diameter patrol
field.  KMOS operates in natural seeing, covers $\rm 0.8-2.4~\mu m$ with four
bandpasses at $R \sim 4000$ each, and is well suited to detect faint, extended
line emission over a wide redshift span.
With $> 2000$ SFGs targeted so far, KMOS has put results from single-IFU work
on a more robust statistical footing \citep[e.g.,][]{Wis15,Wis19,Har16,Har17}.
Importantly, it has also allowed to push into regimes previously unexplored
with IFUs, including line emission of massive sub-MS galaxies \citep{Belli17b},
and continuum for stellar populations and kinematics of massive quiescent field
and cluster galaxies \citep[e.g.,][]{Mendel15,Bei17}.

\begin{figure}[!t]
\centering
\includegraphics[scale=0.55,trim={4.0cm 3.0cm 3.0cm 3.5cm},clip=1,angle=0]
                {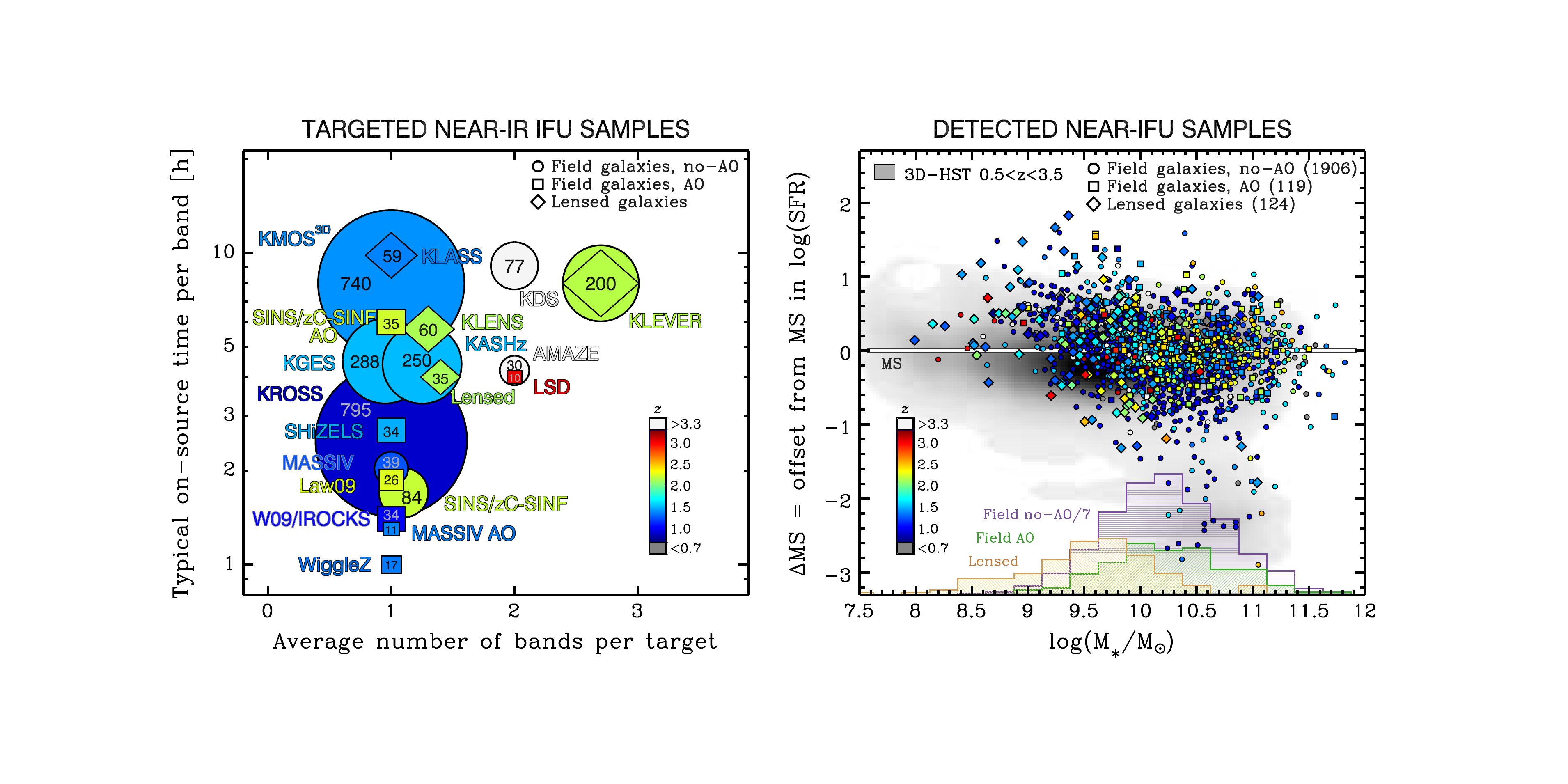}
\renewcommand\baselinestretch{0.85}
\caption{
Overview of near-IR IFU surveys of line emission of $z \sim 1 - 3$ galaxies
in observational and galaxy parameter space
(see list in Supplemental Table~2).
{\it Left:\/} The surveys are represented in terms of the average number of
spectral bands covered per target, median on-source integration time per band
(on a logarithmic scale), and number of objects targeted (with symbol area
proportional to this number).
Samples of field galaxies observed in natural seeing or with AO, and lensed
galaxies in either mode, are plotted as circles, squares, and diamonds,
respectively.  The color coding denotes the median redshift of the samples.
{\it Right:\/} Distribution of detected targets in stellar mass vs. MS offset
for samples where these properties are available.  Different colors and symbols
are used to show the redshift, and differentiate field vs. lensed galaxies,
and no-AO vs. AO data, as labeled in the plot.  The underlying distribution
of galaxies in a similar $z$ range to $H_{160} = 26.5~{\rm mag}$ from the
3D-HST catalog is shown in greyscale for comparison.  The histograms compare
the projected distributions in $M_{\star}$ of field galaxies observed in natural
seeing (scaled by $\times 1/7$, in purple), field galaxies observed with AO
(green), and lensed galaxies (yellow).
}
\label{IFUsurveys.fig}
\end{figure}

Figure~\ref{IFUsurveys.fig} illustrates the observational and galaxy parameter
space of the main near-IR IFU surveys of rest-optical line emission.
The $\log(M_{\star}/{\rm M_{\odot}}) \ga 9.5$ SFG population is extensively
covered; detections also extend to SFRs $\sim 10\times$ or more below the MS,
and to $\log(M_{\star}/{\rm M_{\odot}}) \la 9$ preferentially above the MS.
While optical IFU spectroscopy is more relevant to $z \la 1$ studies, synergies
are increasingly exploited by combining near-IR IFU samples with recent optical
wide-field IFU surveys with VLT/MUSE at intermediate redshifts and large
$z \la 0.15$ samples such as SAMI, CALIFA, and MaNGA observed with panoramic
IFUs.
Together, these surveys enable consistent comparisons and evolutionary links
based on fully resolved kinematic and emission line properties from the same
diagnostics over the past 11~Gyr of cosmic time.

\subsection{Other Wavelengths as Probes of total SFRs, Cold Gas, and AGN}
  \label{surveys_other.sec}

\subsubsection{Infrared observations}
  \label{surveys_other_ir.sec}

\looseness=-2
Space-borne mid- to far-IR photometry with {\it Spitzer\/}/MIPS
(at 24, 70, $\rm 160\,\mu m$) and then with {\it Herschel\/}
(with PACS at 70, 100, $\rm 160\,\mu m$ and SPIRE at 250, 350, $\rm 500\,\mu m$)
revolutionized IR surveys of distant galaxies thanks to their much improved
sensitivity, angular resolution, and mapping speed compared to previous missions.
They opened the window to statistical censuses of the dust-obscured component
of the stellar and AGN radiation output from galaxies in the form of thermal
emission \citep[see][for a review]{Lut14}.
{\it Spitzer\/}/MIPS has delivered the deepest views of the dusty ISM of
cosmic noon galaxies through 24\,\micron\ observations, enabling the detection
of individual galaxies at $z \sim 2$ down to $\rm SFR \sim 10~M_{\odot}\,yr^{-1}$
\citep[e.g.,][]{Whi14}.  At these redshifts, however, 24\,\micron\ data
measures rest-frame 8\,\micron\ light, where warm and transiently heated dust
in \hii\ regions and around AGN, polycyclic aromatic hydrocarbons arising from
photodissociation regions, and absorption by silicate dust contribute.
The conversion to total IR luminosity and SFR is thus prone to important
uncertainties from the SED that needs to be assumed for the large extrapolation
over the far-IR dust emission peak typically around rest-frame 100\,\micron\
(corresponding to a characteristic dust temperature of $\rm \sim 30~K$).
Measurements with submm instruments (e.g., JCMT/SCUBA and APEX/LABOCA) provided
useful constraints on the Rayleigh-Jeans side, where AGN heating is minimized.
The wavelength coverage and sensitivity afforded by {\it Herschel\/} has been
vital in sampling directly the far-IR SED peak, enabling robust calorimetric
estimates of galaxy SFRs (and cold dust properties).

\subsubsection{Submm to mm observations}
  \label{surveys_other_submm.sec}

\looseness=-2
Observations in the submm to mm regimes probe the cold ISM component in galaxies.
Its main constituent, $\rm H_{\rm 2}$, lacks a permanent electric dipole moment
hence relevant emission lines at low excitation
temperatures.\begin{marginnote}[0pt]
 \entry{(Sub)mm}{Submillimeter and~millimeter.}
\end{marginnote}Therefore, the
strong rotational lines of CO (the second most abundant molecule) are used to
trace molecular gas properties and kinematics of galaxies, with mid-$J$
transitions (2-1, 3-2, 4-3) being commonly employed at $z \sim 2$.
Molecular gas masses (hereafter simply $M_{\rm gas}$) are estimated via an
excitation correction to the ground-state 1-0 line and a conversion from CO
line luminosity to $\rm H_{\rm 2}$ mass \citep[e.g.,][]{Bolatto13,Genz15}.
The CO 1-0 transition is typically fainter than mid-$J$ lines and is shifted
into the high-frequency radio bands, accessible for instance with the JVLA.
The cold dust continuum luminosity is a viable and observationally efficient
proxy for the gas mass and spatial distribution
\citep[e.g.,][]{Sco17,Tac18}.

Great strides in cold ISM studies of $z > 1$ galaxies have been made possible
with the IRAM/NOEMA interferometer in the northern hemisphere and ALMA in the
south.  With the gains in sensitivity and angular resolution of these arrays,
studies of the global cold ISM content have shifted from the most luminous
``submm galaxies'' to the more typical MS population, although substantial
integration times are still needed especially for CO line measurements and
the limited primary beam sizes hamper mapping of sizeable areas
\citep[see recent reviews by][]{Com18,Tac20}.
Pointed CO or continuum surveys have been most efficient at assembling
sets of $\sim 10 - 100$ at $z \sim 2$ drawn from well-characterized parent
samples.
Blank field mosaicking surveys have been undertaken to build censuses out to
$z \sim 4$, either optimized for emission line searches through spectral scans
or emphasizing dust continuum emission, yielding so far a few to several 10's
of secure detections and with counterparts in the (deep) optical to mid-IR
imaging available in the survey fields.
Most of the CO and dust continuum measurements at $z \sim 2$ are for massive
$\log(M_{\star}/{\rm M_{\odot}}) \ga 10-10.5$ SFGs.  Detection in less massive
(unlensed) galaxies becomes increasingly difficult as the amount of gas gets
lower, and the ISM metallicity drops leading to more extensive UV
photodissociation of CO and lower dust abundances.
Alternative cold ISM tracers in emission are not practical
because of their weakness, or because their higher frequency make
them difficult or impossible to access from the ground at these redshifts.
An obvious avenue for the future, in reach of NOEMA and ALMA, is more
systematic spatially-resolved sub-arcsec CO and dust mapping at $z \sim 2$,
which is currently limited to fairly small heteregenous sets dominated by
very luminous or massive galaxies
\citep[e.g.,][]{Tac13,Silv15,Bar17b,Tad17a,Tad17b}.

\subsubsection{Radio observations}
  \label{surveys_other_radio.sec}

At longer radio wavelengths, continuum observations probe AGN and star-forming
systems mainly through non-thermal synchrotron emission, free from dust/gas
obscuration.
In SFGs, the synchrotron emission is produced in supernova remnants and,
towards higher frequencies, free-free emission from \hii\ regions
also contributes \citep{Condon92}.
In AGN sources, the origin is more diverse, including jets, hotspots, and
large-scale lobes, which complicates the quantitative relationship between
observed radio emission and AGN luminosity \citep[e.g.,][]{Tadh16}.
Surveys at $\rm 1.4 - 5~GHz$ and lower frequencies down to $\rm \sim 200~MHz$
($\rm \lambda \sim 6 - 150~cm$) with facilities such as the JVLA, VLBA, LOFAR,
GMRT, GBT, and ATCA have been carried out in many cosmological deep fields, with
a range of sensitivities, beam size, and area.  AGN dominate at brighter flux
densities while SFGs become increasingly important at sub-mJy levels.
Given that the tight radio-IR luminosity correlation for SFGs holds
out to at least $z \sim 3$, with fairly well constrained (mild) evolution
($L_{\rm IR}/L_{\rm 1.4\,GHz} \propto (1 + z)^\alpha$, and $\alpha$ in the
range $-0.1$ to $-0.2$), the radio flux density can serve as SFR estimator,
and the radio excess above the correlation can be used as diagnostic for the
presence of an AGN \citep[e.g.,][]{Magn15,Delh17}.
The deepest GHz-regime VLA imaging at $\sim 1$\arcsec\ resolution
\citep[in AEGIS, GOODS-N, COSMOS;][]{Ivi07,Morr10,Smol17} reaches
$5\sigma$ sensitivities of $\rm \sim 10 - 25~\mu\,Jy$, corresponding
to $\rm SFRs \sim 100~M_{\odot}\,yr^{-1}$ at $z \sim 2$.

\subsubsection{X-ray observations}
  \label{surveys_other_xray.sec}

At the other end of the spectrum, observations of X-ray radiation
($\rm 0.5 - 100~keV$) in galaxies trace predominantly nuclear activity
\citep[e.g.,][]{Brandt15}.
Produced in the immediate vicinity of the SMBH via Compton up-scattering in
the accretion-disk corona, in powerful nuclear jets, and via Compton reflection
and scattering interaction processes with matter throughout the nuclear regions,
X-rays are able to penetrate through substantial gas columns (becoming hindered
in the highly Compton-thick regime with
$N_{\rm H} \gg 1.5 \times 10^{24}~{\rm cm^{-2}}$).
Non-AGN X-ray emission in galaxies arises from X-ray binaries and hot gas
but is both less energetic and softer compared to that of (luminous) AGNs.
The most extensive cosmological surveys have been carried out with the
space-borne Chandra and XMM-Newton Observatories, operating since 20 years,
with on-board instruments enabling efficient spectroscopic imaging of wide
areas in soft and hard bands ($\sim 0.2 - 2$ and $\rm 2 - 10~keV$),
while the more recently launched NuSTAR telescope has started to unveil the
distant universe in $\rm \ga 10~keV$ radiation.
The deepest and sharpest views were achieved with Chandra/ACIS through the
cumulative 7~Ms exposure of $\rm \sim 485~arcmin^{2}$ in the Chandra Deep
Field-South (encompassing GOODS-S), yielding nearly 1000 detections to
$z \sim 4.5$.  While AGN dominate the source counts, the more so at higher
$z$ and luminosities, the depth of the data reaches intrinsic rest-frame
$\rm 0.5 - 7~keV$ $\log(L_{\rm X}/{\rm erg\,s^{-1}}) \la 42.5$ out to
$z \sim 3$ \citep{Luo17}, where SFRs from several 100 to
$\rm 1000~M_{\odot}\,yr^{-1}$ can be detected.
Because the rapid variability of AGN emission at these energies and the
potential presence of high absorbing gas columns near the nucleus can bias
X-ray samples, AGN identification benefits from complementary diagnostics
such as high-excitation rest-UV/optical emission lines, radio luminosity,
and mid-IR colors \citep[e.g.,][]{Pad17}.

\subsection{Mass-matching vs.\ Abundance-matching}
 \label{techmeth_popmatch.sec}

In order to bring theoretical models and observations into the same arena
for an apples-to-apples comparison, a common interface needs to be found.
Different approaches can be employed, bringing this interface either very close
to the direct observables (e.g., by treating numerical simulations with radiative
transfer and/or placing them into a lightcone to predict number counts as a
function of observed flux) or alternatively working from the observables backward
to interpret them in terms of physical quantities (e.g., using the spectral
modeling techniques outlined in the Supplemental Text).
Once stellar population properties such as stellar masses are inferred from
the multi-wavelength SEDs, and provided sufficient depth, mass-complete samples
of galaxies can be extracted from the flux-limited parent catalog.  Those in turn
can serve as basis for population-averaged comparisons to models, where for example
the evolution of the SFR, size, metallicity, rotational velocity or other physical
quantity is traced as a function of redshift at fixed stellar mass.  How far back
the population-averaged evolution can be recovered depends on the mass regime
considered, as the parent catalog's flux limit will necessarily impose a
redshift-dependent mass completeness limit.

\looseness=-2
While valuable, such population censuses
do not by themselves reflect the growth histories of individual systems.
Galaxies gain stellar mass through star formation and merging activity,
moving out of the considered mass bin while others move in.  Methods to
empirically reconstruct evolutionary sequences for individual galaxies
from the mass-complete samples, linking their progenitor and
descendant phases, gained significant attention in recent years.  The
most common ansatz is to assume the preservation of mass ranking, in
which case progenitors and descendants are anticipated to live at the same
comoving number density \citep[e.g.,][]{vD10}.  The resulting
evolutionary tracks can then directly be compared to the main progenitor
branch extracted from a galaxy formation model.

The efficacy of this technique relies in part on the infrequent occurrence
of major galaxy mergers, and indeed refinements have been proposed on the basis
of cosmological simulations to account for a non-negligible divergence in growth
rates, in part influenced by merging activity
\citep{vdVoort16, Wellons17, Clauwens17}.  Here, it is of note that slightly
different prescriptions are desired for tracing galaxies backwards vs.\
forward \citep{Torrey17, Wellons17}, and that the technique is designed
primarily to work when the galaxy population is well described as a
one-parameter family characterized by stellar mass.
If considering subpopulations defined by, e.g., mass and color, galaxies may
not only enter a particular mass bin due to their stellar growth, but also
their color evolution, potentially introducing progenitor bias
\citep[e.g.,][]{Carollo13}.
Finally, from the perspective of the flux-limited parent catalog, the abundance
matching technique leaves more of the collected data unused, as higher mass cuts
are adopted at later times to identify progenitors and descendants, whereas the
deepest mass completeness limits are reached at the lowest redshifts.

\begin{textbox}[!t]
\section{STELLAR, GAS, AND STRUCTURAL PROPERTIES}
\noindent
\begin{sansmath}$\bm{M}_{\boldsymbol{\star}}$, $\bm{M_{\scriptsize\textbf{gas}}}$,
  $\bm{M_{\scriptsize\textbf{bar}}}$\end{sansmath}:
  Total stellar, cold molecular gas, and baryonic (stellar$+$gas) masses. \\
{\bf Schechter function:}
  {\looseness=-2 Parametrization of the galaxy number density
  vs.\ stellar mass (or luminosity), with
  $\Phi(M)$$=$$(\phi^{\star}/M^{\star})\,(M/M^{\star})^{\alpha}\,e^{-M/M^{\star}}$
  or
 $\left[(\phi_{1}^{\star}/M^{\star})\,(M/M^{\star})^{\alpha_{1}} +
 (\phi_{2}^{\star}/M^{\star})\,(M/M^{\star})^{\alpha_{2}}\right]\,e^{-M/M^{\star}}$
  (single- or double-Schechter);
  the characteristic value $M^{\star}$ is referred to as the Schechter mass
  \citep{Schechter76}.} \\
{\bf sSFR}: specific star formation rate, ${\rm SFR}/M_{\star}$. \\
$\boldsymbol{\Delta}{\textbf{MS}}$:
  Logarithmic offset in sSFR (or SFR) from the MS,
  $\log({\rm sSFR}/{\rm sSFR_{MS}}(M_{\star},z))$. \\
{\looseness=-2 {\bf SFH}: 
  Star formation history; common forms include an exponential
  $\propto$\,$e^{-t/\tau}$, delayed $\propto$\,$te^{-t/\tau}$, or log-normal
  $\propto$\,$(t\sqrt{2\pi\tau^{2}})^{-1}\,e^{-0.5(\ln(t)-t_0)^{2}/\tau^{2}}$
  SFR where $t$, $\tau$, and $t_{0}$ are the time, timescale, and logarithmic
  delay~time.} \\
{\bf S\'ersic profile:} Frequently used parametrization of the surface
  density distribution of galaxies,
  $\Sigma(r) =$ 
   $\Sigma(R_{\rm e}) \exp$$\left[-b_{n}\left((r/R_{\rm e})^{1/n}-1\right)\right]$,
  where $n$ is the S\'ersic index, and $b_{n}$ is a scaling coupled to $n$ such
  that half of the total light is within $R_{\rm e}$ \citep[e.g.,][]{Graham05}.
  The Gaussian, exponential, and de Vaucouleurs profiles correspond to
  $n$=0.5 and $b_{n}$=0.69, $n$=1 and $b_{n}$=1.68, and $n$=4 and $b_{n}$=7.67,
  respectively. \\
\begin{sansmath}$\bm{R_{\scriptsize\textbf{e}}}$\end{sansmath}:
  Effective radius, enclosing half the total light (or mass). \\
\begin{sansmath}$\bm{R_{\scriptsize\textbf{d}}}$\end{sansmath}:
  Disk scalelength for an exponential profile
  $\Sigma(r)$ $= \Sigma(0)\,\exp(-r/R_{\rm d})$, in which case
  $R_{\rm e} = 1.68\,R_{\rm d}$. \\
\begin{sansmath}$\bm{b\boldsymbol{/}a}$\end{sansmath}:
  Projected minor-to-major axis ratio of an inclined disk (also denoted $q$). \\
\begin{sansmath}$\bm{R_{\scriptsize\textbf{e,circ}}}$\end{sansmath}:
  Circularized effective radius, scaling $R_{\rm e}$ by $\sqrt{b/a}$. \\
\begin{sansmath}$\bm{R_{\boldsymbol{80}}}$\end{sansmath}:
  Radius enclosing 80\% of the total light (or mass). \\
$\boldsymbol{\Sigma_{\star}}$, $\boldsymbol{\Sigma}_{\scriptsize\textbf{gas}}$,
  $\boldsymbol{\Sigma}_{\scriptsize\textbf{SFR}}$:
  Stellar mass, gas mass, and SFR surface densities conventionally within
  $R_{\rm e}$, taking half the total $M_{\star}$, $M_{\rm gas}$, and
  SFR and dividing by $\pi R_{\rm e}^2$. \\
$\boldsymbol{\Sigma_{\scriptsize\textbf{1kpc}}}$:
  Stellar mass surface density within the central 1~kpc, 
  $M_{\star}({\rm <1kpc})\,/\,\pi$$({\rm 1kpc})^2$, where
  $M_{\star}({\rm <1kpc})$ is computed from the best-fit S\'ersic profile
  to the surface density distribution. \\
\begin{sansmath}$\bm{f_{\scriptsize\textbf{gas}}}$\end{sansmath},
 $\boldsymbol{\tau_{\scriptsize\textbf{depl}}}$:
  Gas-to-baryonic mass fraction $M_{\rm gas}/M_{\rm bar}$, and
  gas depletion time via star formation $M_{\rm gas}/{\rm SFR}$.
\end{textbox}

\section{GLOBAL PROPERTIES OF STAR-FORMING GALAXIES AT z ${\mathbf \sim}$ 2}
\label{global.sec}

Along with the cosmically integrated evolution of the SFR, stellar mass, and
SMBH accretion rate density \citep{Mad14}, a key outcome of lookback surveys
was to reveal and establish the existence of scaling relations between global
properties of galaxies out to at least $z \sim 2.5$, and a census of how they
are populated (often quantified by galaxy type).
In what follows, we first address the build-up of stellar mass in galaxies.
Section~\ref{global_MS.sec} considers the scaling relation between the
(in-situ) growth rate (SFR) and its time integral ($M_{\star}$, including
effects of stellar mass loss and merging),
followed by an overview of results on census
(Section~\ref{global_MF.sec}) and a discussion on the interpretation of
these joint observational constraints (Section~\ref{global_MSMF.sec}).
We then expand our scope to include global structural measures
(Section~\ref{global_ReM.sec}), ISM probes
(Sections~\ref{global_gas.sec} - \ref{global_metal.sec})
and nuclear activity (Section~\ref{global_AGN.sec}).

\subsection{The ``Main Sequence'' of Star-forming Galaxies}
 \label{global_MS.sec}

\looseness=-2
Locally, the existence of a strong correlation between the SFR and stellar
mass of galaxies was first established based on the vast number statistics
offered by the Sloan Digital Sky Survey \citep[SDSS;][]{Brinch04}.
Subsequent work on deep lookback surveys revealed that a similarly tight
and near-linear relation, dubbed the ``Main Sequence,''
was already in place since $z \sim 2$ \citep{Noe07,Elbaz07,Dad07}.
Its main change with cosmic epoch is one of rapid zero-point evolution.
For galaxies below $\rm 10^{10}~M_{\odot}$ the specific SFR evolves
as ${\rm sSFR} \propto (1+z)^{1.9}$ whereas more massive galaxies exhibit a
faster pace of evolution, with ${\rm sSFR} \propto (1+z)^{2.2 - 3.5}$ for
$\log(M_{\star}/{\rm M_{\odot}}) = 10.2 - 11.2$ \citep{Whi14}.

\begin{marginnote}[0pt]
 \entry{sSFR}{Specific star formation rate.}
\end{marginnote}

\begin{figure}[!t]
\centering
\includegraphics[scale=0.60,trim={3.5cm 2.5cm 2.0cm 0.0cm},clip=1,angle=0]
                {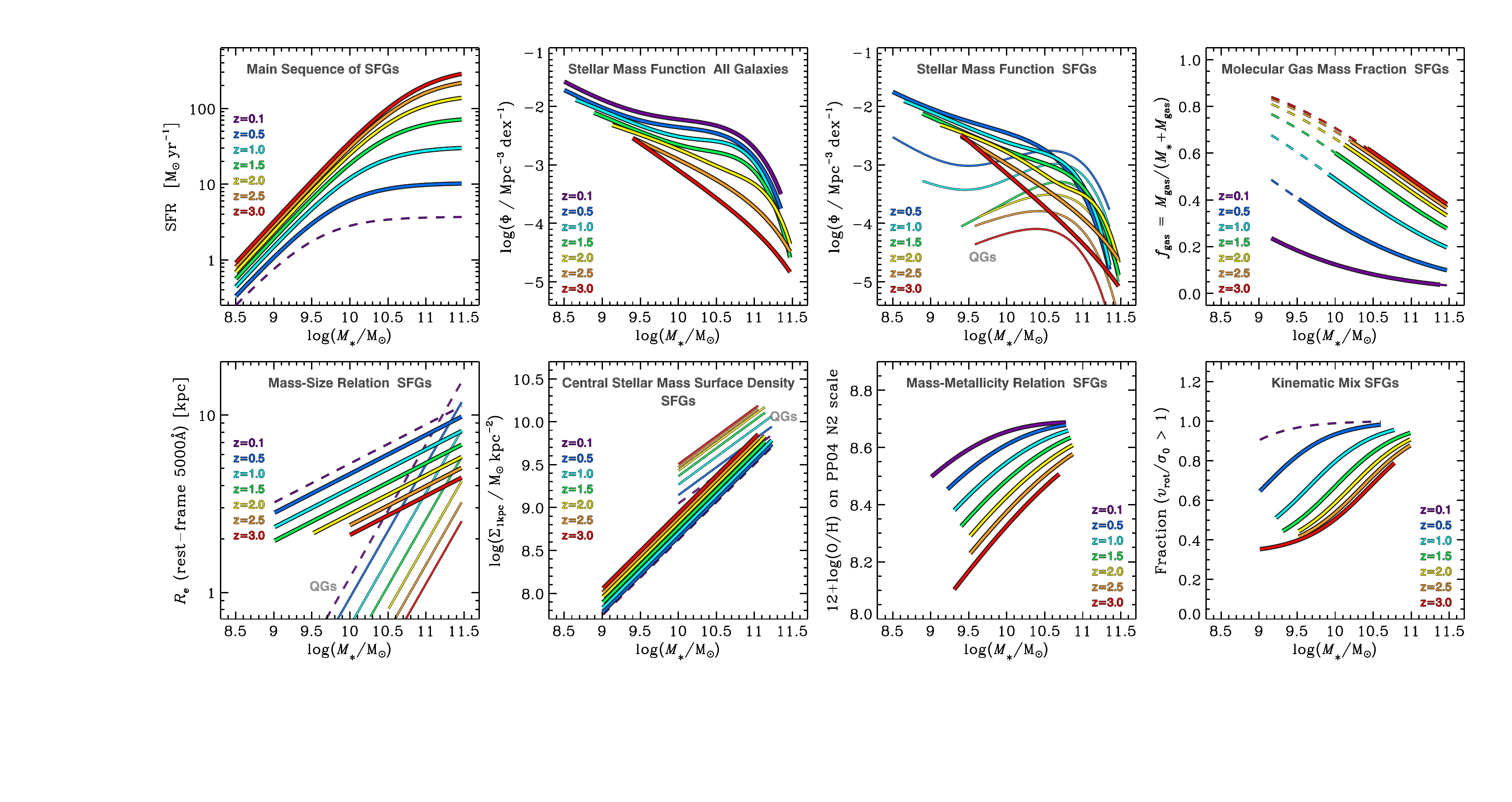} 
\caption{
Evolution of selected galaxy scaling relations and censuses from $z = 3$
to $z = 0.1$ (color-coded from red to purple as labeled in the panels).
{\it Top row:}
The first three panels show the MS of SFGs, the stellar mass function for all
galaxies and for SFGs and quiescent galaxies (QGs) separately (thick and thin
lines), consistently extracted from the same data set \citep{Tom14, Tom16}.
The rightmost panel plots the molecular gas mass fraction from the scaling
relations of \citet{Tac20}.  
{\it Bottom row:}
The two leftmost panels contrast the relationships between stellar mass,
size, and stellar mass surface density within $r < 1\,{\rm kpc}$ for SFGs
and QGs, respectively \citep[thick and thin lines;][]{vdWel14a,Bar17a}.
The next panel plots the stellar mass vs.\
gas-phase metallicity relation estimated via \niiha\ from
\citet{EWuy14} and \citet{Zahid14} (using the \citealt{Pet04} calibration).
The rightmost panel illustrates the evolution of the fraction of
rotation-dominated SFGs (with ratio of intrinsic rotation velocity
to velocity dispersion $> 1$) based on ionized gas kinematics
\citep{Kassin12,Sim17,Turner17,Wis19}.
In the various panels, dashed lines indicate extrapolations in $M_{\star}$
and/or $z$ when no consistent measurements or fits were available.
}
\label{MSMF.fig}
\end{figure}

\looseness=-2
The past few years have seen a consolidation of the MS relationship,
leading to an emerging picture in which (a) the scatter is constant at
$\rm 0.2 - 0.3~dex$ over the full stellar mass and redshift range probed,
(b) the low-mass slope is consistent with unity, and (c) a turnover and
flattening is evident at higher masses, most prominently so towards lower
redshifts and conversely nearly vanishing by $z \sim 2$
\citep[e.g.,][Figure~\ref{MSMF.fig}]{Whi14, Schreiber15, Tom16}.
Some studies favor or adopt single powerlaw fits \citep{Spe14, Pearson18},
then finding its slope to steepen with increasing redshift.

\looseness=-2
Quantitative differences in derived scatter, slope/shape and normalization
can be attributed to a range of reasons, including (1) method and strictness
of SFG selection, (2) dynamic range over which the relation is fit, and
(3) use of different SFR tracers.  We briefly elaborate on these systematics
before highlighting the significance of the MS scaling relation.  

\begin{marginnote}[130pt]
 \entry{\begin{sansmath}$\bm{UVJ}$\end{sansmath}}
       {Rest-frame $U-V$ vs.\ $V-J$ color diagram.}
\end{marginnote}

\looseness=-2
\citet{Whi12} demonstrate how a $UVJ$ color selection vs.\ selecting
only blue star-forming galaxies makes the difference between finding a
sub-linear vs.\ linear slope.  Similarly, \citet{Rod14} and \citet{Joh15}
illustrate how, by adopting different color cuts or selection criteria based
on SED-modeled properties, inferred slopes may vary between $\sim 0.8$ and
$\sim 1$.  Noteworthy also is that, when restricting samples to pure disks or
considering only the disk components of SFGs, a slope of unity is found
\citep{Abr14}.  As we will allude to in Section~\ref{global_MSMF.sec},
galaxies may well lack a bimodality in their sSFR distribution akin to that seen
in their colors, implying that the choice of SFG selection criterion may largely
be arbitrary.  In this case there is no formally correct answer regarding the MS
shape, and inferences on galaxy evolution need to treat the SFG and quiescent
population jointly or at least preserve internal consistency in selection
criteria used.

Unavoidably, the dynamic range in stellar mass over which the MS shape can be
constrained is a function of redshift, with for example the ZFOURGE magnitude
limit of $K_{\rm s} = 25~{\rm mag}$ corresponding to 90\% completeness limits of
$\log(M_{\star}/{\rm M_{\odot}}) \sim 8.5$, 9.5 and 10 at $z \sim 1$, 2 and 3,
respectively \citep{Tom16}.  Particularly if there is curvature to the MS
this can impact recovered parameterizations that adopt a power-law slope.
The finite depth of observed SFR tracers further implies that many studies
rely at least in part on stacking procedures, which may suffer from confusion
biases \citep{Pearson18}.

\looseness=-2
Even with extreme depth and a consistent SFG definition, determinations of the
MS scatter, normalization, and shape will be affected by uncertainties in the
inferred SFRs and stellar masses.  When the two are derived from overlapping
data and sometimes a single modeling procedure, the
uncertainties will be correlated and can potentially conspire to an
artificially tight relation, compensating the opposite and mostly subtle
boosting of scatter due to finite redshift bins that is not always
accounted for.  A comprehensive discussion of systematic uncertainties
affecting estimates of SFR and $M_{\star}$ is presented by
\citet[see also Supplemental Text, which summarizes the ingredients,
assumptions, and challenges of spectral modeling techniques]{Conroy13}.
Possible concerns include the saturation of reddening as a dust attenuation
tracer at the highly star-forming and massive end \citep[e.g.,][]{Wuy11a},
extra extinction towards \hii\ regions which remains difficult to pin down
observationally \citep[e.g.,][]{Red15},
contamination by other sources
of emission such as AGN, circumstellar dust around asymptotic giant branch
(AGB) stars or diffuse cirrus dust heated by old stellar populations,
and unintended biases induced by the choice of adopted
parameterization of and/or prior on the SFH \citep{Carn19, Lej19a}.

Despite the above considerations, the meta-analysis by \citet{Spe14} finds
a remarkable consensus among MS observations, with an interpublication scatter
as small as 0.1~dex.  On an individual galaxy basis different SFR estimates
do of course vary more than that, but may not need to agree in detail either,
as they can probe different timescales.  H recombination lines are the closest
to a measurement of the instantaneous SFR because of the
short lifetime ($\sim 10$ Myr) of Lyman-continuum producing OB stars, while
rest-UV and IR tracers will integrate the contribution of stars with (stellar)
main sequence lifetimes of $\sim 100$ Myr.  As such, differences in MS scatter
inferred from different SFR tracers could in principle encode the
short-term stochasticity of star formation and the timescale on which
galaxies lose ``memory'' of previous activity \citep{Cap19}.
A slightly enhanced scatter around the H$\alpha$-based MS at cosmic noon
relative to the one constructed from UV or UV+IR based diagnostics has
been reported \citep{Shi15, Belli17b}, but systematic uncertainties regarding
dust corrections make the interpretation in terms of star formation timescales
not unique.

\begin{marginnote}[295pt]
 \entry{SFH}{Star formation history.}
\end{marginnote}

\looseness=-2
Setting aside the above caveats, we conclude this Section by noting two
immediate implications of the existence of a MS relation and its observed
evolution with cosmic time.  First, assuming SFGs are located on the MS
at all times we can integrate along the evolving scaling relation to recover
the typical star formation history (SFH).  Doing so, one unambiguously finds
the star formation activity first rises before it falls, as such mimicking
the shape of the cosmic SFR density evolution
\citep[][]{Ren09,Pen10,Lei12,Spe14,Tom16,Cie17}.
In common with findings from the fossil record \citep[e.g.,][]{Thomas05},
these studies also infer SFHs of more massive galaxies to
peak earlier.  A second point of significance is that the tightness of the
MS implies that any large excursions in star formation activity as one
might expect from (major) merging have either very short duty cycles or are
very rare \citep{Rod11}.

\vspace{-1ex}

\subsection{The Stellar Mass Function}
 \label{global_MF.sec}

An extensive body of work has documented the census of galaxies as a function
of their stellar mass over most of cosmic history on the basis of well-sampled
SEDs for deep, near-IR selected samples
\citep[e.g.,][see Figure\ \ref{MSMF.fig}]{Ilb13,Tom14}.
In common between
these studies are the following findings.  First, provided sufficiently deep
stellar mass completeness limits a double-Schechter functional form is favored
over a single-Schechter fit.  This conclusion holds for both the star-forming
and quiescent population individually, and for the combined, total galaxy
stellar mass function.  Second, between $z \sim 2$ and the present day there
is no statistically significant evolution in the characteristic mass $M^{\star}$
of either the total stellar mass function or that of SFGs \citep{Pen10}.
Values quoted in the literature for this characteristic mass vary in the range
$\log(M^{\star}/{\rm M_{\odot}}) = 10.6 - 11$ with the higher results stemming
from single-Schechter and the lower ones from double-Schechter fits
\citep[see, e.g.,][]{Tom14}.
Minor
differences further arise from the adopted fitting method ($1/V_{\rm{max}}$ vs.\
maximum-likelihood), systematics in the determination of redshifts and stellar
mass, and how uncertainties in the latter are accounted for.  A third conclusion
is that little to no evolution in the low-mass slope $\alpha \sim -1.5$ is noted
since cosmic noon, neither for the total nor the star-forming galaxy
population.\footnote{
 In the case of a double-Schechter fit, we here refer to
 the steeper of the two fitted slopes $(\alpha_1, \alpha_2)$, which dominates
 at the low-mass end.}
Most of the evolution over the past 10 Gyr can thus be described by an increase in
$\Phi^{\star}$.  The redshift-invariance of the low-mass slope $\alpha$ is in line
with a MS slope of unity at masses below $\log(M_{\star}/{\rm M_{\odot}}) < 10.5$.
As pointed out by \citet{Pen10}, a sub-linear MS slope would inevitably lead to
a fast steepening of $\alpha$, and only very slight deviations from a unity
slope can be accommodated by merging away low-mass galaxies.
Whereas until recently inconsistencies at the level of $\rm 0.2 - 0.3~dex$
were found between integrating the MS metric and the evolving stellar mass
function \citep{Lej15, Tom16}, which could not all be accounted for by merging,
the latest such exercise with revised SFRs and stellar masses from advanced
spectral modeling shows an improved  internal consistency
\citep{Lej19b}.\footnote{
 All modeling combining measures of SFR and stellar mass accounts for stellar
 mass loss, which reduces the mass present in stars (including remnants)
 compared to the integral over the SFH, by a factor 0.6
 at late times for a Chabrier IMF.}

As illustrated in Figure\ \ref{MSMF.fig}, the quiescent
galaxy mass function looks markedly different.  At all epochs it features a
clear peak around $M^{\star}$, and the quiescent population grows in numbers more
rapidly than the star-forming one.  At masses above $10^{10}~{\rm M_{\odot}}$
quiescent number densities have grown by a factor 6 since $z \sim 2$, whereas
at lower masses there is a $15 - 30\times$ increase.
The mass-dependent growth of the quiescent population,
with quenching of low-mass galaxies happening at later times
\citep{Ilb13, Huang13}, has been attributed to two different quenching channels.
Since the environmentally driven low-mass channel only manifests itself
appreciably after the epoch of cosmic noon, this review focuses on the high-mass
quenching dominant at early times.

\subsection{Interpreting the Observed Stellar Mass Growth}
 \label{global_MSMF.sec}

Whereas observational campaigns of the stellar mass growth across most of
cosmic history have tightened the error bars on its scaling relation (the MS)
and census (the galaxy stellar mass function), perhaps of more debate today is
the interpretation of these observational diagnostics.  That is, what are the
implications of the cross-sectional view of the galaxy population at a range
of epochs for the evolutionary tracks that individual galaxies follow?

In this context two schools of thought have developed which, to use the
nomenclature of \citet{Abr15}, can be described as ``mean-based" and
``dispersion-based" approaches.  The former aim to reconstruct the average SFH
of individual galaxies based on ensemble averages \citep[e.g.,][]{Pen10,Beh13b},
whereas the latter put emphasis on the diversity of SFHs
\citep[e.g.,][]{Gla13,Kel14,Abr15,Abr16}.
Both schools infer characteristic SFHs that first rise
and then fall, but differ in key aspects of the interpretation.

\citet{Pen10} for example adopt the redshift-invariance of $M^{\star}$
as an indication that galaxies live on and grow along the evolving MS
until they reach this critical mass, after which the probability of
quenching increases rapidly $\propto 1-e^{-M/M^{\star}}$.
Proposed mechanisms to explain this ``mass quenching'' include the rapid
expulsion of gas by SMBHs, but there is no consensus yet regarding
its physical cause.  Scatter around the MS is in such models typically
attributed to short timescale ($\rm \sim 10^{7-8}~yr$) variations
in SFR at a given mass, induced by the breathing cycles of star
formation feedback and temporal fluctuations in the rate of gaseous
inflows and/or minor mergers.

\begin{marginnote}[180pt]
 \entry{DM}{Dark matter.}
\end{marginnote}

\looseness=-2
The ``dispersion-based" school on the other hand attributes scatter
around the MS relation as an imprint of SFHs that are differentiated on
Hubble timescales.  In this picture galaxies follow smooth trajectories
that let them pass across the moving MS, rather than at any time being
stochastically scattered around the scaling relation.  There is in such
a scenario no discernable signature of quenching.  That is, no rapid
quenching mode and no specific time (other than arguably the peak in SFH)
at which a shutdown in star formation is triggered.  Along the same vein,
\citet{Eal14} report a continuous distribution of galaxies in the
${\rm sSFR} - M_{\star}$ space, lacking a bimodality in specific SFRs as
undeniably seen in their color distribution.  The color bimodality, they
argue, reveals the peculiarities of stellar evolution (i.e., ageing stellar
populations saturating in color) rather than a signature of galaxy evolution
producing two sharply distinct populations of galaxies.
A common interpretation in these studies is that the SFH shape is set by initial
density conditions intimately related to dark matter (DM) properties such as
the halo formation redshift.  A family of log-normal SFHs,
parameterized by varying peak times and widths, can yield an adequate
description of the relevant observational metrics
\citep{Gla13, Abr15, Abr16, Die17}, although the fact that 
the central limit theorem produces a similar relation between
SFR and stellar mass within a framework in which galaxies grow stochastically
illustrates that this inference is not unique \citep{Kel14, Kel16}.

\citet{Spe14} present a hybrid approach in which average SFHs
are derived by integrating the MS similar to what was done by
\citet{Ren09} and \citet{Pen10}, but its scatter is reproduced by imposing
an initial spread in formation times to the smooth evolutionary tracks as
opposed to adding short-term fluctuations in SFR at a given mass.  Turning
to numerical simulations of galaxy formation where the individual evolutionary
paths of galaxies are by construction known, \citet[][]{Mat19}
argue that the MS scatter contains contributions from (slightly dominant)
short-timescale self-regulation of star formation as well as halo-related
variations on Hubble timescales.  Of course, the precise contribution from
short-timescale fluctuations may depend on the detailed recipes implemented
in the numerical simulation.  

A promising path forward to discriminate between the two schools
of thought is to look for correlations between the offset from
the MS midline and other SFG properties that can be
assumed to vary more slowly over time, such as galaxy structure.  Absence of
a correlation within the MS scatter would then favor a short-timescale origin
whereas a correlation between MS offset and longer lasting features would favor
a Hubble-timescale differentiation.  This requires accurate SFR
measurements, where possible contrasting MS offsets quantified using multiple SFR
tracers \citep[e.g.,][]{Fang18}, ideally with different timescale sensitivities
\citep{Cap19}.  Given challenges posed in this regard by dust treatment,
we conclude that SFRs and stellar masses
by themselves may ultimately prove insufficient to recover the underlying
evolutionary paths of galaxies.  Progress thus entails incorporating the
information provided by spatially resolved studies of the build-up of galaxies
in all their baryonic components (stars, gas, metals), tied with kinematic
tracers of the full gravitational potential (i.e., including DM)
and of the feedback processes at play.  In the remainder of this Section,
we cover the global structure, ISM, and accretion scaling relations, to delve
more into resolved properties in Section~\ref{resolved.sec}.

\subsection{The Mass-Size Relation}
 \label{global_ReM.sec}

Following initial work with {\it HST\/} on the sizes of the UV-selected
subpopulation of SFGs \citep{Gia96,Ferg04}, size evolution of mass-complete
samples since cosmic noon was first explored in large numbers using ground-based
near-IR surveys \citep{Fra08,Williams09}, to be transformed by rest-optical
imaging at high resolution for statistical samples after the installment of
WFC3 onboard {\it HST\/}.
This Section focuses on stellar light-weighted sizes.  Insights gained
from a multi-tracer analysis combining stellar mass-weighted sizes and
radial distributions of star formation, gas, and dust are covered in
Section\ \ref{resolved_axisymm.sec}.

\begin{marginnote}[275pt]
 \entry{PSF}{Point spread function.}
\end{marginnote}

\looseness=-2
Even when concentrating on a single tracer/wavelength,
multiple definitions of galaxy size are possible, and increasingly explored
alongside one another.  Different methods classify broadly as parametric and
non-parametric.  By far the most common approach entails fitting of a parametric
(usually S\'{e}rsic) functional form convolved with the point spread function
to the two-dimensional (2D) surface brightness distribution,
and adopting the radius enclosing 50\% of the light (a.k.a.\ the effective radius)
as size measure, either defined along the major axis or in circularized form
($R_{\rm{e,circ}} = \sqrt{b/a}~R_{\rm{e}}$).  Variations include
quantifying galaxy size based on a different percentile (e.g., $R_{80}$) or
decomposing the light distribution in multiple components (e.g., bulge and disk)
with a size associated to each.  Non-parametric approaches range from curve of
growth analyses to quantifying the pixel area above a given surface brightness
threshold.  The former requires a center and aperture definition, whereas the
latter is designed to function well also for highly irregular morphologies but
requires accounting for cosmological surface brightness dimming and luminosity
evolution.  Unlike the parametric approach that applies forward modeling of
point spread function (PSF) smearing,
the finite resolution is to be accounted for a posteriori in these
non-parametric measures, typically using a simulation-based lookup table as
correction factors are size and profile shape dependent.  Here, we outline the
main inferences from conventional S\'{e}rsic fitting, but note in passing
how some conclusions change, even on a qualitative level, when adopting an
alternative definition of size.

\looseness=-2
The sizes of star-forming and quiescent galaxies both show a tight
($< 0.2$ dex intrinsic scatter) but distinct scaling with galaxy stellar mass
\citep[][see Figure\ \ref{MSMF.fig}]{vdWel14a}.
SFGs are larger than their quiescent counterparts at all masses over the
$0 < z < 3$ range.  Their size-mass relation exhibits a non-evolving slope of
$\frac{d \log R_e}{d \log M_{\star}} = 0.22$ compared to the steeper slope of
$\frac{d \log R_e}{d \log M_{\star}} = 0.75$ for early-type galaxies.  Considering
the redshift dependence of the intercept, a slower evolution in the average size of
the population at fixed mass is quantified for SFGs ($R_e \propto (1+z)^{-0.75}$)
compared the quiescent systems, which as a population show dramatic growth from
compact red nuggets at cosmic noon to the large ellipticals in today's Universe
($R_e \propto (1+z)^{-1.48}$).  Of note is that the above characterizes the
evolution in the size distribution of the population, not by itself the
evolutionary tracks of individual galaxies.  Connecting progenitor-descendant
sequences based on their constant cumulative number density as outlined in
Section\ \ref{techmeth_popmatch.sec}, information from the evolving galaxy
stellar mass function can be folded in together with the size measurements to
infer that:
(1) the progenitors of present-day Milky Way mass galaxies have evolved,
on average, along individual growth tracks of
$\frac{\Delta \log R_e}{\Delta \log M_{\star}} = 0.27 - 0.3$ (i.e., an inside-out
growth track slightly steeper than the slope of the star-forming size-mass relation
at any epoch; \citealt{vD13,vD15}); and
(2) the most massive galaxies have experienced much steeper size growth with
individual tracks following $\frac{\Delta \log R_e}{\Delta \log M_{\star}} = 2$,
consistent with scenarios where an early dissipative core formation phase is
followed by the build-up of profile wings through dissipationless, predominantly
minor, mergers \citep{vD10, Patel13a}.

\begin{marginnote}[170pt]
 \entry{SMHM}{Stellar mass -- halo~mass~relation.}
\end{marginnote}

\looseness=-2
The formation of galactic disks is inherently linked to the DM halos
that host them.  In its simplest form, disk scalelengths are expected to scale
with the virial radii of their host halos as:
\begin{equation}
R_{\rm d} = 
 \frac{1}{\sqrt{2}} \left(\frac{j_{\rm d}}{m_{\rm d}}\right)\lambda\,r_{200},
\label{mo98.eq}
\end{equation}
which boils down to a linear scaling with the virial radius $r_{200}$ provided
the accreting baryons retain the specific angular momentum of their host halo
($j_{\rm d}/m_{\rm d} = 1$; \citealt{Mo98}).
The width of the log-normal distribution in
spin parameters $\lambda$ obtained from N-body simulations in a $\rm \Lambda CDM$
cosmology \citep{Bullock01} is sufficient to account for the observed scatter in
the size-mass relation.  Such a scenario predicts an evolution in size at fixed
halo mass following $R \propto H(z)^{-2/3}$, in agreement with the observed
evolution for late-type galaxies by \citet{vdWel14a}, who note that a
parameterization as a function of $H(z)$ is marginally favored over that with
the scale factor $(1+z)$.
Adopting the stellar mass - halo mass (SMHM) relation inferred
from abundance matching, the observed size - mass relation can be converted to a
galaxy size - halo size relation \citep{Kra13, Huang17, Som18}.  Applied to
observations at $0 < z < 3$, such analyses reveal a linear relation
between $R_{\rm e}$ and $r_{200}$ and hence evidence for homologous growth between
galaxies and their host halos.  At least at $0.5<z<3$ the normalization for
late-type galaxies is consistent with expectations from simple disk formation
models (see also Section\ \ref{resolved_dyn_angmom.sec} for kinematic evidence
of specific angular momentum retention in an ensemble-averaged sense).
The effective radii of early-type galaxies on the other hand lie below the
relation at all epochs.  \citet{Mowla19a} however
suggest that expressed in $R_{80}$ quiescent galaxies and SFGs occupy a single
size-mass relation, with these outer size measurements exhibiting
a close relationship to the host halos for the full population.
Whereas observations and simulations agree on a general linear
relation of the form $R_d = A\ r_{200}$, recent theoretical work has
called into question whether the proportionality constant $A$, and hence the
variation in galaxy size at fixed mass, is set by the halo spin parameter
$\lambda$ as in equation\ \ref{mo98.eq}, halo concentration \citep{Jiang19},
or a combination of both \citep{Som18}.

\begin{textbox}[!t]
\section{DARK MATTER HALO AND RELATED PROPERTIES}
\noindent
\begin{sansmath}$\bm{r_{\boldsymbol{200}}}$\end{sansmath}:
  {\looseness=-2
  Virial radius of a DM halo, usually the radius within which the
  mean mass density is 200 times the critical density for closure of the
  Universe at the redshift of interest; also denoted $R_{\rm vir}$
  \citep{Mo98}.} \\
$\boldsymbol{\lambda}$:
  Spin parameter of a DM halo \citep{Bullock01}. \\
\begin{sansmath}$\bm{M_{\scriptsize\textbf{DM}}}$,
                $\bm{J_{\scriptsize\textbf{DM}}}$,
                $\bm{j_{\scriptsize\textbf{DM}}}$\end{sansmath}:
  Mass of a DM halo, and its total and specific angular momentum at the
  virial radius (with $j_{\rm DM} = J_{\rm DM}/M_{\rm DM}$). \\
\begin{sansmath}$\bm{m_{\scriptsize\textbf{d}}}$,
                $\bm{j_{\scriptsize\textbf{d}}}$\end{sansmath}:
  Mass and angular momentum of the baryonic disk galaxy expressed as fractions
  of the host DM halo mass and angular momentum
  (such that $M_{\rm bar}=m_{\rm d}\,M_{\rm DM}$,
   $J_{\rm bar}=j_{\rm d}\,J_{\rm DM}$).
\end{textbox}

\begin{marginnote}[140pt]
 \entry{LBG}{\looseness=-2 Lyman break galaxy, selected based on its
             characteristic rest-UV spectral break.}
\end{marginnote}

Key in the above results is that they are based on mass-complete samples of
galaxies.  Individual sub-populations may differ in their growth rate.
\citet{Allen17} report a significantly faster size growth for
Lyman break galaxies (LBGs; $R_{\rm e} \propto (1+z)^{-1.2}$)
than for the underlying full SFG population since $z \sim 7$,
a trend also seen in previous studies spanning a more modest redshift range,
implying that LBGs represent a special subsample of highly star-forming and
compact galaxies.
Population differences aside, \citet{Ribeiro16} report for the same sample of
spectroscopically confirmed SFGs at $2 < z < 4.5$ differences in size evolution
at fixed mass ranging from $R_{\rm e} \propto (1+z)^{-1.4}$ using conventional
S\'{e}rsic profile fits to no size evolution at all over the considered 2 billion
years leading up to cosmic noon when adopting a non-parametric measure of size
quantified based on the pixel count above a threshold surface brightness.
They attribute this to galaxies in their earliest phase of assembly being quite
extended and irregular, and poorly described by a single S\'{e}rsic profile.
An example at later epochs where alternative size definitions change trends in
a qualitative manner includes work by \citet{Carollo13} who adopt curve-of-growth
sizes with a posteriori PSF correction factors to conclude, at odds with
\citet{vdWel14a}, that there is no decline in number densities of compact
quiescent galaxies since $z < 1.5$,
thus placing more emphasis on progenitor bias than individual
galaxy growth as an explanation of the observed size evolution of
early-type galaxies.

\subsection{Cold Gas Content}
 \label{global_gas.sec}

The cold gas reservoir of galaxies lies at the core of their evolution, fueling
their star formation activity and SMBH growth, and efficiently mediating mass,
angular momentum, and energy transfer.  CO line or far-IR to $\rm \sim 1\,mm$
dust continuum observations have accumulated ample evidence that SFGs at cosmic
noon have copious amounts of molecular gas \citep[see reviews by][]{Com18,Tac20}.
A recent focus has been on scaling relationships described in relation to the MS,
facilitating the interpretation in the framework of galaxy evolution and providing
well-calibrated recipes to estimate $M_{\rm gas}$ in the absence of actual cold
ISM measurements \citep[e.g.,][]{Genz15,Sco17,Tac18}.
These analyses showed that over $z \sim 0 - 4$ the depletion time
$\tau_{\rm depl} = M_{\rm gas}/{\rm SFR}$ depends primarily on redshift and MS
offset $\Delta{\rm MS} = \log({\rm sSFR}/{\rm sSFR}_{\rm MS}(M_{\star},z))$,
and so does the ratio of molecular gas to stellar mass $\mu_{\rm gas}$
with an additional dependence on $M_{\star}$.
In the updated derivation by \citet{Tac20}, unifying CO and dust continuum-based
gas mass estimates including most recent NOEMA and ALMA data, and adopting the
\citet{Spe14} MS parametrization,
$\log(\tau_{\rm depl}) = 0.21 - 0.98\log(1+z) - 0.49\,\Delta{\rm MS} +
     0.03\,(\log(M_{\star}/{\rm M_{\odot}}) - 10.7)$.
Accordingly, the depletion time for MS SFGs at fixed $M_{\star}$ increases by
a factor of $\sim 3$ from $z = 2$ to the present day while the gas fraction
$f_{\rm gas} = M_{\rm gas}/(M_{\star}+M_{\rm gas})$ drops by a factor of
$\sim 10$ (Figure\ \ref{MSMF.fig}).
It also follows from these gas scalings, the near-linear MS and
its evolution \citep[from][]{Spe14}, and the size-mass relation for SFGs
\citep[from][]{vdWel14a}, that the gas mass surface density at fixed
$M_{\star}$ evolves strongly over $0 < z < 2$ as
$\Sigma_{\rm gas} \propto (1 + z)^{a}$ with $a \sim 4$,
and more slowly at $2 < z < 4$ with $a \sim 2$.

At all epochs, the average gas depletion time is nearly ten times shorter than
the Hubble time, requiring sustained replenishment of the galactic cold gas
reservoirs to maintain the SFG population as a whole on the tight observed MS.
As summarized by \citet{Tac20}, this argument is a cornerstone of the
``equilibrium growth'' model, and favors that the bulk of SFGs are fed by
smoother gas accretion modes via cold streams along the cosmic web and minor
mergers rather than major mergers.
At fixed $M_{\star}$ and $z$, the gas scaling relations imply that the enhanced
SFRs well above the MS are driven by both higher gas fractions and higher star
formation efficiencies ($1/\tau_{\rm depl}$), plausibly reflecting increased
gas accretion and concentration as, e.g., in a major merger event.
On the MS, the star formation efficiency is roughly constant but \fgas\
decreases towards higher masses, along with the sSFR, suggesting that a lack
of fuel (resulting from, e.g., suppressed accretion or gas removal) sets
quenching on rather than reduced efficiency (from, e.g., gas stabilization
against fragmentation by a massive bulge or ISM heating mechanisms).  Setting
tighter constraints on these scenarios through measurements of the cold ISM in
sub-MS galaxies
at $z > 1$ is very challenging, and the very few results published to date
are inconclusive \citep[e.g.,][]{Bez19}.

\looseness=-2
Gas scaling relations at $z > 1$ are most firmly established at
$\log(M_{\star}/{\rm M_{\odot}}) \ga 10$, where high-$z$ samples probe well
the SFG population and where the luminosity-to-gas mass calibrations are best
constrained.
The more extensive data sets now available do not support a significant
dependence of the CO-$\rm H_{2}$ conversion on $\rm \Delta MS$ \citep{Tac20}.
In contrast, there is a strong variation of CO-$\rm H_{2}$ and of the
dust-to-gas ratio with metallicity \citep[e.g.,][]{Genz12,Bolatto13},
which is folded in the scaling relations given above.
At $z \ga 0.5$, the atomic gas contribution to $M_{\rm gas}$ on galactic
scales is generally neglected (though a 36\% correction for He is applied)
since most of the hydrogen is expected to be in molecular form at the high
densities inferred ($\rm > 10~M_{\odot}\,pc^{-2}$) and Damped 
Lyman\,$\alpha$ Absorbers
studies indicate a slow evolution in H{\footnotesize I} gas density
\citep[$\propto (1+z)^{0.57}$;][]{Per20}.

\vspace{-1ex}

\subsection{Metallicity and ISM Conditions}
 \label{global_metal.sec}

The metal content of galaxies is a sensitive probe of the baryon cycle,
carrying the imprint of gas accretion, stellar nucleosynthesis, galactic
winds, and internal gas mixing.
Observational constraints for $z \sim 2$ SFGs have largely come from strong
rest-optical nebular emission lines, interpreted through empirical and/or
theoretical calibrations in terms of the gas-phase oxygen abundance (O/H).
These lines also depend on the nebular conditions and structure, and on the
excitation sources, affecting calibrations.
The reviews by \citet{Mai19} and \citet{Kew19} discuss in detail the strengths
and limitations of various indicators, and stress the importance of combining
multiple diagnostics, of adopting the same method(s) to reduce the impact of
systematic differences in calibrations, and of using consistent approaches
in deriving galaxy properties ($M_{\star}$, SFR, ...) used to establish
scaling relations.

\begin{marginnote}[480pt]
  \entry{\begin{sansmath}$\bm{n_{\scriptsize\textbf{e}}}$\end{sansmath}}
        {Local electron density, the number of electrons per unit volume
         of an ionized nebula.}
\end{marginnote}

\looseness=-2
Offsets in the location of (non-AGN) $z \sim 2$ SFGs relative to the $z \sim 0$
excitation sequences in line ratio diagrams have long been known
(e.g., in \nii$\rm \lambda 6584 / H\alpha$ and
          \sii$\rm \lambda\lambda 6716,6731 / H\alpha$ vs.\
          \oiii$\rm \lambda 5007 / H\beta$, and
 (\oii$\lambda\lambda 3726,3729 + $\oiii$\lambda\lambda\,4959,5007$)$\rm / H\beta$
 vs.\ \oiii$\lambda 5007 / $\oii$\lambda\lambda 3726,3729$).
The growing near-IR spectroscopic data sets at $z \sim 2$ have
enabled a more systematic exploration of the origin of the observed offsets,
providing evidence for evolving conditions of the ionized gas in terms of a
harder ionizing radiation, elevated N/O abundance ratio, higher electron
density and ISM pressure, and higher ionization parameter, at fixed O/H
abundance \citep[e.g.,][]{Masters16,Ste16,Strom18,Kash19}.
Other factors may be at play such as the presence of weak AGN activity,
galactic-scale outflows and shocks, and diffuse ionized gas ---
the importance of which varies with redshift --- as well as sample selection,
and aperture and weighting effects where spectra of high- and low-$z$ galaxies
may encompass different physical regions and span a range of excitation
\citep[e.g.,][]{Sha15,Sha19,Kaa17,San17}.
Constraints on the electron density of ionized gas have also been
obtained from the \oii\ and \sii\ doublet ratio, pointing to an increase
with redshift, with $n_{\rm e}$ in the range $\rm 100 - 400~cm^{-3}$ for
$z \sim 2$ SFGs compared to $\rm \sim 25~cm^{-3}$ for $z \sim 0$ galaxies
\citep[e.g.,][]{San16a,Kaa17}.
These estimates may be somewhat inflated by emission from denser gas in the
ubiquitous galactic winds at $z \sim 2$ (Section\ \ref{resolved_outflows.sec})
in the single-component line fits commonly performed.

\looseness=-2
Turning to metallicity, while the ``strong line'' methods based on nebular
rest-optical emission can lead to systematic differences in $\rm \log(O/H)$
by up to $\rm \sim 0.7~dex$, relative estimates based on the same calibration
are more accurate.  The general shape and evolution of the mass-metallicity
relation (MZR) agree qualitatively among various studies out to $z \sim 3.5$,
with lower metallicities at lower $M_{\star}$, an overall decline in
metallicity at earlier times, and a stronger evolution in the low-mass regime,
in agreement with the (scarcer) results from rest-UV metallicity-sensitive
features in young stars \citep[see][]{Mai19}.
Among several proposed parametrizations, the form
$12 + \log({\rm O/H}) = {\rm Z_{0}} + 
  \log\left[1 - \exp(-(M_{\star}/{\rm M_{0}})^\gamma)\right]$
is physically motivated based on considerations of the chemical yields
in the presence of inflows and outflows.  It describes well the bending
shape of the MZR up to $z \sim 2.5$, where $\rm Z_{0}$ is the asymptotic
value at high mass and $\rm M_{0}$ is the evolving turnover mass (with
$\rm M_{0} \propto (1+z)^\beta$ where $\beta \sim 2.6 - 2.9$) below
which the relation follows a power law of index $\gamma \sim 0.4 - 0.6$
\citep[e.g.,][and references therein; see Figure\ \ref{MSMF.fig}]{Zahid14,EWuy14}.

\begin{marginnote}[95pt]
  \entry{MZR}{Stellar mass - metallicity relation.}
  \entry{FMR}{\looseness=-2 Fundamental metallicity relation,
              linking stellar mass, metallicity, and SFR.}
\end{marginnote}

A secondary dependence on the SFR, ultimately tied to the gas fraction, is
expected in a theoretical framework, where accretion of metal-poor gas dilutes
the galactic gas-phase metallicity while increasing the gas reservoir fueling
star formation.
Based on the large set of SDSS local galaxy spectra and first results at high
$z$, \citet{Man10} proposed a redshift-invariant fundamental metallicity relation
(FMR) between $\rm \log(O/H)$, $M_{\star}$, and SFR, parametrized in terms of
$\log(M_{\star}) - \alpha\,\log({\rm SFR})$.
While subsequent work at high $z$ has led to mixed results possibly due to the
limited dynamic range and uncertainties in SFRs, a consensus is now emerging
for the detection of a FMR albeit with hints of a modest evolution with lower
$\rm \log(O/H)$ at fixed $M_{\star}$ and SFR to $z \sim 2.5$, and possibly
stronger evolution at $z \ga 3$ \citep[e.g.,][]{San18,Mai19}.
Such an evolution may reflect a progressive increase of the mass loading
factor $\eta$ of galactic winds (the ratio of mass outflow rate to SFR)
and/or decrease of the metallicity of inflowing gas with lookback time;
beyond $z \sim 3$, infall rates of more pristine gas may overwhelm metal
production through stellar nucleosynthesis, resulting in stronger dilution.
Theoretical models and numerical simulations that match the observed
MZR, FMR, and evolution thereof underscore the role of stellar feedback in the
chemical evolution of galaxies, requiring an increasing $\eta$ in lower-mass
galaxies and winds removing gas at roughly the same rate as consumed by star
formation around $\log(M_{\star}/{\rm M_{\odot}}) \sim 10$
\citep[e.g.,][]{Erb08,Lil13,Mur15,Dav17}.
More direct observational constraints on $\eta$ at $z \sim 2$ will be
discussed in Section\ \ref{resolved_outflows_sfout.sec}.

\subsection{AGN Demographics}
 \label{global_AGN.sec}

\looseness=-2
The link between the growth of galaxies and their SMBHs, deduced from local
scaling relations and the co-evolution in cosmic SFR and black hole accretion
rate densities, has motivated an abundant literature on AGN activity and
feedback across cosmic time
\citep[e.g.,][for reviews]{Fab12,Hec14,Lut14,Brandt15,Pad17}.
We summarize key aspects on the demographics of radiative-mode AGN at high $z$.

\begin{marginnote}[300pt]
 \entry{\begin{sansmath}$\bm{L_{\scriptsize\textbf{X,AGN}}}$\end{sansmath}}
       {\looseness=-2
        X-ray AGN luminosity, generally computed in the rest-frame hard
        $\rm 2 - 10~keV$ band and corrected for absorption.}
\end{marginnote}

AGN, identified at X-ray and other wavelengths, are preferentially found in
higher mass galaxies, which, for an underlying positive correlation between
AGN luminosity and host mass, reflects flux limits in the data from which
AGN are identified.  Comparisons of the host properties of X-ray-selected
AGN with those of mass-matched samples of inactive galaxies showed that AGN
reside mainly in MS SFGs with little correlation between X-ray luminosity
$L_{\rm X,AGN}$ and SFR, are rarely associated with disturbed morphologies,
but are more prevalent in hosts with denser stellar cores
\citep[e.g.,][]{Silv09,Koc12,Koc17,Mull12a,Sant12}.
The lack of correlation between $L_{\rm X,AGN}$ and SFR is understood in
terms of the short-term $\rm \la 10^{6}~yr$ variability of AGN compared
to the $\rm \ga 10^{8}~yr$ timescales of galactic star formation processes
\citep[e.g.,][]{Hic14}.  X-ray stacking analyses, effectively averaging over
time, revealed a closer connection between inferred SMBH accretion rate and
host SFR \citep[e.g.,][]{Mull12b}.
The ratio of average SMBH accretion rate to SFR appears to be largely independent
of galaxy stellar mass, and so is the distribution of specific $L_{\rm X,AGN}$
(often taken as a proxy for the Eddington ratio;
e.g., \citealt{Aird12,Aird18}).
While the distribution in specific $L_{\rm X,AGN}$ shifts to higher values
towards higher $z$, a mass-independent distribution at fixed $z$ implies that
a wider range of $L_{\rm X,AGN}/M_{\star}$ is probed at higher host mass.
AGN selected by rest-optical and mid-IR diagnostics are less prone to
variability effects but susceptible to similar biases related to ``dilution''
by host galaxy emission \citep[e.g.,][]{Pad17}.
A longer term connection between $L_{\rm X,AGN}$ and SFR, coupled with evidence
from morphologies, is consistent with a picture in which $z \sim 2$ AGN are
fueled by stochastic accretion, and secular processes (rather than major mergers)
within the gas-rich hosts promote the growth of both SMBH and a central bulge
\citep[e.g.,][]{Mull12b}.
The exception might be for the most luminous and most obscured mid-IR-selected
AGN, underrepresented in X-ray surveys, whose morphologies are significantly
more frequently disturbed or indicative of merging \citep{Don18}.

\looseness=-2
Observations, as well as theoretical models and cosmological simulations
\citep[e.g.,][]{Som15,Naab17} support a link between AGN and star formation
quenching at high masses.  Causality, however, remains so far elusive.
Empirical connections through galactic structure and outflows are discussed
in Sections\ \ref{resolved_axisymm.sec} and \ref{resolved_outflows.sec}.

\section{RESOLVED PROPERTIES OF STAR-FORMING GALAXIES at z ${\mathbf \sim}$ 2}
\label{resolved.sec}

Our understanding of the processes driving the evolution of the global
galaxy properties discussed above has greatly benefitted from the growing
amount of data resolving individual galaxies.  A key finding was that high $z$
SFGs are predominantly disks, albeit more turbulent than local spirals.
The growth and evolution of disks as derived from stellar light, star
formation, and kinematic tracers is first discussed
(Sections\ \ref{resolved_axisymm.sec} - \ref{resolved_rotdisk.sec}),
followed by emerging dynamical constraints on the interplay between
baryons and DM on galactic scales (Section\ \ref{resolved_dyn.sec})
and deviations from disk rotation (Section\ \ref{resolved_perturb.sec}).
Non-gravitational motions (i.e., gas outflows) are then addressed, as
direct probe of feedback in action (Section\ \ref{resolved_outflows.sec}).

\subsection{Star-Forming Galaxies as Axisymmetric Systems}
 \label{resolved_axisymm.sec}

\subsubsection{Morphological disk settling and the emerging Hubble sequence}
   \label{resolved_axisymm_diskhubble.sec}

Many key features regarding the structural build-up of SFGs can be captured
in a framework where we consider them as flattened, axisymmetric structures.
This approach also fundamentally underpins semi-analytical models where any
structural evolution is only described radially.  Intrinsic 3D shapes inferred
from projected axial ratio distributions illustrate how at any given epoch there
is a tendency of increased fractions of SFGs with prolate (i.e., elongated)
shapes in the low-mass regime,
whereas the fraction of oblate (i.e., disky) systems increases with
mass and toward later times \citep{vdWel14b, Zhang19}.
This downsizing pattern for morphological disk settling
finds its counterpart in kinematic surveys, which show similar mass
and redshift dependencies for orderly rotating disk fractions, with
dispersion-dominated systems gaining in prevalence toward lower masses
(see Section\ \ref{resolved_rotdisk.sec} and Figure\ \ref{MSMF.fig}).
This Section discusses the radial characteristics of SFGs with an emphasis
on relatively massive ($\rm \ga 10^{10}\,M_{\odot}$) systems for which the
axisymmetric disk framework is most appropriate.
In the next Section we discuss how or where the actual
morphology deviates from axisymmetry.

\looseness=-2
Salient features of the size-mass relation of SFGs were discussed in
Section~\ref{global_ReM.sec}.  The same {\it HST\/} surveys also shed light
on surface brightness profile shapes, often quantified parametrically with
a S\'{e}rsic model.  For the MS population, exponential disk
profiles of $n \sim 1$ are the norm \citep{Wuy11b}, in line with the
disk-like nature inferred from axial ratios and kinematics.
Exceptions arise more frequently at the very tip of the MS, and
among the rare population of starbursting outliers above the MS, which on
average are characterized by more centrally concentrated profiles.  
We can conclude that the overall structure quantified from rest-optical/UV light
and colors correlates with location in the $M_{\star} - {\rm SFR}$
plane, with most star formation happening in disks while quiescent
galaxies feature cuspier profiles.
A Hubble sequence, where the dominant morphology and stellar populations are
intimately tied, can thus be said to be in place already since at least
$z \sim 2.5$.

While it is most straightforward to compare sizes and profile shapes
across epochs at fixed mass, individual galaxies build up stellar mass over
time through star formation and mergers.   Applying the cumulative comoving
number density technique outlined in Section\ \ref{techmeth_popmatch.sec},
progenitor-descendant sequences have been reconstructed to reveal the
growth in size and build-up of extended profile wings around central cores
for galaxies at the most massive end and to recover the structural growth
history of Milky Way progenitors
\citep[][see also \citealt{Patel13b}]{vD10, vD13}.
The latter feature a more modest size growth and at least at $1<z<2.5$ a more
self-similar evolution in profile shape than the most massive galaxies which
increase rapidly in S\'{e}rsic index.

Other than by disentangling the population growth from growth of individual
systems, major advances in our understanding of structural evolution are
arising from comparing multiple tracers.  Initially, this focused on rest-UV to
rest-optical stellar emission, but increasingly this is complemented by resolved
probes of ionized and molecular gas as well as reprocessed emission by dust.

\begin{marginnote}[420pt]
 \entry{M/L}{Mass-to-light ratio.}
\end{marginnote}


\subsubsection{Stellar mass distributions}
   \label{resolved_axisymm_massmaps.sec}

\begin{figure}[!t]
\centering
\includegraphics[scale=0.70,trim={-0.5cm 5.0cm -0.5cm 7.0cm},clip=1,angle=0]{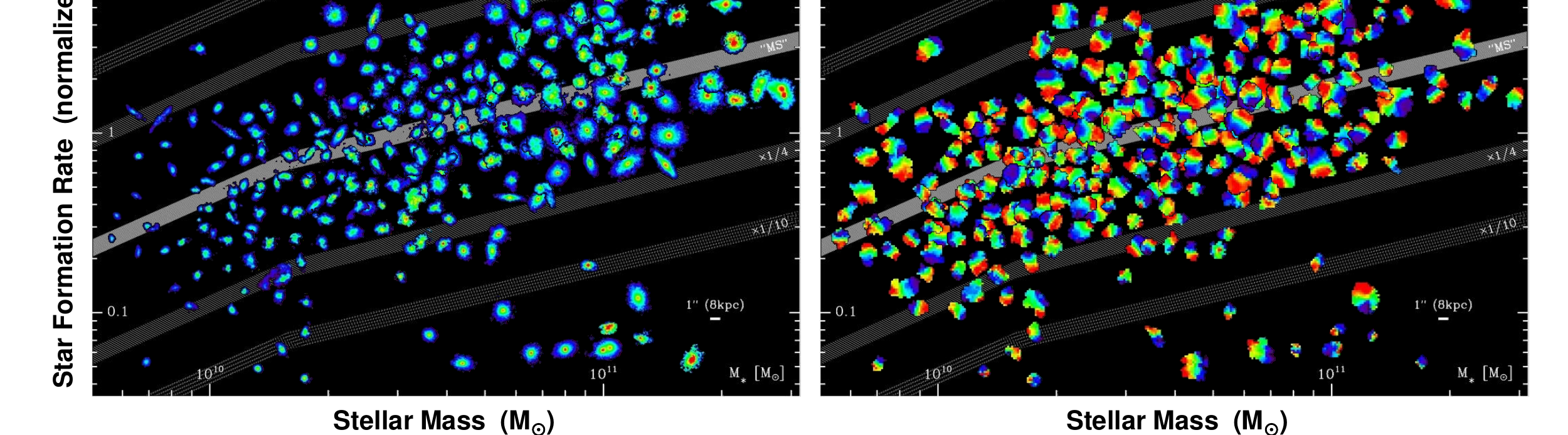}
\renewcommand\baselinestretch{0.85}
\caption{
Stellar structure and kinematics of 250 massive SFGs at cosmic noon.
{\it Left:\/}
Stellar mass maps derived from spatially-resolved SED modeling of multi-band
{\it HST\/} imaging \citep{Lang14}.
Blue to red colors correspond to increasing stellar mass surface density,
with the same range shown for all objects.
{\it Right:\/}
Velocity fields from H$\alpha$ obtained with the VLT/KMOS multi-IFU instrument as
part of the \kmostd\ survey \citep{Wis19}, at a resolution of $\rm \sim 4~kpc$.
Blue to red colors correspond to the maximum blueshifted to redshifted velocities
relative to systemic, adjusted individually for each object.
The galaxies are plotted as a function of their $M_{\star}$ and SFR (normalized
to the SFR of the MS at $\log(M_{\star}/{\rm M_{\odot}}) = 10.5$ at the
redshift of each object, shown as light grey thick line).
The stellar mass maps are relatively smooth and show an increasing bulge prominence towards higher masses.  The kinematics show an overall majority of rotating disks.
}
\label{SFGoverview.fig}
\end{figure}

\looseness=-2
With resolved imaging sampling the distribution of stellar light below and
above the Balmer/4000\AA\ break out to $z \sim 2.5$ a picture has emerged in
which negative color gradients (i.e., redder centers than outskirts) become
increasingly prominent towards the high-mass end and at later times
\citep[e.g.,][]{Liu17, Liu18}.
The age-dust degeneracy in a space of mass-to-light (M/L) ratio vs.\
rest-optical color \citep[e.g.,][]{Bell01} allows for a relatively robust
translation of the multi-band light maps to a stellar mass distribution
\citep[][see Figure\ \ref{SFGoverview.fig}]
{Wuy12,Szomoru13,Tacch15,Wang17,Suess19}.
In common to such studies
is the finding of more compact and centrally concentrated stellar mass profiles
compared to those observed in light, especially at lower redshifts and higher
masses.  Carrying out bulge-disk decompositions on stellar mass maps,
\citet{Lang14} find that while SFGs are well described by exponential disks
at low masses, once crossing the Schechter mass they already contain 40-50\%
of their stars in a bulge component, even prior to their eventual quenching.
Overall, taking both SFGs and quiescent galaxies together, it is now well
established that measures of bulge prominence or central surface density
\citep[e.g., $\Sigma_{1\rm{kpc}}$;][see Figure\ \ref{MSMF.fig}]{Cheung12,Bar17a}
form much more reliable
predictors of quiescence than stellar mass by itself.  However, the origin
of this strong correlation, in particular its interpretation in terms of a
causal connection, remains debated \citep{Lil16, Abr18}.
Building on the increased prevalence of AGN with host central
stellar mass density and the empirical inference that quenching sets on when
the cumulative radiative energy of SMBHs reaches $\sim 4\,\times$ the halo
binding energy, \citet{Chen20} recently put forward a phenomenological model
that strengthens the role of AGN in quenching by explaining naturally the
structural differences between star-forming and quenched galaxies.

\subsubsection{Observed H$\alpha$ profiles}
   \label{resolved_axisymm_Hamaps.sec}

\looseness=-2
WFC3 grism surveys such as 3D-HST \citep{Mom16}
give access to the H$\alpha$ surface brightness distributions
on kpc scales for galaxies out to $z \sim 1.5$.
Such observations have illustrated that the H$\alpha$ emission of MS galaxies,
tracing the unobscured instantaneous star formation, follows on average
exponential disk profiles \citep{Nel13}, and that there is a resolved
equivalent of the MS, a correlation between the local star formation and
stellar surface density (\citealt{Wuy13}; see also \citealt{Wang17}).
Deviations from this relation are seen in the centers of -- particularly massive --
SFGs, with also asymmetric features such as clumps contributing to the scatter
(Section\ \ref{resolved_devsymm.sec}).
Stacking H$\alpha$ and $H_{140}$ maps of 3200 $z \sim 1$ SFGs \citet{Nel16b}
find the (unobscured) star formation to be slightly more extended than the
stellar continuum emission, with a weak dependence on mass:
$R_{\rm e,H\alpha}/R_{{\rm e},H} = 
 1.1 \left(M_{\star}/ 10^{10}\ {\rm M_{\odot}}\right)^{0.054}$.
Translated to H$\alpha$ equivalent widths (EWs) this results in centrally
dipping profiles, with the central depression in H$\alpha$ EW being most
prominent at the high-mass end.  AO-assisted IFU surveys were able to push
resolved H$\alpha$ EW measurements out to $z \sim 2.5$,
resulting in qualitatively similar findings \citep{Tacch15}.
With such\begin{marginnote}[130pt]
 \entry{EW}{Equivalent width, for an emission line equal to the ratio
            of line flux to continuum flux density.}
\end{marginnote}numbers
at present limited to a few dozen (fewer when considering the
high-mass end alone) and accumulated at a rate of $\sim$one 8-m telescope
night per object, significant progress on number statistics here is anticipated
from grism observing modes on {\it JWST\/}.  Already with existing ground-based
(yet seeing-limited) instrumentation, however, larger samples with consistent
continuum and H$\alpha$ size measurements over the full $0.6<z<2.6$ range
can be compiled.  Doing so, \citet{Wilman20} find an average size ratio of
$\frac{R_{\rm e,H\alpha}}{R_{{\rm e},F160W}} = 1.26$,
without significant dependence on the
redshift, mass and star formation activity.  Adopting the observed size ratio
as an upper limit to $\frac{R_{\rm e,SF}}{R_{{\rm e},M_{\star}}}$ (a limit due
to the possible presence of differential extinction and dust gradients), they
infer the associated size growth due to star formation alone to proceed along
a vector of $\frac{d \log R_{\rm e}}{d \log M_{\star}} \sim 0.26$, consistent
with results from constant comoving number density arguments and only slightly
steeper than the observed slope of the size-mass relation at any epoch.  Other
processes than simply adding new stars, such as feedback, angular momentum
redistribution, (minor) mergers and the preferential quenching of more compact
SFGs may need to be invoked to reconcile the relatively slow growth due to star
formation with the observed size evolution of SFGs.

\subsubsection{Attenuation gradients}
   \label{resolved_axisymm_dustgrad.sec}

\looseness=-2
In the absence of dust all of the above radial profiles, size differences, and
red centeredness would be attributed most straightforwardly in terms of stellar
population age (or sSFR) gradients consistent with a picture of inside-out disk
growth.  SFGs at cosmic noon however are far from dust free, particularly in
the massive and highly star-forming regime where most of the internal color
dispersion \citep[e.g.,][]{Boada15} and radial gradients are seen.  With only
a single rest-optical color, the effects of age and dust are fully degenerate.
While this enables a robust estimate of spatial M/L ratio variations, explaining
the origin of these variations (spatially inhomogeneous SFH vs.\ extinction) is by
the same token a challenging task.  Several approaches have been pursued to pin
down to which degree levels of extinction vary across galaxy disks.
\citet{Wuy12} used resolved SED modeling of 7-band ACS+WFC3 photometry to
constrain the stellar populations of individual pixel bins in $0.5<z<2.5$ SFGs.
\citet{Nel16a} were able to extract a more direct probe of extinction for
$z \sim 1$ SFGs in the form of the Balmer decrement (H$\alpha$/H$\beta$),
although relying on stacked profiles for relatively broad bins in stellar
mass.  Complementary broadband approaches further include use of the
dust-sensitive UV slope $\beta$ \citep{Tacch18} and a rest-frame $UVI$
color-color diagram \citep{Wang17, Liu17}.  

\looseness=-2
While quantitative differences remain and direct comparisons are complicated
by differences in applicable redshift range and technique (e.g., individual
galaxies vs.\ stacking), a converging picture is emerging.  Galaxies do feature
radial gradients in extinction, with the amount of central enhancement increasing
with stellar mass, reaching $\sim 2$ magnitudes of central extinction at the
high-mass end.  Propagating this knowledge to the reconstruction of sSFR profiles
yields on average surprisingly flat profiles over the full radial range for
intermediate mass galaxies.  Only among the most massive galaxies central
drops in the star formation activity remain present after dust correction,
a trend that is interpreted as a signature of inside-out quenching.
For example, \citet{Tacch18} exploit near-IR AO-assisted IFU data
at $z \sim 2$ to find a radially constant mass-doubling timescale of
$\sim 300$ Myr for SFGs below $\rm \la 10^{11}\,M_{\odot}$, and central
star formation suppression by a factor of $\sim 10$ above this mass.  At
$z \sim 1$, \citet{Wang17} report qualitatively similar results with flat
sSFR profiles for SFGs below $\rm 10^{10.5}\,M_{\odot}$ and central declines
of 20-25\% above this mass.  \citep[see also][]{Liu16}.
The flat inferred sSFR profiles of intermediate mass galaxies are seemingly
at odds with the inside-out growth inferred from constant comoving number
density arguments (Section\ \ref{resolved_axisymm_diskhubble.sec}).
Possibly the stellar build-up proceeds more rapidly outside the inner
$\sim 2\,R_{\rm e}$ within which most stellar population and dust gradients
have been quantified, but it has also been argued that the mass-weighted size
growth may be more modest than the observed light-weighted one \citep{Suess19}.
Resolved $UVJ$ diagrams and direct measurements\footnote{
 It should be noted that at the $R \sim 130$ resolution of the WFC3/G141 grism
 H$\alpha$ and [NII] are blended and additionally underlying stellar absorption,
 especially to H$\beta$, should be accounted for.}
of radial Balmer decrement profiles of individual galaxies will undoubtedly
play a vital role in progressing our understanding of where within SFGs stars
are formed, and are within reach of {\it JWST}'s imaging and (grism)
spectroscopic capabilities.  That said, we caution that
the central effective $A_V$ of $\sim 2$ magnitudes inferred for massive SFGs
under a
foreground screen approximation may well conceal total dust
column densities that are several times higher, depending on the dust geometry,
its clumpiness and albedo \citep{Seon16}.  We thus conclude that dust modeling
at present poses a key challenge to quantifying galaxy sizes and SFR distributions
at the massive end.

\subsubsection{Compact dusty cores and implications for SFR profiles}
  \label{resolved_axisymm_dustcores.sec}

\looseness=-2
Having highlighted the significant role of dust, it is important to
underline the potential offered by far-IR to radio observations to complement our
view of where star formation is happening (as seen reprocessed by dust) and where
within the disks cold gas, the fuel for star formation, resides (as revealed by
CO line emission).  Here, ALMA, NOEMA and the JVLA with their recently enhanced
sensitivities and long baselines are making major contributions.
In low-$J$ CO transitions, MS galaxies feature a similar extent as observed
in the (rest-)optical.
This appears to be the case both at $z \sim 1$ \citep{Tac13} and
$z \sim 2$ \citep{Bolatto15}, although numbers in the higher redshift bin
remain limited.  A different picture is painted when considering continuum
probes of star formation.
Comprised predominantly of non-thermal synchrotron radiation from charged
particles accelerated within supernova remnants, 1.4\,GHz continuum
emission serves as a dust-unbiased SFR tracer \citep{Condon92}.
Using a {\it uv}-stacking algorithm to trace the 1.4\,GHz continuum
size evolution of $\sim 1000$ MS SFGs with
$\rm 10^{10} - 10^{11}~M_{\odot}$ spanning $0 < z <3$, \citet{Lindroos18}
find the measured radio sizes to be typically a factor of two smaller than
those measured in the rest-optical.  Likewise, focusing on thermal dust emission
from a sample of normal MS galaxies at $\log(M_{\star}/{\rm M_{\odot}}) \ga 11$,
\citet{Tad17a} combined ALMA $\rm 870\,\mu m$ observations in compact and extended
configuration to infer that the dust sizes of their targets were more than a
factor of two smaller than those observed at rest-optical (and even more so
H$\alpha$) wavelengths.  

\looseness=-2
These results are in contrast to what naively would be anticipated given
the typical centrally declining H$\alpha$ EWs in the same high-mass regime
at $z \sim 2$, even after dust corrections.  Star formation happening in such
centrally concentrated cores could in several hundred Myr build up a central
bulge with $\Sigma_{1\rm{kpc}} > 10^{10}\ {\rm M_{\odot}\ kpc^{-2}}$, akin to
the central densities of lower redshift quiescent systems.
Resolved maps at a second, higher frequency IR wavelength are needed to rule
out or reveal any negative gradients in dust temperature that may bias the
inferred half-SFR sizes to low values.  In the handful of objects
where resolved CO and dust continuum measurements are both available, authors
also noted the smaller dust compared to CO sizes \citep{Spilker15, Tad17b}.
\citet{Rujo16} on the other hand found 5\,cm and 1.3\,mm sizes of somewhat lower
mass ($\langle \log(M_{\star}/{\rm M_{\odot}}) \rangle \sim 10.7$) SFGs at the same
redshift to both be comparable to the extent of the stellar mass maps.  Enhancing
the robustness of multi-tracer structural measurements and interpreting the
relative sizes of dust, stellar, and H$\alpha$ emission as a function of mass
stands as an important challenge for future studies.  This applies especially
to reconciling the apparently inconsistent findings from cold ISM and H$\alpha$
observations at the massive end.  At present, ambiguity remains whether this
is due to uncertain dust corrections or instead differences between samples that
fit into a common evolutionary sequence where massive galaxies undergo compaction
events triggering nuclear starbursts (responsible for the compact dust sizes)
followed by a phase of inside-out quenching (responsible for the centrally
declining H$\alpha$ EWs; e.g., \citealt{Tacch16}).

\subsection{Deviations from Axisymmetry}
 \label{resolved_devsymm.sec}

\subsubsection{Shapes and morphologies}
   \label{resolved_devsymm_morph.sec}

\looseness=-2
Thus far, we discussed the structural properties of high-z SFGs in terms of
sizes and radial profiles.  Skewed axial ratio distributions for low-mass
($\log(M_{\star}/{\rm M_{\odot}}) < 10$) SFGs at cosmic noon suggest that a
framework of flattened axisymmetric disks may be inappropriate for young
systems that have not had the time to settle into an equilibrium disk
configuration \citep{Law12b,vdWel14b}.  Modeling the joint distribution of
projected axis ratios and sizes, and accounting for the finding that smaller
SFGs are systematically rounder, \citet{Zhang19} argue that prolate and/or
spheroidal shapes may in fact be even more common than inferred by
\citet{vdWel14b}, also for the $\log(M_{\star}/{\rm M_{\odot}}) = 10-10.5$ regime.
They report young, low-mass galaxies in the VELA set of high-resolution
hydrodynamical cosmological zoom-in simulations to be prolate as well.
Kinematics reveal a qualitatively similar trend, with a threshold mass
for disk settling that decreases with decreasing redshift
(Section\ \ref{resolved_rotdisk.sec}).

\looseness=-2
Even above $\rm 10^{10}\,M_{\odot}$, the morphological appearance of high-$z$
SFGs often looks markedly different from that of the relatively smooth disk
population in the local Universe.  Rising fractions of irregular morphologies
were first noted in early {\it HST\/} observations probing the rest-UV
\citep{Griffiths94, Windhorst95, Abraham96}, and later quantified using
the larger samples provided by rest-optical legacy surveys such as CANDELS
\citep[e.g.,][]{Con14, Huert16}.  Methods to quantify the evolving morphological
mix through cosmic time encompass non-parametric morphological measures
(e.g., concentration, asymmetry, Gini, M$_{20}$; \citealt{Con03}; \citealt{Lotz04})
or Principal Component Analysis thereof \citep{Peth16},
visual classifications by experts or citizen scientists
\citep[e.g.,][]{Kart15, Simmons17}, and increasingly deep-learning
techniques using visually classified training sets \citep{Huert15} as well
as unsupervised machine learning \citep{Hocking18}.  Cross-comparisons often
show a good concordance between these approaches.  This does however not
mean that the physical origin of the irregular morphologies, often featuring
asymmetries in the form of off-center clumps, can be readily interpreted.
While historically frequently used alongside pair counts to quantify the
evolution in merger rates, the clumpy morphologies are nowadays more often
interpreted as massive star-forming regions originating in marginally stable,
gas-rich disks.  The tightness of the MS, kinematic evidence for
ordered rotation, average surface brightness profiles and axial ratio
distributions as well as probes of the cold gas reservoirs all contributed
to this paradigm shift.  In addition, the wavelength dependence of clumpy
morphologies (more prominent in the rest-UV where they are identified)
attributed to spatial variations in the SFH and/or dust
extinction also imply that the underlying mass distribution is smoother than
the galaxies appear in light, unlike what may be expected from mergers
\citep[e.g.,][]{Wuy12, Cib15}.

Indeed, several studies have addressed the ability to identify mergers at
cosmic noon by exploiting mock observations of galaxies extracted from
simulated cosmological volumes where their (non-)merger state is intrinsically
known \citep[e.g.,][]{Snyder15, Thompson15}.  This exercise reveals
dependencies of completeness and contamination fraction of the selected
mergers on merger stage, viewing angle, and depth of observation, yielding
in some of the simulations that reproduce realistically high gas fractions
at $z \sim 2$ results that are no better than a random guess \citep{Abruzzo18}.

This does not imply that mergers do not happen, nor that all clumps share the
same formation process.  Targeting mostly higher redshifts ($2<z<6$),
\citet{Ribeiro17} find the most massive clumps ($\rm \sim 10^9\,M_{\odot}$) to
typically reside in galaxies featuring just 2 clumps, arguably interpretable
as a merger, whereas less massive clumps ($\rm < 10^9\,M_{\odot}$) occur in
galaxies featuring a larger number of them, consistent with disk fragmentation.
The distinction between ex-situ and in-situ clumps, with the former featuring
higher masses and older stellar ages, is also seen in hydrodynamical simulations
\citep{Mandelker17}.

\subsubsection{Clump properties}
   \label{resolved_devsymm_clumps.sec}

\looseness=-2
Turning to the properties of individual clumps, a first realization stemming
from multi-band stellar population analyses of these features is that, while
striking in appearance, they do not dominate the integrated UV emission of
the galaxies, let alone add up to a major contribution of the star formation,
and even less so account for a substantial fraction of the overall
stellar mass.  While a precise breakdown depends on details of sample
selection, clump selection (e.g., threshold depth and wavelength),
and whether and how underlying diffuse disk emission is accounted for
\citep[e.g.,][]{FS11, Guo18}, different censuses report clump contributions
(i.e., summed over all clumps) to the overall UV emission, SFR, and stellar mass
of mass-selected SFGs at cosmic noon of $\sim 20\%$, $\sim 5 - 18\%$ and
$\la 7\%$, respectively \citep{Wuy12, Guo15, Soto17}.  The fraction of SFGs
that appear clumpy is itself a function of both mass and redshift.  While
$\sim 60\%$ of low-mass ($\log(M_{\star}/{\rm M_{\odot}}) < 9.8$) SFGs features
clumpy UV morphologies over the full $0.5 < z < 3$ range, the clumpy fraction for
intermediate and high-mass SFGs drops from 55\% to 40\% and from 55\% to 15\%
over the same $z$ range, respectively \citep{Guo15}.

\looseness=-2
The characteristic scales of giant star-forming clumps reported in the
literature are on the order of a kiloparsec, with corresponding stellar
masses ranging up to a few $\rm 10^9\ M_{\odot}$ \citep[e.g.,][]{FS11, Guo18}.
These scales are in accordance with the Toomre scale and mass anticipated
for gravitational instabilities within gas-rich turbulent disks
\citep{Elm09, Genz08, Genz11, Dek09a}.  It is worth noting though that
structures on these scales are only marginally resolved in studies of field
galaxies, and may correspond to conglomerations of blended clumps of
smaller physical scales.  Samples of a handful of lensed galaxies reaching
spatial resolutions of $\rm 20 - 100~pc$ do indeed reveal progressively
smaller clump sizes as the resolution is enhanced with respect to blank
field observations \citep{Dess17, Rigby17}.  This is illustrated perhaps
most convincingly in the analysis of multiple lensed images of the same
object at different magnifications \citep{Cava18}.
In this light, zoom-in simulations of turbulent
gas-rich disks resolving the multi-phase ISM on parsec scales will prove useful
in tracing fragmentation below the Toomre scale and interpreting the higher
resolution observations that will become feasible with {\it JWST\/} and ultimately
the extremely large telescopes.
Already, first attempts on lensed samples are made to characterize
the clump mass functions \citep{Dess18}, yielding results consistent with a
power-law slope of $-2$ anticipated for fragmentation due to a turbulent
cascade \citep{Chandar14, Adamo17}.

Typical stellar ages inferred for the star-forming clumps are on the order
of $\rm 100 - 200~Myr$ \citep{FS11, Wuy12, Guo12, Guo18}.  A single massive
clump consisting almost entirely of line emission (i.e., massive in gas,
but an order of magnitude lower in stellar mass) was discovered by
\citet{Zan15}, for which they estimate an age of $\rm < 10~Myr$, confirming
the in-situ formation by gravitational collapse as origin of the clump
phenomenon.  Mimicking the azimuthally averaged radial trends of stellar
population tracers discussed in Section\ \ref{resolved_axisymm.sec},
clumps themselves also feature redder rest-optical colors, lower H$\alpha$
EWs, and -- inferred from those -- older ages (by a few 100 Myr) and lower sSFRs
towards the galaxy centers \citep{FS11, Adamo13, Guo12, Guo18, Soto17}.
The gradients steepen with increasing stellar mass and decreasing redshift,
and are found to be overall steeper than the radial gradients observed for
the intra-clump regions \citep{Guo18}.  As a caveat we note that in most
of these studies radial gradients are quantified on the basis of ensembles
of clumps collected from multiple galaxies within relatively coarse bins of
mass and redshift, as the number of detectable clumps in individual systems
remains limited.

The longevity of clumps forms an outstanding question with significant
implications for the subsequent structural evolution of the
galaxies that host them.  If remaining intact and surviving internal
stellar feedback for a few hundred Myr, their inward migration due to
dynamical friction is predicted to be an efficient mode of in-situ bulge
growth \citep[e.g.,][]{Bournaud07, Elm08, Cev10}.  On the other hand,
simulations with stronger feedback implementations such as FIRE
\citep{Okl17} and NIHAO \citep{Buck17} feature shorter clump lifetimes
($\la 50$ Myr) and substantially less inward migration.  Despite their
differences both flavors of simulations claim to reproduce the observed
phenomenology of wavelength dependent clump prominence, their characteristic
stellar ages and even radial gradients \citep[e.g.,][]{Okl17, Mandelker14}.
A duty cycle argument relating the existence of a very young clump as found
by \citet{Zan15} to the abundance of equally massive clumps that are older
supports long inferred clump lifetimes ($\rm \sim 500~Myr$).
Measured ages of the stellar populations in clumps may not necessarily match
the timescale of clump survival as clumps are in constant interaction with
their surrounding disk due to outflows, tidal stripping, and continued accretion
\citep{Bournaud14}.  Perhaps the observable with most discriminating power
between the different suites of simulations will prove to be the gas fraction,
on an individual clump basis, but even already at the galaxy-integrated level.

\subsection{Star-Forming Galaxies as Rotating Turbulent Disks}
 \label{resolved_rotdisk.sec}

Near-IR IFU observations, mainly of H$\alpha$ but also \oiii\ or \oii\ line
emission, have provided the most comprehensive and detailed censuses of the
kinematic properties of $z \sim 2$ SFGs, and the most convincing evidence for the
prevalence of disks among them.  Mitigating M/L variations that can complicate the
interpretation of morphologies, especially at $z > 1$, kinematics trace the full
underlying mass distribution and are a sensitive probe of a system's dynamical
state.  Spatially-resolved kinematics of cold gas line emission from (sub)mm
interferometry are still scarce for typical $z \sim 2$ MS SFGs, and while
near-IR slit spectra have also been exploited to derive emission line kinematic
properties, they give spatially limited information with larger uncertainties 
related to slit placement relative to the galaxy center and kinematic axis.
Stellar kinematics at $z > 1$ are still restricted to quiescent galaxies, 
absent of young hot stars filling in absorption features, and in all
but a few cases are limited to galaxy-integrated velocity dispersions.

The first step in exploiting 3D kinematic data is to identify the nature
of the galaxies.  Different procedures are followed but they conceptually
rely on similar criteria based on 2D maps, and on the main derived parameters of
maximum rotation velocity \vrot\ and local velocity dispersion \sigo\ corrected
as appropriate for spatial and spectral resolution and for galaxy inclination
(extraction methods are summarized in the Supplemental Text).
The basis is encapsulated in the following set of disk criteria adopted in several
studies, motivated by expectations for an ideal rotating disk, and increasingly
stringent and demanding of the data \citep[e.g.,][]{FS18,Wis19}:
\begin{enumerate}
\item a smooth monotonic velocity gradient across the galaxy, defining the
   kinematic axis;
\item a centrally peaked velocity dispersion distribution with maximum at
   the position of steepest velocity gradient, defining the kinematic center;
\item dominant rotational support, quantified by the $v_{\rm rot}/\sigma_{0}$
   ratio;
\item co-aligned morphological and kinematic major axes (a.k.a. kinematic
   misalignment);
\item spatial coincidence of the kinematic and morphological centers.
\end{enumerate}
Application of these criteria is usually done from measurements of the
parameters and visual inspection, or through comparisons to disk models.
Kinemetry, an approach based on harmonic expansion along ellipses of the
moment maps of the line-of-sight velocity distribution,
has also been used in some studies to quantify the degree of asymmetry
in velocity and dispersion maps, either as main classification or in
support of the criteria above.
Details on disk modeling and kinemetry can be found in the Supplemental Text.
It is increasingly common to supplement the kinematic criteria with information
on galaxy morphology and possible companions, e.g., from {\it HST\/} imaging,
for a more complete characterization.

\begin{textbox}[!t]
\section{KINEMATIC PROPERTIES}
\noindent
{\bf Rotation curve (RC):}
  Rotation velocity $v$ vs.\ galactocentric radius $r$.  For a ``Freeman'' thin
  disk with exponential surface density distribution, scale length $R_{\rm d}$,
  and $y \equiv r/2R_{\rm d}$,
  $v^{2}(r) = 4\pi G\Sigma_{0}R_{\rm d}y^{2}\,[I_{0}(y)K_{0}(y)-I_{1}(y)K_{1}(y)]$,
  where $G$ is the gravitational constant, $\Sigma_{0}$ is the central surface
  density, and $I_{i}$ and $K_{i}$ are the modified Bessel functions of order $i$.
  At fixed mass profile $M(r)$, thick disks
  (scale height $h$\,$\sim$\,$0.2$\,$-$\,$0.3$\,$R_{\rm d}$)
  have a $\sim 8\%$ lower $v$ peak reached at $\sim 10\%$ larger radius,
  while in the spherical approximation the peak is $\sim 15\%$ lower and at
  $\sim 20\%$ smaller radius
  \citep{Free70, Bin08, Noor08}. \\
\begin{sansmath}$\bm{v_{\scriptsize\textbf{rot}}}$\end{sansmath}:
  Maximum intrinsic rotation velocity (i.e., corrected for beam smearing and
  galaxy inclination when measured from observations), with $R_{\rm max}$
  denoting the radius where it is reached in intrinsic space. \\
\begin{sansmath}$\bm{v_{\scriptsize\textbf{2.2}}}$\end{sansmath}:
  Intrinsic rotation velocity at $r = 2.2\,R_{\rm d}$, where a Freeman disk
  RC peaks (corresponding to $1.3\,R_{\rm e}$).  For $n \neq 1$ profiles,
  $v_{2.2}$ differs from the peak $v_{\rm rot}$. \\
$\boldsymbol{\sigma_{0}}$:
  Local intrinsic velocity dispersion (i.e., corrected for beam smearing when
  derived from observations); it is assumed to be isotropic and constant across
  disks (Section\ \ref{resolved_rotdisk_sig0.sec}). \\
\begin{sansmath}$\bm{v_{\scriptsize\textbf{c}}}$\end{sansmath}:
  Circular velocity, here as a measure of the potential well.  For a thin
  disk, $v_{\rm c} = v_{\rm rot}$; for a thick disk with non-negligible
  turbulent pressure gradient,
  $v_{\rm c}^{2}(r) = v_{\rm rot}^{2}(r) + 2\,\sigma_{0}^{2}\,(r/R_{\rm d})$
  \citep[e.g.,][]{Bur16}. \\
\begin{sansmath}$\bm{S_{\scriptsize\textbf{0.5}}}$\end{sansmath}:
  Alternative kinematic estimator for a spherically symmetric system in an
  isothermal potential, defined as
  $S_{0.5}^{2} = 0.5v_{\rm rot}^{2} + \sigma_{0}^{2}$ \citep[e.g.][]{Wei06a}. \\
\begin{sansmath}$\bm{M_{\scriptsize\textbf{dyn}}}$\end{sansmath}:
  Enclosed dynamical mass.  For a spherical distribution,
  $M_{\rm dyn}(r) = r\,v_{\rm c}^{2}/G$;
  for a Freeman disk,
  $M_{\rm dyn}(r) = 2\pi \Sigma_{0}R_{\rm d}^{2}
   \left[1 - e^{-r/R_{\rm d}}(1 + r/R_{\rm d})\right]$
  \citep{Bin08}. \\
\begin{sansmath}$\bm{f}_{\boldsymbol{\star}}$, $\bm{f_{\scriptsize\textbf{bar}}}$,
  $\bm{f_{\scriptsize\textbf{DM}}}$\end{sansmath}:
  Ratio of stellar, baryonic, and DM mass to dynamical mass. \\
\begin{sansmath}$\bm{j_{\scriptsize\textbf{d}}}$\end{sansmath}:
  specific angular momentum of a (disk) galaxy, $\propto v(r) \times r$.
\end{textbox}

\looseness=-2
The outcome of the morpho-kinematic classification scheme depends on how well the
galaxies are resolved and how sensitive the data are.  It is usually adequate
to provide a first-order description of the system and the basis for quantitative
interpretation of the measurements.  Deeper data detecting fainter extended
emission and/or higher resolution (AO-assisted vs.\ seeing-limited) set better
constraints on the nature of the galaxies and can reveal additional interesting
features (Section\ \ref{resolved_perturb.sec}).
The choice of $v_{\rm rot}/\sigma_{0}$ threshold varies from 1 to 3 between
different studies, with the intermediate value of $\sqrt{3.36}$ corresponding
to equal contribution from rotation and random motions to the dynamical support
of a turbulent disk.
Several efforts have been devoted to assess the reliability of kinematic
classification based on mock observations of template data, encompassing
nearby systems to high resolution cosmological simulations.
Low misclassification fractions of $\sim 10\%-30\%$ are generally obtained
for disks and major mergers alike, with the range reflecting the specific
criteria employed, and data resolution and S/N \citep{Sha08,Epi10,Bello16}.
Using zoom-in simulations from the VELA suite of $z \sim 2$ isolated galaxies
and mergers over many sightlines to create $\sim 24000$ mock-observed data sets
in $0\farcs 6$ seeing, \citet{Sim19} conclude that disks are identified with
high confidence, while misclassification of mergers as disks varies widely
but, unsurprisingly, is lowest ($\la 20\%$) when applying all critera above
and folding in {\it HST\/}-like morphological information.

\subsubsection{Disk fractions}
  \label{resolved_rotdisk_fdisk.sec}

\looseness=-2
Recent large kinematic surveys have confirmed the findings from earlier smaller
samples that up to $z \sim 2.5$ a large proportion of massive SFGs are fairly
regular disks, albeit with higher velocity dispersions than present-day spirals.
The largest and most complete surveys, comprising hundreds of SFGs on/around the
MS at $9 \la \log(M_{\star}/{\rm M_{\odot}}) \la 11.5$ with resolved kinematics
from KMOS, find $\sim 70\% - 80\%$ of rotation-dominated galaxies
\citep[i.e., satisfying criteria 1-3 above, with $v_{\rm rot}/\sigma_{0} > 1$;][]
{Wis15,Wis19,Sto16},
a result borne out by deep AO-assisted SINFONI data of 35 $z \sim 1.5 - 2.5$
SFGs in the same mass range \citep{FS18}.  Imposing all criteria reduces the
disk fractions $f_{\rm disk}$ to $\sim 50\% - 60\%$.
Significant trends in the kinematic mix of SFGs are emerging from $z \ga 0.6$
IFU surveys, with lower $f_{\rm disk}$ at earlier epochs and, at fixed $z$,
towards lower masses \citep[e.g.,][]{Wis19,Turner17}.
\begin{marginnote}[0pt]
 \entry{\begin{sansmath}$\bm{f_{\scriptsize\textbf{disk}}}$\end{sansmath}}
       {Fraction of galaxies classified as disks.}
\end{marginnote}These results strengthen
and extend out to $z \sim 3.5$ findings from optical and near-IR slit
spectroscopy over $z \sim 0.2 - 2.5$ \citep[e.g.,][]{Kassin12,Sim17}.
The dependence on $M_{\star}$ and $z$ of the fraction of
rotation-dominated galaxies is illustrated in Figure\ \ref{MSMF.fig} 
(where the curves are adjusted to match the binned data presented by
\citealt{Sim17}, \citealt{Turner17}, and \citealt{Wis19}).\footnote{
A sigmoid function in $M_{\star}$ and lookback time $t$ is used in
Figure\ \ref{MSMF.fig} so that the fraction is bounded, and because this
functional form better reproduces the data in $t$ than in $z$.
The curves broadly match the trends implied by the linear fits in $t$
for different $M_{\star}$ bins given by \citet{Sim17}.
}
The trends reflect primarily those with \vsigo, with the evolution of \sigo\
largely driving the $z$ variation and the connection between \vrot\ and
galaxy mass (via the Tully-Fisher relation) dominating the $M_{\star}$
dependence (see Sections\ \ref{resolved_rotdisk_sig0.sec} and
\ref{resolved_dyn_tfr.sec}).
They also account for the range in $f_{\rm disk}$ ($\sim 25\% - 75\%$) reported
by various $z \sim 0.5 - 3.5$ studies based on samples of $\sim 10 - 60$ SFGs
probing different mass and redshift ranges, in addition to other factors such
as S/N, resolution, details and strictness of the classification procedure
\citep[e.g.,][]{Law09,Epi12,Liv15,Mie16,Mas17,Gir18a,Gill19}.

\looseness=-2
The variation of disk fraction and \vsigo\ with galaxy mass and redshift
has been interpreted in a ``disk settling'' scenario \citep{Kassin12}.
Massive SFGs settled earlier into more rotationally-dominated ``mature'' disks,
gradually followed by lower-mass galaxies at later times and with more massive
disks being dynamically colder at all epochs.  This evolution is reflected in
the trends between mass, morphology, and specific angular momentum of disks
(discussed in Section\ \ref{resolved_dyn_angmom.sec}).
It also finds its counterpart in the structure of the stellar component from
{\it HST\/} imaging inferred from the projected axial ratio distributions
(Section\ \ref{resolved_devsymm.sec}),
and is qualitatively reproduced by the recent high-resolution TNG50
cosmological simulation \citep{Pil19}.

Sections\ \ref{resolved_rotdisk_sig0.sec} and
\ref{resolved_dyn.sec} focus on the properties of disks identified
as described above and interpreted in an ideal disk framework,
Section\ \ref{resolved_perturb.sec} comments on deviations thereof.
The mass dependence of the disk fraction implies that disk samples
preferentially probe, on average, higher mass SFGs compared to
mass-selected samples.

\subsubsection{Disk turbulence}
  \label{resolved_rotdisk_sig0.sec}

\looseness=-2
The elevated gas velocity dispersion of $z \sim 2$ disks is well established
and implies that they are geometrically thick,\footnote{
For a disk of finite intrinsic thickness $q_{0} = (b/a)_{0}$,
the inclination $i$ is obtained via $\sin^{2}(i) = (1 - q^{2})/(1 - q_{0}^{2})$;
for $q_{0} \sim 0.2$ suggested by axial ratio (and $v_{\rm rot}/\sigma_{0}$)
distributions at $z \sim 2$, the difference in inclination correction compared
to $q_{0} = 0$ amounts to $\sim 2\%$ or less.}
as observed in {\it HST\/} imaging \citep[e.g.,][]{Elm05,Elm17}.
At the level of beam smearing of high-$z$ observations ($\rm \sim 4 - 5~kpc$
in natural seeing, and $\rm \sim 1 - 2~kpc$ using AO), unresolved noncircular
motions induced by deviations from axisymmetry of the gravitational potential
(e.g., massive clumps, bars) or related to outflows may
contribute to the measured \sigo\ along with local turbulent gas motions.
The agreement in \sigo\ between no-AO and AO data sets (after beam
smearing corrections) suggests that potential noncircular motions on
$\rm \ga 1~kpc$ scales have little impact on the measurements.
For simplicity, \sigo\ is usually referred to as ``turbulence.''

Typical dispersions measured in ionized gas are $\rm \sim 45~km\,s^{-1}$ at
$z \sim 2$, compared to $\rm \sim 25~km\,s^{-1}$ at $z \sim 0$, varying
as 
$\sigma_{0} \approx 23 + 10$$z$~$\rm km\,s^{-1}$;
cold atomic and molecular gas
measurements at $z > 0$ are scarcer but follow a similar evolution albeit
with $\rm \sim 10 - 15~km\,s^{-1}$ lower dispersions
\citep[][and references therein]{Ueb19}.
The \sigo\ evolution is consistent with that of the galactic gas mass fractions
in the framework of marginally-stable $Q \sim 1$ gas-rich disks in which
$v_{\rm rot}/\sigma_{0} \propto f_{\rm gas}^{-1}$
\citep[e.g.,][]{Genz11,Wis15,John18}.
At fixed redshift, the scatter in \sigo\ is substantial and there is evidence
that an important part of it is intrinsic to the galaxy population, but only a
weak or no trend is found between \sigo\ and global galaxy parameters such as
$M_{\star}$, SFR, $f_{\rm gas}$, mass and SFR surface densities, or inclination
\citep[e.g.,][]{Jones10a,Genz11,John18,Ueb19}.
Reasons could include limited ranges and uncertainties in properties in a given
$z$ slice, complex dependence of \sigo\ on more than one parameter, or possible
accretion-driven variations on short $\rm \la 100~Myr$ timescales as recently
proposed by \citet{Hung19} based on FIRE high-resolution numerical simulations.
In high S/N AO data of well resolved disks, no convincing trend on
spatially-resolved $\rm \sim 1 - 2~kpc$ scales has been seen either between
\sigo, $\rm \Sigma_{SFR}$, or even galactocentric radius
in the outer disk parts (away from where beam smearing corrections become
large and more uncertain; \citealt{FS18,Ueb19}).  Given the lack of clear
variations, the disk dispersions are thus taken as isotropic and radially
constant.

\looseness=-2
Constraining the physical driver(s) of the gas turbulence at high $z$ thus
still proves difficult.  This supersonic turbulence would rapidly decay within
a crossing time ($\rm \sim 10^{7}~yr$) if not continuously powered, and gas
accretion from the cosmic web, disk instabilities, and stellar feedback
have been proposed as energy sources
\citep[see, e.g., summaries by][]{Kru18,Ueb19}.
Theoretical models and numerical simulations make different
predictions as to the generated amount of gas turbulence
\citep[e.g.,][]{Aumer10,Hopkins12,Gatto15,Gold15,Gold16}.
The impact of stellar feedback varies a lot depending on the inclusion/treatment
of radiation pressure and the location where feedback is injected into the ISM,
although a general conclusion is that it can maintain galaxy-wide turbulence of
$\rm \sim 10 - 20~km\,s^{-1}$ (and is necessary to reproduce various other galaxy
properties and scaling relations).
In contrast, gravitational processes including gas transport and instabilities
within the disks appear to more easily match the observed range of \sigo\ under
the conditions prevailing at higher redshifts.
Plausibly, both forms of drivers are present as in the unified model of
\citet{Kru18}, with gravity-driving dominating at earlier cosmic times and
a gradual transition to feedback-driving at later times.
Further insights will benefit from more direct estimates of cold gas masses
in individual galaxies, and maps of the gas, SFR, and kinematics at high
spatial and velocity resolution.

\subsection{Mass and Angular Momentum Budget}
 \label{resolved_dyn.sec}

\looseness=-2
Constraints from resolved kinematics have been used to investigate
the mass budget and angular momentum of high $z$ SFGs.
At $z \sim 2$, it is important to account for the significant contribution of
gas to the baryonic component, and of the random motions to the dynamical support
In the turbulent disk framework, the circular velocity $v_{\rm c}$
(as a measure of the potential well) at radius $r$ can be computed through
$v_{\rm c}^{2} = v_{\rm rot}^{2} + 2\sigma_{0}^{2}(r/R_{\rm d})$.
Corrections can be applied for deviations from $n \approx 1$
profiles (e.g., when a massive bulge is present), and for disk ``truncation''
appreciably reducing $R_{\rm e}/R_{\rm d}$ in strongly dispersion-dominated
cases $v_{\rm rot}/\sigma_{0} \la 2$ \citep{Bur16}.
The enclosed dynamical mass can be estimated, for instance at $R_{\rm e}$,
through $M_{\rm dyn} = R_{\rm e}\,v_{\rm c}^{2}\,/\,G$, where $G$ is the
gravitational constant.
This expression is for the spherical approximation; for an infinitely thin
Freeman disk the values would scale down by $\times 0.8$.
Other methods to derive dynamical masses have been used, including a two-pronged
approach applying the rotating disk estimator neglecting pressure support for
rotation-dominated disks (i.e., taking $v_{\rm c} = v_{\rm rot}$)
and through the virial mass estimator with the integrated dispersion for
dispersion-dominated sources
($M_{\rm dyn} = \alpha R_{\rm e}\sigma^{2}/G$ with $\alpha$ in the
range $\sim 3 - 5$ typically adopted).
Forward modeling accounting for disk thickness and turbulence, and fitting
simultaneously the velocity and dispersion profiles, incorporates
self-consistently all relevant effects though comparisons with simpler
approaches as outlined above indicate overall agreement within $\rm \sim 0.2~dex$
\citep[e.g.,][]{FS18}.

\subsubsection{Dynamical vs.\ baryonic mass estimates} 
  \label{resolved_dyn_mdyn.sec}

\looseness=-2
In the most straightforward approach to constraining the mass budget, global
dynamical mass estimates are compared to stellar and gas mass estimates.
Studies based on near-IR IFU or slit spectroscopy data generally concur on
overall elevated baryonic mass fractions
$f_{\rm bar} = \left(M_{\star} + M_{\rm gas}\right)/M_{\rm dyn}$, with
large scatter, among $z \sim 2$ SFGs \citep[e.g.,][]{FS09,Sto16,Pri20}.
Modeling deep H$\alpha$ kinematic data over a wide $M_{\star}$ range across
$z \sim 0.7 - 2.7$ from the \kmostd\ survey in legacy fields providing detailed
constraints on galaxy stellar and size properties, \citet{Wuy16b} found a large
rise in \fbar\ derived within the central $1\,R_{\rm e}$ regions from $\sim 45\%$
at $z \sim 0.9$ to $\sim 90\%$ at $\sim 2.3$ and a modest increase in stellar
mass fraction $f_{\star} = M_{\star}/M_{\rm dyn}$ from $\sim 30\%$ to $\sim 40\%$,
reflecting the $f_{\rm gas}$ evolution.  The scatter at fixed $z$ is driven by
positive correlations with average stellar and gas mass surface densities at
$< R_{\rm e}$.  These trends hold when accounting
for mass incompleteness or considering only progenitors of $z = 0$
$\log(M_{\star}/{\rm M_{\odot}}) \geq 10.7$ galaxies, and are fairly robust
to SED modeling assumptions or gas scaling relations among plausible choices.
At $z \sim 2$, the \mdyn-based
results thus leave little room for DM mass contribution ($f_{\rm DM}$) within
the $\sim 1 - 2\,R_{\rm e}$ probed by the observations.
Noting that the analyses above are for a Chabrier IMF, more bottom-heavy
galaxy-wide IMFs such as a Salpeter slope down to $\rm 0.1~M_{\odot}$
would also be disfavored.

\begin{marginnote}[95pt]
 \entry{TFR}{Tully-Fisher relation, linking measures of a galaxy
             mass~and~kinematics.  Various forms are considered,
             involving $M_{\star}$ or $M_{\rm bar}$ as a function of
             \vrot, $v_{2.2}$, \vcirc, or $S_{0.5}$.}
\end{marginnote}

\subsubsection{Tully-Fisher relation} 
  \label{resolved_dyn_tfr.sec}

The Tully-Fisher relation (TFR) relates measures of galaxy mass to the full
potential well; it is thus sensitive to the galactic baryonic content and
can place powerful constraints on cosmological disk formation models
\citep[e.g.,][among many others]{Mo98,Dut07,Som08,McG12}.
Kinematic
studies agree on the existence of a TFR out to $z \sim 3$ and on the
reduced scatter about the relation when accounting for pressure support in the
turbulent high $z$ disks but with mixed outcome as to the evolution, ranging
from none over $z \sim 0 - 1$ \citep[e.g.,][]{Kassin07,Miller12,Tiley19a} to
significant in the sense of lower disk mass at fixed velocity for $z \sim 0.6-3.5$
samples \citep[e.g.,][]{Cre09,Turner17,Ueb17}.
The conclusions hinge on several interrelated factors including the adopted
form and parametrization of the relation, galaxy sample properties, and choice
of reference $z \sim 0$ TFR \citep{Ueb17,Tiley19a}.
The range in galaxy parameters spanned by the high $z$ data sets generally
hampers reliable fits to the slope of the relation, such that the evolution
is usually quantified in terms of the zero-point (ZP) obtained assuming a
non-evolving slope.  The magnitude of the ZP offsets also depends on whether
the relation is constructed from the stellar or the baryonic mass, and from
\vrot, $v_{2.2}$, \vcirc, or $S_{0.5}$.

\looseness=-2
Exploiting the wide $0.7 < z < 2.7$ baseline from \kmostd, the study of
\citet{Ueb17} provided the most self-consistent constraints across cosmic noon
based on H$\alpha$ kinematics from IFU observations, identical analysis method,
selection through uniform data quality, galaxy parameters, and stringent disk
criteria, with resulting $\log(M_{\star}/{\rm M_{\odot}}) > 10$ subsamples well
matched in $M_{\star}$ and location around the MS and mass-size relations.
Focussing on (fixed-slope) relations in terms of $v_{\rm c}$, the stellar TFR
shows no significant ZP evolution from $z \sim 2.3$ to $\sim 0.9$ while the
baryonic TFR ZP exhibits a negative evolution (lower \mbar\ at fixed \vcirc),
and both relations imply similar positive evolution since $z \sim 0.9$ compared
to published $z \sim 0$ TFRs.
In the latter redshift interval, \citet{Tiley19a} found instead little, if any,
evolution in terms of $M_{\star} - v_{2.2}$ using matched data quality, methods,
and samples over $\log(M_{\star}/{\rm M_{\odot}}) \sim 9 - 11$ from the KROSS
and local SAMI IFU surveys of H$\alpha$.
The persisting discrepancies around $z \sim 1$ underscore the importance of
disentangling observationally- and physically-driven effects in order to
establish firmly the evolution and explore the residuals of the TFR across
all of $0 < z < 3$.

\subsubsection{Outer disk rotation curves} 
  \label{resolved_dyn_rcout.sec}

Constraining the mass distribution from the shape of the rotation curve (RC)
alleviates the uncertainties of global M/L conversions for the baryonic
components.  This approach is challenging at $z \sim 2$ with current
instrumentation, as tracing emission line kinematics beyond
$\sim 1 - 2\,R_{\rm e}$ requires very long integrations.

Recent results from very sensitive H$\alpha$ IFU data of a handful of
massive $z \sim 1-2.5$ star-forming disks extending to $r \sim 10 - 20~{\rm kpc}$
showed significant and symmetric drops in the individual RCs beyond their peak
\citep{Genz17}.  Similar falloffs in stacked H$\alpha$ RCs reaching
$\sim 3.5 - 4\,R_{\rm e}$ constructed from high quality IFU data of $\sim 100$
typical $\log(M_{\star}/{\rm M_{\odot}}) \ga 10$ star-forming disks suggested
that this behaviour may be widespread at high $z$, and on average more pronounced
towards higher $z$ and lower \vsigo\ disks \citep{Lang17}.  The outer slopes for
these samples are nearly Keplerian, and in stark contrast to the flat or rising
RCs of local spirals.
The falloffs can be naturally explained as the imprint of baryons strongly
dominating the mass over the regions probed by the kinematics together
with sizeable levels of pressure support maintained well past the RC peak.
The more detailed constraints from the individual extended RCs and dispersion
profiles, simultaneously fit with turbulent disk $+$ bulge $+$ DM halo models,
yield $f_{\rm DM}(R_{\rm e}) \la 20\%$ with the 3/6 galaxies at $z > 2$ having
the lowest fractions.
In turn, the stacked RC is best matched by models with a high fraction of total
bayonic disk mass to DM halo mass $m_{\rm d} \sim 0.05$, in line with analysis
of the angular momenta of a larger sample of $z \sim 0.8 - 2.6$ SFGs \citep{Bur16}
and consistent with abundance matching results once accounting for the high
\fgas\ at high $z$ \citep[$M_{\star}/M_{\rm halo} \sim 0.02$;][]{Mos13,Beh13a}.
Although the exact numbers depend on details of the distribution of the mass
components, the implications of low central DM fractions and an overall high
disk to DM halo mass ratio were shown to be fairly robust to the assumptions
within plausible ranges.

\begin{figure}[!t]
\centering
\includegraphics[scale=0.53,trim={4.0cm 3.0cm 1.0cm 1.5cm},clip=0,angle=0]{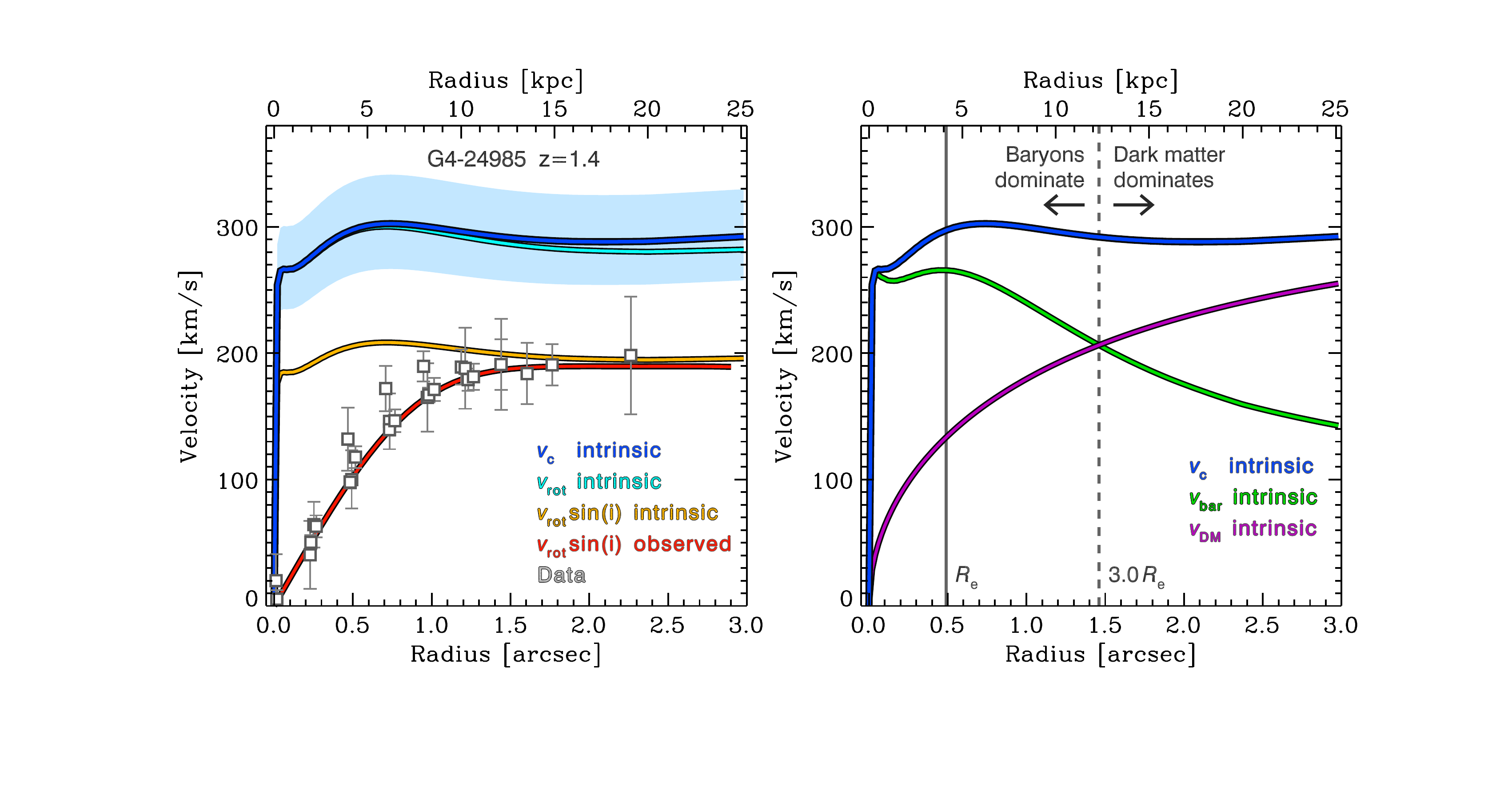} 
\renewcommand\baselinestretch{0.85}
\caption{
Example kinematic modeling of a massive $z$$=$1.4 SFG with sensitive H$\alpha$
and CO 3-2 observations, a bulge-to-total mass ratio of $\sim 0.25$, and
large $v_{\rm rot}/\sigma_{0} \sim 10$ (from \citealt{Ueb18}).
{\it Left:\/} RC in observed and intrinsic space.
The observed, folded H$\alpha$ and CO velocity curve (grey squares) extends
to 18~kpc.
The best-fit model curve of the circular velocity ($v_{\rm c}$) in intrinsic
space is plotted as blue line
(with blue shading showing the $1\sigma$ uncertainties of the inclination
correction).  The other lines show, successively, the effects of pressure
support (i.e., the $v_{\rm rot}$ curve; cyan line) that are minimal
in this galaxy, the effects of inclination ($v_{\rm rot} \times \sin(i)$;
yellow line), and the resulting beam-smeared velocity curve in
observed space (red line).
{\it Right:\/} The relative contribution to the model $v_{\rm c}$
in intrinsic space (blue line) from the baryons and from the DM halo
(green and purple lines, respectively).
Baryons strongly dominate within the half-light radius while DM starts to
dominate the mass budget beyond $\rm \approx 12~kpc$ or $\approx 3\,R_{\rm e}$
(vertical solid and dashed lines).
}
\label{kinmod.fig}
\end{figure}

\begin{figure}[!ht]
\centering
\includegraphics[scale=0.65,trim={41.0cm 6.5cm 39.0cm 6.5cm},clip=1,angle=0]
                {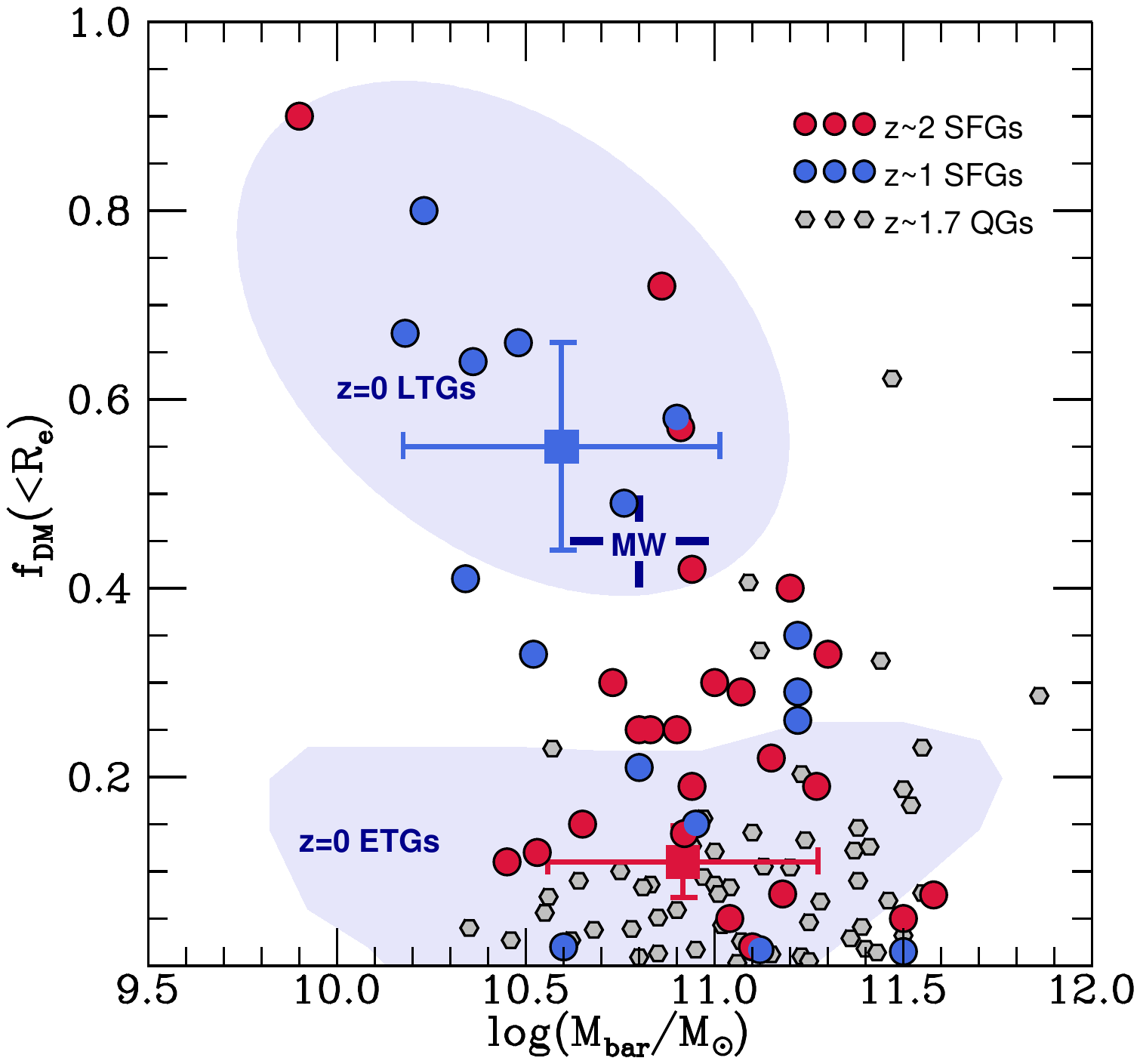}
\renewcommand\baselinestretch{0.85}
\caption{
Location of 40 massive $z \sim 0.7 - 2.7$ SFGs in a diagram of the baryonic
mass vs.\ inner ($< R_{\rm e}$) DM mass fraction.
The $f_{\rm DM}(R_{\rm e})$ is derived from kinematic modeling (as shown in
Figure\ \ref{kinmod.fig}) of high quality H$\alpha$
and CO RCs (and velocity dispersion
profiles) extending out to radii in the range $\rm 10 - 20~kpc$ from the samples
of \citet{Genz17,Genz20} and \citet{Ueb18}
(red and blue circles for $z \sim 2$ and $z \sim 1$ disks), with
typical uncertainties of $\pm 0.1-0.15$ for individual $f_{\rm DM}(R_{\rm e})$
estimates.
Median $f_{\rm DM}(R_{\rm e})$ from modeling the inner regions kinematics
of larger SFG samples at $z \sim 2.3$ (red square) and $z \sim 0.9$
(blue square) from \citet{Wuy16b} are overplotted, as well as results from
quiescent galaxies at $z \sim 1.7$ based on stellar velocity dispersions
presented by \citet{Mendel20}.
Approximate areas occupied by $z = 0$ massive early-type
and late-type galaxies (ETGs, LTGs; from \citealt{Cap13} and \citealt{Mart13})
and the Milky Way \citep{BH16} are indicated with blue shading and the
cross-hair symbol.
}
\label{Mbar_fDM.fig}
\end{figure}

These findings spurred a number of follow-up studies, reporting mixed results.
For instance, \citet{Tiley19b} concluded that the averaged H$\alpha$ outer
RCs at $z \sim 0.9 - 2.5$ are flat or rising, in contrast to \citet{Lang17}.
As noted by both groups, stacking methodology matters.
Re-scaling the data according to the observed radius $R_{\rm max}$ and velocity
at the RC peak as favored by \citeauthor{Lang17} is more sensitive to the relative
concentration of baryons vs.\ DM
and, although it relies on detecting a change of slope in the inner velocity
gradient, possible biases against recovery of flat or rising RCs were shown
to be unlikely.
Normalizing instead with the radius and velocity in observed space corresponding
to $3\,R_{\rm d}$ based on $n = 1$ fits to the morphologies as favored by
\citeauthor{Tiley19b} probes the baryonic to DM content on more global scales
but any spread in $R_{\rm max}/R_{\rm d}$ from a range in S\'ersic indices
would smear the peak in the composite RC.
Importantly, the stacked samples are different, with the \citeauthor{Lang17}
stricter disk selection resulting in higher mass and \vsigo\ ranges compared
to \citeauthor{Tiley19b}.
Comparisons are therefore not straightforward but given the large variations
in $f_{\rm DM}$ with radius, the conclusion of \citet{Tiley19b} that within
$6\,R_{\rm d} \approx 3.6\,R_{\rm e}$ high $z$ SFGs are DM-dominated is not
necessarily incompatible with them being strongly baryon-dominated within
$1\,R_{\rm e}$, even when displaying a flat outer RC in high \vsigo\ cases
\citep[Figure\ \ref{kinmod.fig} and, e.g.,][]{Ueb18}.
On-going extensions to several tens of $z \sim 0.7 - 2.7$ disks with high
quality individual kinematics data are revealing ever more clearly a dependence
with galaxy mass, redshift, and measures of central baryonic mass concentration
\citep{Genz20}, which were apparent in some previous outer RC studies.
These trends echo the findings from the global mass budget 
(Section\ \ref{resolved_dyn_mdyn.sec}), account for the strong baryon
dominance to $r \sim 8~{\rm kpc}$ reported by \citet{vD15} for 10 compact
massive SFGs $2 < z < 2.5$ from the declining composite RC inferred from
integrated H$\alpha$ line widths, and explain the range of conclusions
from different outer RC samples.

\subsubsection{Angular momentum}
  \label{resolved_dyn_angmom.sec}

\looseness=-2
The connection between $z \sim 2$ SFGs and their host DM halos has
been further explored via measurements of the specific angular momentum
$j_{\rm d} \propto v_{\rm rot} \times R_{\rm e}$.
The inferred halo scale angular momenta are broadly consistent in mean and
scatter with the theoretically predicted lognormal distribution of halo spin
parameters $\lambda$, and the $j_{\rm d}$ estimates scale approximately as
$\propto M_{\star}^{2/3}$ similar to the theoretical
$j_{\rm DM} \propto M_{\rm DM}^{2/3}$ \citep[e.g.,][]{Bur16,Har17,Swi17}.
The long made assumption that {\it on average\/}
$j_{\rm d}/j_{\rm DM} \sim 1$, expected if disks retain most of the
specific angular momentum acquired by tidal torques in their early formation
phases and shown to hold for local spirals \citep[e.g.,][]{Fall13}, thus appears
to be borne out by observations up to $z \sim 2.5$.  Even in a population-wide
sense, this finding is not trivial given that (i) infalling baryons can lose and
gain angular momentum from the virial to the disk scale \citep[e.g.,][]{Dan15},
(ii) angular momentum can be efficiently redistributed in and out of galaxies
\citep[e.g.,][]{Dek09a,Ueb14,Bournaud16}, and (iii) $< 15\%$ of the cosmically
available baryons are incorporated into the stellar component of galaxies
\citep[$m_{\rm d,\star} \approx 0.02$; e.g.,][]{Mos13,Beh13a} and at most
30\% when including gas at $z \sim 2$ \citep{Bur16}.
Although simulations and semi-analytical models are now able to produce
realistic distributions of galaxy size, specific angular momentum, and
stellar-to-halo mass ratios, there is no consensus yet on how various
mechanisms interact to preserve {\it net\/} angular momentum
\citep[e.g.,][]{Genel15,Zav16,Jiang19}.

The observed scatter in specific angular momenta has an intrinsic component
at all epochs.  The low $j_{\rm d}$ tail encompasses massive early-type spirals
and ellipticals at $z \sim 0$, and more dispersion-dominated (and unstable) as
well as more centrally concentrated star-forming disks in the high $z$ samples.
These correlations reflect an underlying mass-spin-morphology relation that
likely underpins the Hubble sequence \citep[e.g.,][]{Obr14} and may suggest that
``disk settling'' with cosmic time (see Section\ \ref{resolved_rotdisk.sec}) is
driven at least in part by angular momentum evolution \citep[e.g.,][]{Swi17}.
Noting that central mass concentration increases with galaxy mass and
thus ``disk maturity'' (see Sections\ \ref{resolved_axisymm.sec} and
\ref{resolved_rotdisk.sec}), \citet{Bur16} found in their $z \sim 0.7 - 2.6$
sample of star-forming disks a significantly weaker anticorrelation between
$\lambda \times (j_{\rm d}/j_{\rm DM})$ and central stellar surface density
$\Sigma_{\rm 1kpc}$ than with the galaxy-averaged $\Sigma_{\star}$ and
$\Sigma_{\rm gas}$ --- a result suggesting that accumulation of (low angular
momentum) material in the galaxy centers may be decoupled from the processes
that set the global disk scale and angular momentum.

\subsubsection{Interpreting the mass and angular momentum budget}
  \label{resolved_dyn_disc.sec}

A consistent picture appears to be emerging in which
$\log(M_{\star}/{\rm M_{\odot}}) \ga 10$ star-forming disks at $z \sim 2$
are typically baryon-rich on the physical scales probed by the data, with
lower DM mass contribution at $< 1 R_{\rm e}$ among more massive, centrally
denser, and higher $z$ galaxies.
These trends are qualitatively reproduced by matched populations
(in $M_{\star}$, SFR, $R_{\rm e}$) in recent cosmological numerical
simulations \citep[e.g.,][]{Wuy16b,Swi17,Lov18,Tek18}.
While the role of the evolving gas content can be easily understood,
the trends in mass fractions and ZP of the TFRs point to differences in the
relative distribution of baryons vs.\ DM on galactic scales among SFGs of
comparable masses at different cosmic epochs.
These have been ascribed to a combination of
(i) disk size growth at fixed mass, where the baryons at lower $z$ extend
further into the surrounding DM halo,
(ii) evolving DM halo profiles, with shallower inner profiles at earlier times
\citep[e.g., less concentrated, or more cored;][]{Mart12,Dut14}, and
(iii) efficient dissipative processes in the gas-rich environments at higher
$z$ concentrating baryons in the central regions \citep[e.g.,][]{Dek14,Zol15}.
The weaker coupling between $\lambda \times (j_{\rm d}/j_{\rm DM})$ and
$\Sigma_{\rm 1kpc}$ vs.\ $\Sigma_{\star}$ and $\Sigma_{\rm gas}$ \citep{Bur16}
would naturally result from inward radial gas transport through the latter
processes.

\looseness=-2
The kinematically inferred low $f_{\rm DM}(R_{\rm e})$ of massive $z \sim 2$
star-forming disks overlaps with the range for $z \sim 0$ massive early-type
galaxies --- their likely descendants.  This echoes evolutionary links
based, for instance, on the stellar sizes and central mass densities
(Section\ \ref{global_ReM.sec} and \ref{resolved_axisymm.sec}), and on the
fossil record \citep[e.g.,][]{Cap16}.
Current $z \sim 2$ results are summarized in Figure\ \ref{Mbar_fDM.fig}
(following \citealt{Genz17,Ueb18}) incorporating an expanded sample of
individually modeled RCs \citep{Genz20}.
The inverse dependence of $f_{\rm DM}(R_{\rm e})$ with galaxy mass (and mass
concentration) is reminiscent of the trends observed in local disks captured
by the unified picture of \citet{Cou15}.  In this picture, the outward moving
transition from baryon-dominated center to DM-dominated outskirts (relative
to a fiducial $2.2\,R_{\rm d} \approx 1.3\,R_{\rm e}$ for $n \sim 1$) in more
massive systems can be tied to the disk size -- circular velocity -- stellar
mass scaling relations, with scatter in $f_{\rm DM}$ attributed at least in
part to size variations at fixed $v_{\rm c}$.
The differentiation in $f_{\rm DM}(R_{\rm e})$ at fixed mass seen between local
early- and late-type galaxies \citep[e.g.,][]{Cou15} also appears to be present
at cosmic noon \citep[e.g.,][]{Mendel20}, which plausibly is rooted in the same
processes that lead to the distinction between SFGs and quiescent galaxies in
their stellar structural properties \citep[e.g.,][]{Lang14,vdWel14a}.

By necessity, the kinematics of $z \sim 2$ star-forming disks are interpreted
in a simplified axisymmetric framework with circular orbital motions.
Observations of local disks indicate 
frequent deviations from this simple assumption caused, for instance, by
interactions, warps and other such dynamical instabilities, and radial motions,
which are difficult to constrain at the typical resolution and S/N of high-$z$
data.  Signatures of the latter are discussed in the next Section.
Bending instabilities, such as warping or buckling, may be
expected to be suppressed or short-lived in gas-rich turbulent disks
\citep[see the discussion by][]{Genz17}.
Minor interaction-induced perturbations may not be ruled out but the exclusion
of galaxies with potential companions wherever possible should reduce their role
in disk samples.
The validity of the disk framework for low-mass objects may be called into
question in light of the increasing prevalence of prolate and/or triaxial
systems towards lower masses and higher $z$ suggested by statistical studies
of the morphological axial ratios (Section\ \ref{resolved_devsymm.sec}),
although this may be a lesser concern when applying the morpho-kinematic disk
criteria (notably the requirement of kinematic and morphological major axes
alignment; e.g., \citealt{Fra91}).
Furthermore, the generally small and spatially flat residuals in velocity and
dispersion maps compared to axisymmetric disk models (resulting from the disk
selection criteria employed in most studies) suggest that the potential impact
of minor merger perturbations and prolateness/triaxiality is small in
the kinematic analyses.

Cosmological simulations are useful to assess the validity of assumptions
made in interpreting data under more realistic high redshift environments.
For instance, \citet{Wellons20} quantified the effects of pressure gradients,
noncircular motions, and asphericity in the gravitational potential on the
rotation velocity and \mdyn\ estimates in high-resolution FIRE numerical
simulations of a range of massive turbulent disks at $1 \la z \la 3$ based on
the mass particle distributions, finding that pressure support usually makes
the largest impact and that when it is accounted for, kinematically-derived
mass profiles agree with the true enclosed mass within $\sim 10\%$ typically
over the $r \la 10 - 20~{\rm kpc}$ range explored.
Realistically replicating observables and empirical methodologies from
simulations is not straightforward and subject to various limitations
(numerical resolution, sub-grid recipes, radiative transfer, ...) but the
informative potential is motivating a growing number of investigations to
improve on both the simulation ingredients and data interpretation.

\subsection{Deviations from Disk Rotation}
  \label{resolved_perturb.sec}

In kinematics data of $z \sim 2$ SFGs, modest deviations from regular patterns
are seen in a subset of galaxies otherwise consistent with global disk rotation.
Interpreting such kinematic asymmetries is not trivial in high $z$ data but 
can plausibly be ascribed to internally- or externally-induced in-disk inflows,
or to outflows.
As will be discussed further in the next Section, the emission associated with
the latter has a broad velocity distribution but low amplitude, and should have
a negligible effect on the single-component line profile fitting that is usually
performed in extracting 2D kinematic maps, unless the outflow is particularly
strong \citep{FS18}\footnote{
 In integrated or slit spectra, the presence of outflows can however inflate
 the line widths and lead to overestimates of $M_{\rm dyn}$ based thereupon
 \citep{Wis18}.
}.
The gas-rich environments prevailing at $z \sim 2$ are expected to naturally
promote perturbations in the marginally-stable $Q \sim 1$ gas-rich disks,
with fragmentation and efficient transport of material towards the center
via, e.g., inward gas streaming and clump migration, while the gas reservoirs
of galaxies are continuously replenished by anisotropic accretion via streams
and minor mergers.
Material streaming inwards can induce twists in the isovelocity contours
and off-center peaks in the dispersion map at the level of a few tens of
$\rm km\,s^{-1}$
($v_{\rm rad} \sim 2 \times \sigma_{0} \times (\sigma_{0}/v_{\rm rot})$),
and differences in magnitude and orientation of the angular momentum between
inner and outer regions expected to remain even after bulge growth slows
\citep[e.g.,][]{vdK78,Cap16}.
Characteristic signatures thereof are indeed identified in some of the $z \sim 2$
galaxies with high S/N, high resolution AO-assisted observations, along with
morphologically identified bar- and spiral-like features in some cases
\citep[e.g.,][]{Genz06,Law12a,FS18}.
These processes may be important in bulge and SMBH buildup, and
concurrent disk growth through angular momentum redistribution
\citep[e.g.,][]{Bournaud14,Dek14,Zol15}.
The ubiquity of dense stellar cores and large nuclear concentrations of cold gas
in massive $z \sim 2$ SFGs, and the weak correlation between disk-scale angular
momentum with $\Sigma_{\rm 1kpc}$ call for further sensitive and high resolution
kinematics data to more directly assess the role of radial gas transport at cosmic
noon vs.\ alternative scenarios such as inside-out galaxy growth
\citep{vD15,Lil16}.

Strong kinematic distortions are generally interpreted as indicative of
major merging.  Assuming very simplistically that all SFGs not identified
as rotation-dominated disks according to the classification scheme of
Section\ \ref{resolved_rotdisk.sec} are major mergers,
the fractions thereof would be $\sim 25\% - 40\%$ at $z \sim 1 - 2.5$ and
$\log(M_{\star}/{\rm M_{\odot}}) \ga 10.5$ (depending on the exact set of
criteria and $z$; \citealt{Wis19}).  These fractions are comparable to those
inferred from morphologies and close pair statistics in a similar mass range
\citep[e.g.,][]{Con14,Lop13,Rod18}, and consistent with cosmological
simulations \citep[e.g.,][]{Genel08,Snyder17}.
Taking the major merger fraction as $1 - f_{\rm disk}$ is obviously an
oversimplification.  Shallower data are more biased towards high surface
brightness regions that may partly and unevenly sample the full system and
result in apparently disturbed kinematics, exacerbated for clumpy morphologies
\citep[see Fig.\ 9 of][]{FS18}.  A poorly resolved, low \vsigo\ object does not
necessarily imply it is a major merger \citep{New13}.  More face-on disks may
also be more difficult to identify because of the resulting small projected
velocity gradient, reduced central dispersion peak, and possible clumps biasing
the determination of morphological position angle and center \citep{Wuy16b}.
As is the case for morphologies, kinematic signatures of
interactions depend strongly on the
system's orbital configuration, the properties and mass ratio of the progenitor
galaxies, the sightline, and the merger stage
\citep[e.g.,][]{Bello16,Sim19}, introducing uncertainties in identifying major
mergers.  Despite these uncertainties, the kinematic mix among
$\log(M_{\star}/{\rm M_{\odot}}) \ga 10$
SFGs at $z \sim 2$ suggests a dominant timescale in a disk configuration,
consistent with several other lines of evidence from
scaling relations of galaxy properties pointing to the importance of processes
other than major merger events in building up stellar mass and structure.

\subsection{Galactic-scale Outflows}
 \label{resolved_outflows.sec}

\looseness=-2
Galactic winds are thought to play a critical role in the evolution of
galaxies by regulating their mass build-up, size growth, star formation, and
chemical enrichment, by redistributing angular momentum, and by mediating the
relationship between SMBHs and their host galaxies.
Stellar feedback at low galaxy masses expels gas from the shallow
potential wells, reducing the reservoirs fueling star formation and
keeping a low galactic metal content \citep[e.g.,][]{Dek86,Dav17}.
Above the Schechter mass $\log(M_{\star}/{\rm M_{\odot}}) \sim 10.7$
(or $\log(M_{\rm halo}/{\rm M_{\odot}}) \ga 12$), accreting SMBHs are thought
to be important in suppressing star formation through ejective ``QSO mode''
feedback driving powerful winds during high Eddington ratio phases that sweep
gas out of the host galaxy, and subsequent preventive ``radio mode'' feedback
maintaining galaxies quenched by depositing kinetic energy into the halo that
inhibits cooling alongside virial shocks \citep[see review by][]{Fab12}.

\begin{marginnote}[270pt]
 \entry{QSO}{Quasi-stellar object (quasar).}
 \entry{IS}{Interstellar.}
\end{marginnote}

Galactic winds should be particularly effective at the peak epoch of star
formation and SMBH accretion rates.
The most easily accessible diagnostics at high $z$ are rest-UV to optical
interstellar (IS) absorption features and nebular emission lines, which probe
neutral and warm ionized gas phases.  Winds are identified through their
kinematic imprint: centroid velocity offsets and broad wings of blueshifted
IS absorption relative to the systemic redshift (e.g., from stellar features),
redshifted Ly$\alpha$ profile (accessible at $z > 2$), and broad line emission
typically underneath a narrower component arising from star-forming regions
in the galaxy.\footnote{
While common, a blueshift for emission line tracers is not necessary;
depending on outflow geometry and extinction by dust in the galaxy,
backside receding gas can be detected.
}
Alongside understanding the physical drivers of outflows, a major goal of
studies at high $z$ is to assess their role in galaxy evolution.
To this aim, population-wide censuses are essential to reveal the global
and time-averaged impact of outflows, reducing biases from selection
on properties that would be closely linked to the strongest activity.
Such censuses have been greatly facilitated with the advent of optical and
near-IR MOS and IFU instruments.  IFU observations have proven particularly
powerful, by (i) locating the launching sites and constraining
the extent of outflowing gas, and (ii) facilitating the separation between
large-scale gravitationally-driven and outflow-related motions that both
contribute to velocity broadening in integrated spectra.

\subsubsection{Outflow Demographics at $z \sim 2$}
  \label{resolved_outflows_demo.sec}

\begin{figure}[!t]
\centering
\includegraphics[scale=0.63,trim={0.0cm 2.4cm 0.0cm 1.3cm},clip=0,angle=0]
                {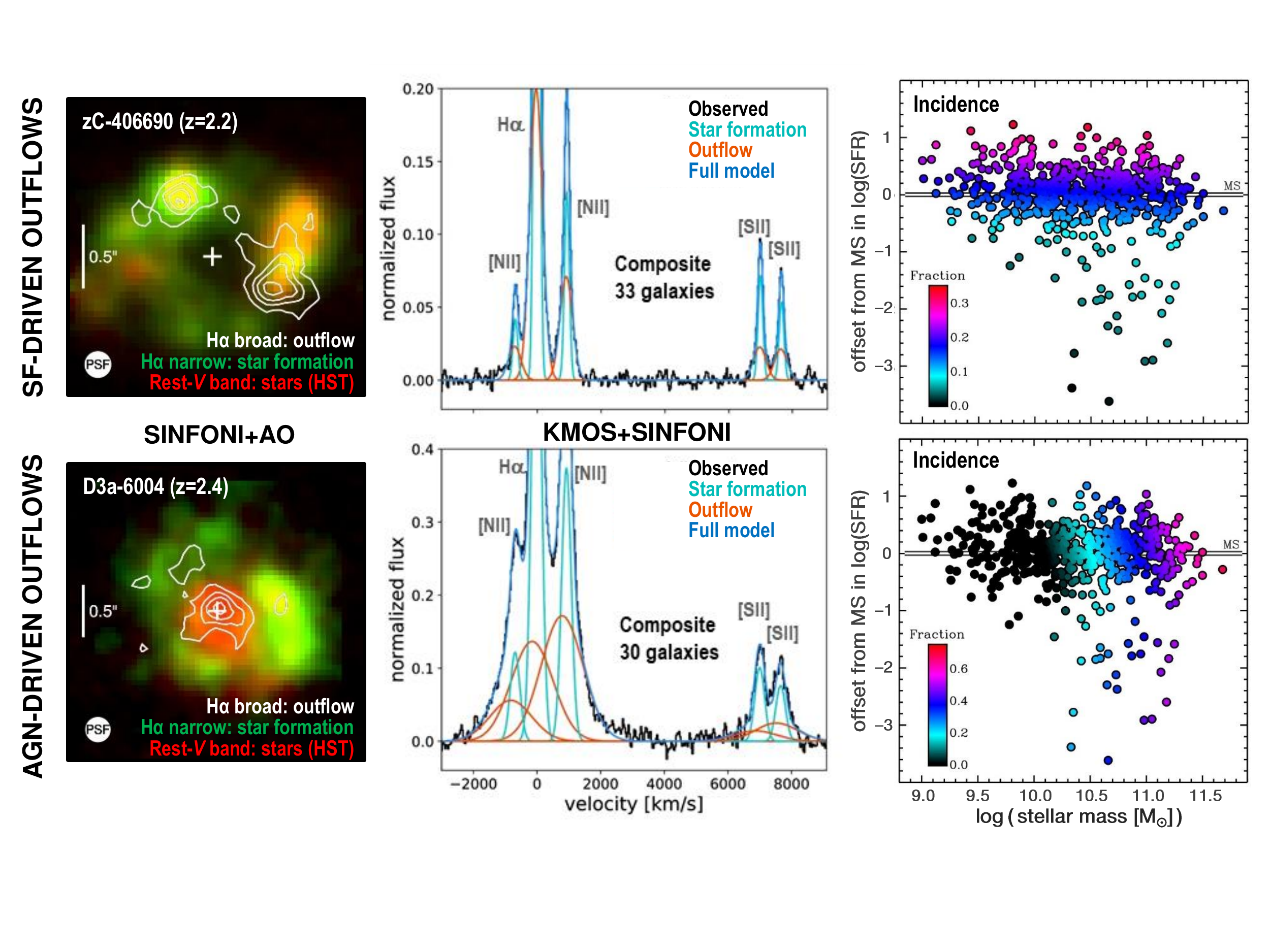}
\renewcommand\baselinestretch{0.85}
\caption{
\looseness=-2
Distinction between star formation- and AGN-driven ionized gas winds at
$z \sim 1 - 3$ (top and bottom rows, respectively), in terms of spatial,
spectral, and demographic properties (left to right).
The maps show two galaxies observed with SINFONI$+$AO and {\it HST\/} at
FWHM resolution of $\rm \sim 1.8~kpc$, with the stellar rest-optical
light and narrow H$\alpha$ emission from star-forming regions shown in red and
green colors, and the broad H$\alpha$$+$[N{\small II}] outflow emission shown
in white contours.
The composite spectra are constructed from near-IR IFU observations with
KMOS and SINFONI, where the continuum was subtracted and large-scale orbital
motions were removed based on the narrow H$\alpha$ velocity maps prior to
stacking.
The demographic trends are based on the fraction of individual objects
exhibiting the spectral signatures of SF- and AGN-driven outflows.
Figure based on data published by \citet{New12a} and \citet{FS14, FS19}.
}
\label{outflows.fig}
\end{figure}

\looseness=-2
Much like in the nearby Universe \citep[e.g.,][]{Vei05}, SF- and AGN-driven
winds at high $z$ are distinguished on the basis of their velocities,
spatial origin, and excitation properties (Figure~\ref{outflows.fig}).
SF-driven outflows with velocities up to several $\rm 100~km\,s^{-1}$ are detected
from shifted IS absorption and Ly$\alpha$ emission \citep[e.g.,][]{Sha03,Wei09},
and from broad $\rm FWHM \sim 400 - 500~km\,s^{-1}$ emission in H$\alpha$, \nii,
and \sii\ on galactic and sub-galactic scales\begin{marginnote}[10pt]
 \entry{FWHM}{Full width at half maximum.}
\end{marginnote}\citep[e.g.,][]{Genz11,New12a,New12b}.
In deep $\rm \sim 1 - 2~kpc$ resolution IFU$+$AO observations, the broad emission
arises from extended regions across the galaxies and is often enhanced near bright
star-forming clumps.  The line excitation properties are consistent with dominant
photoionization by young stars and possibly modest contribution by shocks.
Faster AGN-driven winds with velocities up to a few $\rm 1000~km\,s^{-1}$ in
$z \sim 2$ galaxies hosting luminous $\log(L_{\rm AGN}/{\rm erg\,s^{-1}}) > 45$
AGN are identified from various rest-UV/optical tracers
\citep[see reviews by][]{Fab12,Hec14}.
In near-IR observations, spatially extended broad emission with typical
$\rm FWHM \sim 1000 - 2000~km\,s^{-1}$ is detected in Balmer as well as
forbidden \nii, \sii, and \oiii\ emission
\citep[precluding significant
 contribution from high-density broad-line region gas;]
 []{Nes08,Can12,Genz14b,Cre15}.
It typically originates from the center of galaxies, can
extend over $\rm 5 - 10~kpc$ for luminous QSOs, and both broad and narrow
component line ratios indicate high excitation.

SF- and AGN-driven winds follow distinct demographic trends, most clearly
revealed in a recent near-IR IFU study of a sample of $\sim 600$ primarily
mass-selected galaxies at $0.6 < z < 2.7$, covering a wide range in both
mass and star formation activity levels
($9.0 < \log(M_{\star}/{\rm M_{\odot}}) < 11.7$ and $\rm -3.6 < \Delta MS < 1.2$,
see Figure\ \ref{outflows.fig}; \citealt{FS19}).
SF-driven outflows are observed at all masses, with an incidence that correlates
mainly with star formation properties, and more specifically the MS offset,
specific and absolute SFR, and $\Sigma_{\rm SFR}$.
In contrast, the incidence of AGN-driven outflows (identified based on the
combination of rest-optical line profiles and multi-wavelength AGN diagnostics)
depends strongly on stellar
mass and measures of central stellar mass concentration, irrespective of the
level and intensity of star formation activity.
The strong differentiation in resulting stacked spectra and decoupling in
incidence trends suggest little cross-contamination between {\it dominant\/}
SF- and AGN-driven winds.

Several aspects are important in interpreting demographics and comparing between
studies.  In both nebular emission and IS absorption tracers, the ability to
detect an outflow depends on the strength of the wind signature (along with
S/N and spectral resolution of the data), such that the trends in incidence
partly reflect trends in outflow properties.
Slower or weaker winds are more difficult to detect, especially in nebular lines
because of the blending with emission from star formation, which underscore the
advantage of IFU data in enabling the removal of large-scale orbital motions of
the host galaxy.
IS absorption features integrate along the line of sight, are sensitive to
outflowing material over a wider range and to lower gas densities, and probe
material over physical scales up to tens of kpc hence plausibly average over
longer timescales.
In turn, the emission line technique is more sensitive to denser material
closer to the launching sites (as evidenced by high-resolution IFU maps),
making it a more instantaneous probe of outflows.
Differences in spatial scales probed, along with possibly less collimated
winds in ``puffier'' higher $z$ galaxies \citep{Law12c}, may lead to different
dependences with galaxy inclination.
Given the trends with galaxy properties discussed below, results will also
depend on the sample selection and parameter space coverage.

For SF-driven winds, qualitatively similar trends in incidence, or in strength
and velocity width of the wind signature, with measures of star formation
activity have been found in many other studies.  Quantitatively, there are some
notable differences especially between studies using different techniques.
For instance, among the full near-IR sample studied by \citet{FS19}, the global
fraction of SF-driven outflows is $\sim 11\%$, reaches $\sim 25\% - 30\%$ at
$\rm \Delta MS \ga 0.5~dex$ or
$\rm \Sigma_{SFR} \ga 5 - 10~M_{\odot}\,yr^{-1}\,kpc^{-2}$; no strong trend
with galaxy inclination is found \citep[see also, e.g.,][]{New12b}.
These fractions are lower than the $\ga 50\%$ based on the occurrence of
blueshifted IS absorption lines after accounting for anisotropic geometry
(trends with inclination in these studies are stronger) and clumpiness
of the outflowing gas \citep[e.g.,][]{Wei09, Kor12}.
These differences in incidence are consistent with different physical and time
scales of outflows probed by each technique and possibly reflect differences
in sample selection (mass- vs.\ UV-selected).
The interdependence between SFR, $M_{\star}$, and $z$, and the choice
of criteria employed to identify/exclude AGN, may introduce residual trends
with $M_{\star}$ \citep[e.g.,][]{Wei09,Free19,Swi19}.
In general, SF-driven outflows appear to become most prominent above
$\rm \Sigma_{\rm SFR} \ga 0.5 - 1~M_{\odot}\,yr^{-1}$, suggesting a higher
threshold for wind breakout that may be related to the geometrically thicker,
denser, and more turbulent ISM in high $z$ galaxies \citep[e.g.,][]{New12b,Dav19}.

\looseness=-2
Turning to AGN-driven outflows, near-IR studies considering the full galaxy
population highlighted a steep increase in incidence towards higher masses,
most pronounced above $\log(M_{\star}/{\rm M_{\odot}}) \sim 10.7$
\citep[e.g.,][]{Genz14b,FS19,Leu19},
qualitatively tracking the behavior for AGN fractions identified in flux-limited
surveys (e.g., in X-rays; see Section~\ref{global_AGN.sec}).
The tight positive trends with measures of central stellar mass concentration
(such as $\Sigma_{\star}$ and $\Sigma_{\rm 1kpc}$;
\citealt{FS19}, see also \citealt{Wis18}) may not be surprising in light of
the observed correlations between these properties and $M_{\star}$,
and the elevated fraction of AGN among compact SFGs in
$\log(M_{\star}/{\rm M_{\odot}}) \ga 10$ samples
\citep[e.g.,][]{Bar14a,Ran14,Koc17}.
Among AGN, the frequency and/or velocities of outflows appears to increase with
$L_{\rm AGN}$, consistent with simple expectations where more luminous AGN can
drive more powerful winds \citep[e.g.,][]{Har12,Har16,Bru15a,Leu19}.
In terms of absolute fractions, most studies imply fairly large outflow fractions
among AGN, in the range $\sim 50\% - 75\%$ except for \citet{Leu19}, who report a
lower $17\%$ incidence (but similar {\it trends\/} with galaxy and AGN properties).
\citet{Leu19} noted in their sample that the outflow fraction among AGN is
roughly constant with $M_{\star}$, and so is $L_{\rm AGN}$, concluding that
AGN can drive an outflow with equal probability irrespective of the host galaxy
mass and that observed trends among the full galaxy population reflect those
in AGN luminosity coupled with a (mass-independent) Eddington ratio distribution.
Detailed comparisons between all studies are still hampered by the heterogeneity
in sample size, selection, AGN and outflow identification, data sets (IFU vs.\
slit spectra) and S/N, but broadly support the picture that more powerful
AGN-driven outflows become common in the most massive galaxies.

\begin{textbox}[!t]
\section{PROPERTIES OF OUTFLOWS AND THEIR POWER SOURCES}
\noindent
\begin{sansmath}$\bm{v_{\scriptsize\textbf{out}}}$\end{sansmath}:
  Outflow velocity, estimated from the profile of the emission or absorption
  line wind tracer.  Methods based on the centroid or median velocity shift
  relative to the systemic value probe the bulk of outflowing gas.  Other
  methods including the line width at a fraction of peak amplitude or of
  cumulative flux/absorption probe the wind velocity distribution. \\
\begin{sansmath}$\bm{R_{\scriptsize\textbf{out}}}$\end{sansmath}:
  Outflow radial extent, most easily and directly obtained from maps of emission
  tracing the wind gas. \\
\begin{sansmath}$\bm{n_{\scriptsize\textbf{e,out}}},
                 \bm{N_{\scriptsize\textbf{H}}}$\end{sansmath}:
  Local electron density and hydrogen column density of the outflowing gas. \\
\begin{sansmath}$\bm{M_{\scriptsize\textbf{out}}}$\end{sansmath}:
  Mass of outflowing material; it is $\propto L_{\rm br}\,n_{\rm e,out}^{-1}$
  for ionized gas emission tracers where $L_{\rm br}$ is the luminosity of the
  broad outflow-related line component, and
  $\propto N_{\rm H}\,R_{\rm out}\,v_{\rm out}$ for IS absorption tracers. \\
{\textit{\textbf{\.{M}}}}$_{\scriptsize\textbf{out}}$:
  Mass outflow rate, estimated as $M_{\rm out}\times (v_{\rm out}/R_{\rm out})$. \\
$\boldsymbol{\eta}$:
  Mass loading factor, the ratio $\dot{M}_{\rm out}/{\rm SFR}$. \\
{\textit{\textbf{\.{E}}}}$_{\scriptsize\textbf{out}}$,
{\textit{\textbf{\.{p}}}}$_{\scriptsize\textbf{out}}$:
  {\looseness=-2
  Outflow energy and momentum rates,
  $\frac{1}{2}\dot{M}_{\rm out}\,v_{\rm out}^{2}$ and
  $\dot{M}_{\rm out}\,v_{\rm out}$; the ratio with 
  stellar or AGN luminosity $L$ and momentum rate $L/c$ constrains the
  wind power source and driving~mechanism.} \\
\begin{sansmath}$\bm{{L}_{\scriptsize\textbf{SFR}}},
                 \bm{{L}_{\scriptsize\textbf{AGN}}}$\end{sansmath}:
  Bolometric luminosity of the stellar population, dominated by young
  massive stars such that $L_{\rm SFR} \sim 10^{10}\,{\rm SFR}$, and from the
  AGN, estimated from, e.g., X-ray, SED modeling, or nebular line emission.
\end{textbox}

\subsubsection{Properties of star formation-driven winds}
  \label{resolved_outflows_sfout.sec}

\looseness=-2
Outflow velocity, mass, momentum, and energy properties across the galaxy
population are essential to constrain the physical drivers of winds and impact
of stellar feedback on the evolution of galaxies \citep[e.g.,][]{Dut09, Dav17}.
By necessity, many simplifications are involved in interpreting the data of
high $z$ galaxies, usually in the context of idealized models consisting of
a conical or spherical geometry, with the velocity distribution, extent,
and gas mass being the main parameters.
In the theoretical framework, the outflow velocity is generally assumed to
be close to the escape velocity, such that $v_{\rm out} \propto v_{\rm c}$.
Energy and momentum conservation arguments lead to mass loading factors
$\eta \propto v_{\rm c}^{-2}$ for energy-driven winds and
$\eta \propto v_{\rm c}^{-1}$ for momentum-driven winds, where
$\eta = \dot{M}_{\rm out}/{\rm SFR}$ and $\dot{M}_{\rm out}$ is
the mass outflow rate \citep[e.g.,][]{Mur05, Opp06}.
With $v_{\rm c} \propto M_{\rm bar}^{1/3}$ or $\propto M_{\star}^{1/3}$
\citep[e.g.,][]{Mo98} and $M_{\star} \propto {\rm SFR}$ on the MS, $\eta$
is expected to follow a power-law with stellar mass and SFR with index
$-2/3$ or $-1/3$ for energy- or momentum-driven winds, respectively.
There is a strong predicted differentiation in
$v_{\rm out} \propto \Sigma_{\rm SFR}^{\alpha}$, with $\alpha \sim 0.1$
for energy-driven winds and $\alpha \sim 2$ for momentum-driven
winds \citep[e.g.,][]{Str04, Mur05, Mur11}.
These scalings are consistent with recent cosmological zoom simulations
of high $z$ galaxies \citep[e.g.,][]{Hopkins12, Mur15}.

Measurements of $v_{\rm out}$ rely on parametrizations of the observed line
profiles, and various approaches have been followed depending on the diagnostic
and the data set (e.g., based on the FWHM or full width at zero power of
emission tracers, centroid or fractional absolute or cumulative absorption
for IS lines).  Despite these differences,
studies generally find results consistent with $v_{\rm out}/v_{\rm c}$ ratios
within a factor of a few around unity, with a linear or slightly sub-linear
$v_{\rm out} - v_{\rm c}$ trend \citep[e.g.,][]{Wei09,Erb12,Dav19,FS19,Swi19}.
Given that the escape velocity $v_{\rm esc} \approx 3\,v_{\rm c}$ for realistic
halo mass distributions \citep{Bin08}, these results indicate that the higher
velocity tail of the outflowing gas may escape from galaxies, and more easily
so in lower-mass galaxies, but that recycling may not be negligible.

\looseness=-2
The broad H$\alpha$ emission is well suited to estimate mass outflow rates
and energetics.  Assuming case B recombination and an electron temperature
$T_{\rm e} = 10^{4}~{\rm K}$, the mass of ionized gas in the outflow can
be estimated via
$M_{\rm out} \propto 
  L_{\rm br,0}({\rm H\alpha})\,n_{\rm e,out}^{-1}$
where $L_{\rm br,0}({\rm H\alpha})$ is the intrinsic luminosity in the
broad emission component and $n_{\rm e,out}$ is the local electron density,
from which the mass outflow rate can be computed as
$\dot{M}_{\rm out} = M_{\rm out}\,(v_{\rm out}/R_{\rm out})$ where
$v_{\rm out}$ and $R_{\rm out}$ are the outflow velocity and extent
\citep[e.g.,][]{Genz11,New12a}.
Calculations typically assume H dominates the mass and apply a $36\%$
mass correction for He.
Based on these relationships, $\eta$ estimates in the range $\sim 0.1$ up to
above unity were derived on galactic and sub-galactic scales
\citep[e.g.,][]{Genz11,New12a,New12b,FS19,Dav19,Free19,Swi19}.
While details in assumptions and samples vary among studies, a key difference
lies in the adopted value for $n_{\rm e,out}$, which ranges between $\sim 50$
and $\rm \sim 400~cm^{-3}$.
No significant or mildly positive trends of $\eta$ with stellar mass were found
in the larger samples spanning $9.0 \la \log(M_{\star}/{\rm M_{\odot}}) \la 11$,
assuming a constant $n_{\rm e,out}$ \citep[e.g.,][]{FS19,Free19,Swi19}.
Estimates based on IS absorption features depend more importantly
on geometrical factors, as the absorption strength traces the gas columns
along the line of sight, with
$M_{\rm out} \propto C_{\Omega}\,C_{f}\,N_{\rm H}\,R_{\rm out}\,v_{\rm out}$
(where $C_{\Omega}$ and $C_{f}$ are the angular and clumpiness covering
fractions, and $N_{\rm H}$ is the column density), as well as ISM chemistry
and radiative transfer effects on the line profiles \citep[e.g.,][]{Vei05}.
Under reasonable assumptions, $\eta \sim 1$ were found in outflow studies
of SFG samples employing this technique \citep[e.g.,][]{Wei09,Kor12}.
Comparing wind momentum and energy rate,
$\dot{p}_{\rm out} = \dot{M}_{\rm out}\,v_{\rm out}$ and
$\dot{E}_{\rm out} = 0.5\,\dot{M}_{\rm out}\,v_{\rm out}^{2}$,
to the momentum and luminosity input from star formation
$\dot{p}_{\rm rad} = L_{\rm SFR}/c$ and $L_{\rm SFR}$, most results are
in the ranges
$\dot{p}_{\rm out}/\dot{p}_{\rm rad} \sim 0.1-1$ and
$\dot{E}_{\rm out}/L_{\rm SFR} \sim 10^{-4} - 10^{-3}$ and thus
consistent with momentum-driven winds powered by the star formation activity
\citep[e.g.,]
  [but see \citealt{Swi19} for a contrasting result]{Genz11,New12a,FS19}.
Trends of $v_{\rm out} \propto \Sigma_{\rm SFR}^{0.2-0.4}$ found in other
studies from emission and IS absorption diagnostics suggest a possible
mixture of momentum- and energy-driving \citep[e.g.,][]{Wei09,Dav19}.

\looseness=-2
Estimates of $n_{\rm e,out}$ through the density-sensitive but weak
[S\,{\small II}]\,$\lambda\lambda$6716,6731 doublet have long been hampered
by S/N limitations.  A first reliable broad$+$narrow Gaussian decomposition
in very high S/N stacked spectra
\citep[Figure\ \ref{outflows.fig};][]{FS19} yielded
$n_{\rm e,out} \sim 380~{\rm cm^{-3}}$ for the outflowing gas
(and $n_{\rm e,HII} \sim 75~{\rm cm^{-3}}$
for the narrow star formation-dominated component).
These results suggest the outflowing gas may experience compression,
supported by enhanced broad component \niiha\ ratios in the same stacks
as well as from multiple diagnostic (total) line ratios for some bright individual
star-forming clumps \citep{New12a} and samples with multi-band near-IR spectra
\citep{Free19}.  Different outflow gas densities adopted in the literature can
account for much of the differences in $\eta$ and other outflow properties, as
the observables themselves (broad-to-narrow H$\alpha$ flux ratio, $v_{\rm out}$,
and $R_{\rm out}$) are fairly comparable.  With the new evidence suggesting
higher $n_{\rm e,out}$, a lower range of $\eta$ ($< 1$) in the warm ionized
gas phase would seem favored.

\looseness=-2
Taken at face value, low mass loading factors and the lack of
evidence for an anticorrelation with galaxy stellar mass appear to be in
tension with theoretical expectations and numerical simulations, for which
$\eta \ga 0.3 - 1$ at $\log(M_{\star}/{\rm M_{\odot}}) \sim 10$ and
$\eta \propto M_{\star}^{\alpha}$ with $\alpha$ in the range $-0.35$ to $-0.8$
\citep[e.g.,][]{Lil13,Mur15}.
The tension is compounded by the $v_{\rm out}$ results suggesting that some
fraction of the outflowing gas may not be
able to escape from the galaxy's potential (reducing the effective $\eta$).
Notwithstanding all the simplifications made and large uncertainties, the
mass outflow, momentum, and energy rates discussed above almost certainly
represent lower limits as they miss potentially important wind phases, as
seen in local starburst galaxies where the neutral and cold molecular phases
dominate the mass and the hot phase dominates the energetics
\citep[e.g.,][]{Vei05,Hec17}.

\subsubsection{Properties of AGN-driven winds}
  \label{resolved_outflows_agnout.sec}

The role of ejective AGN feedback through ``QSO mode'' has been much debated in
the recent observational literature.  At high $z$, while individual luminous AGN
may drive sufficiently massive and energetic outflows to suppress star formation
in their host \citep[e.g.,][]{Can12,Cre15,Car16,Kak16},
QSOs are rare, such that their impact on the massive galaxy
population as a whole and in the long run has remained unclear.
The more recent studies based on rest-optical emission lines of larger
$z \sim 2$ samples, encompassing unbiased (mass-selected) populations
and/or AGN selected in deep X-ray surveys, both covering broader ranges
in AGN luminosities
(in some cases down to $\log(L_{\rm AGN}/{\rm erg\,s^{-1}}) \sim 42.5 - 43$)
are shedding new light on this issue
\citep[e.g.,][]{FS14,FS19,Genz14b,Har16,Tal17,Leu19}.

A first general conclusion is that with typical high velocities
$\rm \sim 1000~km\,s^{-1}$, AGN-driven winds are in principle able to escape
the galaxies and even the halo.  The outflow velocity appears to depend on
$L_{\rm AGN}$ but otherwise not on galaxy properties such as $M_{\star}$
or SFR, consistent with the AGN as main power source.
Double-Gaussian fits to high S/N stacked spectra suggest dense gas
with $n_{\rm e,out} \sim 1000~{\rm cm^{-3}}$ from the \sii\ doublet
\citep[Figure\ \ref{outflows.fig};][]{FS19} albeit with significant
uncertainties because of the important blending for the broad emission of
the fast AGN-driven winds and the doublet ratio reaching towards the
high-density limit.
Elevated \niiha\ ratios of $\sim 1 - 2$ in broad and narrow emission alike
for a significant subset of this sample suggests an important contribution
from shock excitation.
Keeping in mind all the uncertainties involved, different assumptions adopted by
different authors, and large scatter among galaxies, there is overall agreement
that on average the momentum and energy rates of AGN-driven outflows exceed those
that could be produced by star formation alone, and are consistent with
energy-driving contributing or even dominating \citep{FS19,Leu19}, as also
suggested by the $v_{\rm out}$ dependence on $L_{\rm AGN}$ \citep{Tal17,Leu19}.
Mass outflow rates (compared to the SFRs) are found to be modest to low
($\eta \la 1$) on average among SFGs, and possibly higher towards the sub-MS
regime.

\looseness=-2
While AGN-driven winds may expel ionized gas at modest rates compared to the
SFRs (similarly to the SF-driven outflows), they carry substantial amounts of
momentum and energy ($\sim 10$ times or more than the SF-driven winds).
If more mass, momentum, and energy are contained in other wind phases
(or if $n_{\rm e,out}$ estimates are lower than adopted), all estimates
would increase.  Measurements in other phases are still scarce at $z \sim 2$;
CO observations suggest $\eta \sim 1$ in two MS SFGs hosting AGN, one of which
is a QSO \citep{Her19, Bru18}.
Even if not substantially depleting the gas reservoirs of their host, the
high-velocity and energetic AGN-driven winds escaping from the galaxies may
interact with halo gas, reach high temperatures with long cooling time,
and help prevent further gas infall together with virial shocks.
The rapid increase in the incidence of AGN-driven winds among the galaxy
population at around the Schechter mass
echoing the decline in specific SFR and molecular gas mass fractions
\citep{Whi14,Tac18} is suggestive of a connection between AGN-driven winds and
quenching, although it may not be sufficient alone to establish a causal link.
Given the wide range in AGN luminosities and inferred Eddington ratios for the
larger samples discussed above, the results appear to be qualitatively in line
with suggestions based on recent cosmological simulations that kinetic feedback
from SMBHs accreting at low Eddington ratio may be more efficient at quenching
star formation through preventive feedback in the
circumgalactic medium \citep{Bow17,Nels18,Pil18a}.

\section{OTHER z $\mathbf{\sim}$ 2 STAR-FORMING POPULATIONS}
\label{otherpops.sec}

We here briefly discuss specific subpopulations among SFGs that have been the
focus of dedicated analyses for reasons of their extreme starburst nature
and/or their role as candidate immediate progenitors to the accumulating
population of quiescent galaxies at cosmic noon.  Salient physical features
of the latter class of galaxies are summarized as well.

\subsection{``MS outliers,'' and Submm Galaxies}
 \label{otherpops_outliers.sec}

Whereas normal MS galaxies are predominantly disks, at all epochs a population
of starbursting outliers exists that may well result from merging activity.
At $z \sim 2$ such starburst galaxies, defined by their SFR being more than four
times higher than on the MS, represent only 2\% of the mass-selected SFGs,
accounting for only 10\% of the cosmic SFR density at this epoch \citep{Rod11}.
Modeling the SFR distribution of SFGs at fixed mass with a double gaussian
reveals a similar, constant or only weakly redshift-dependent, starburst
contribution of $8\% - 14\%$ to the overall SFR budget \citep{Sargent12}.

\looseness=-2
Structurally, there are indications that above-MS outliers exhibit smaller
effective radii and cuspier light profiles than their exponential disk
counterparts along the MS ridgeline.  This is seen for nearby populations,
but in rest-UV/optical and radio observations at cosmic noon as well, albeit
with significant scatter and only when collecting samples over wide areas to
sample the poorly populated high-SFR tail of the galaxy population
\citep{Wuy11b, Elbaz11}.  Splitting the SFG population in below-, on- and
above-MS subsets \citet{Nel16b} find the above-MS SFGs to feature enhanced
H$\alpha$ $\Sigma_{\rm SFR}$ at all radii.
Only for $\log(M_{\star}/{\rm M_{\odot}}) > 10.5$ is the enhancement particularly
seen in the center.  It should be noted though that extreme outliers ($8 \times$
above the MS) have 90\% of their star formation revealed only in the far-IR
and often are optically thick even in H$\alpha$ \citep{Pug17}.

\looseness=-2
Beyond structural properties, a systematic increase in gas fraction
\citep[e.g.,][]{Tac20}, dust temperature \citep{Magn14}, and ratio of
total IR to rest-8$\mu$m luminosities \citep{Elbaz11, Nordon12}
is seen as one moves across the MS towards higher SFRs.
Not only does the amount of obscuration by dust increase \citep{Wuy11b},
the resulting effective attenuation law as imprinted in the IRX-$\beta$
relation also varies systematically with position in SFR-mass space
\citep{Nordon13}.\footnote{The non-universality of
 attenuation law shapes at cosmic noon has also been reported by \citet{Sal16}
 and \citet{Red18}, with increasingly shallower slopes towards the more enriched
 and dustier regime, rooted in changing star -- dust geometries and possibly grain
 size distributions.}
All of these trends between MS offset and physical diagnostics suggest that
the observed scatter around the MS is real, and cannot be fully attributed to
measurement uncertainties associated with the various SFR tracers employed.
Confirming this point more directly, \citet{Fang18} demonstrate
that independent $\Delta {\rm MS}$ measurements based on 24 $\mu$m and
UV-to-optical diagnostics correlate significantly.

\begin{marginnote}[250pt]
 \entry{SMG}{Submm galaxy.}
\end{marginnote}

\looseness=-2
Predating the terminology of a main sequence and orthogonal to the
historical background of rest-UV/optical lookback surveys is the rare
population of very luminous high-$z$ Submm Galaxies (SMGs), first discovered
in the late 1990s through $\rm 850\,\mu m$ observations with SCUBA on the
JCMT \citep[$\ga 15$\arcsec\ beam; ][]{Smail97}.
Since then, higher resolution far-IR observations have refined our understanding
of the nature of SMGs, identifying multi-component morphologies in some and very
compact cores with large velocity ranges in other cases \citep{Tac06,Tac08}.
These results point to merger-driven short-lived ($\rm \sim 100~Myr$)
``maximum starburst'' events.  ALMA 1\,mm observations demonstrated that
multiplicity of single-dish sources becomes increasingly common towards the
bright end, with 28\% of $\rm > 5\,mJy$ sources and 44\% of $\rm > 9\,mJy$
sources being identified as blends \citep{Stach18}.
Using spectroscopic follow-up of individual components for modest samples
\citep{Hayward18} or a statistical approach based on photometric redshifts
for samples of several dozen SMGs \citep{Stach18} it is further apparent that
both chance alignments and physically associated components make up a
significant fraction of the blends, with physically associated pairs adding
up to at least 30\%.

Accounting for multiplicity hence reduces the inferred SFRs for some of the
brightest SMGs, relieving some tension with models and bringing them closer
to the MS.  Their MS offset is further reduced when allowing for multi-component
SFHs, which have a tendency of increasing the inferred stellar
mass.  For this reason, \citet{Mich14} argue that SMGs reside predominantly at the
high-mass tip of the MS rather than being positioned above, consequently also
questioning their merger nature.

The rarity of above-MS outliers and SMGs can be interpreted in terms of
short duty cycles preceding a quenching event.  For example, \citet{Wuy11b}
contrast the number density of $\rm \Delta MS > 0.5$
outliers to the growing number density of quiescent galaxies at
cosmic noon inferring timescales of order $\rm \sim 100~Myr$ for the
starbursting phase.  \citet{Toft14} take a different approach, in which they
contrast the inferred formation redshifts of compact quiescent galaxies at
$z \sim 2$ to the redshift distribution of the $3 < z < 6$ SMG population,
finding a good match that is further underlined by their similar positions
in size-mass space and consistently high characteristic velocities.
Assuming an evolutionary connection, they can reconcile their relative space
densities by invoking an SMG lifetime of $\rm \sim 42~Myr$.
The relatively short timescales found in the above studies are
consistent with the duration of the final merger phase and peak starburst
around coalescence of dissipative major mergers
\citep[e.g.,][]{Mihos94, Hopkins06}.

\subsection{Compact Star-Forming Galaxies}
 \label{otherpops_csfgs.sec}

\begin{marginnote}[185pt]
 \entry{cSFG}{Compact star-forming galaxy.}
\end{marginnote}

In order to reveal evolutionary connections between galaxies before and after
quenching, a selection on the basis of similar structural properties (i.e.,
identifying SFGs in the compact corner of size-mass space where high-$z$
quiescent galaxies reside) has become a popular approach
\citep[e.g.,][]{Bar13, Bar14a}.  After $z \sim 1.8$ the number density of
these compact star-forming galaxies (cSFGs) are dropping precipitously,
while the number density of compact quiescent galaxies is still rising.
Duty cycle arguments akin to those described in the previous Section yield
typical lifetimes for this cSFG phase of $\rm \sim 500 - 800~Myr$, dependent
on the precise compactness and star formation selection criteria imposed
\citep{Bar13, vD15}.  cSFGs thus represent a longer-lasting phase than
the one discussed in Section\ \ref{otherpops_outliers.sec}, which is also
reflected in their larger abundance and larger range in star formation
activities, from above to on and below the MS.

A salient feature of the cSFG population is that both X-ray and line ratio
diagnostics reveal a very high AGN fraction ($\ga 40\%$ based on X-rays and
up to $\sim 75\%$ when folding in line ratio diagnostics).  This enhancement
in AGN activity is highly significant relative to quiescent galaxies but also
compared to similar-mass SFGs that are more extended
\citep{Bar14a, Koc17, Wis18}.  They are further found to be highly obscured,
with dust cores even smaller than their stellar extent \citep{Bar16} and
galaxy-integrated ionized gas velocity dispersions (and in one case a
measurement of a stellar velocity dispersion) of several
$\rm 100~km\,s^{-1}$, consistent with those measured using stellar tracers
in compact quiescent galaxies. The implied dynamical masses of cSFGs are comparable with
their stellar mass content \citep{Nel14, Bar14b, vD15}.
Resolved gas kinematics of cSFGs
have revealed that the large galaxy-integrated linewidths can to a large
degree be attributed to unresolved disk rotation \citep{Bar17b,Wis18}.
While their stellar distributions are by definition compact, the ionized gas
disks are often more extended \citep{vD15, Wis18}.  Even when modeled with
rotating disks and accounting for inclination and beam-smearing effects the
resulting stellar-to-dynamical mass ratios of the more compact SFGs are close
to unity and larger than that of extended SFGs \citep{vD15, Wuy16b, Wis18}.
These dynamical measurements support a picture that cSFGs are in their last
stretch of star formation with already dwindling gas fractions and short
depletion times.  \citet{Spilker16} and \citet{Popping17} have come to a
similar conclusion on the basis of molecular line measurements for this
sub-population.

Several lines of evidence highlight the resemblance in dynamical
terms between cSFGs and the quiescent population to which they are candidate
immediate progenitors.  Compact quiescent galaxies at cosmic noon exhibit more
flattened projected shapes than anticipated for a pressure supported population
\citep{vdWel11, Chang13}, their $M_{\rm dyn}/M_{\star}$ ratios calculated from
galaxy-integrated stellar velocity dispersions using a virial mass estimator
are higher for systems with flatter projected axis ratios \citep{Belli17a},
and in four gravitationally lensed cases stellar velocity curves reveal
unambiguously their rotationally supported nature
\citep{Newman15, Newman18b, Toft17}, consistent with a highly dissipational
formation process \citep{Wuy10, Wellons15}.

\section{THEORETICAL PICTURE AND ADVANCES IN NUMERICAL SIMULATIONS}
\label{disc.sec}

Models of galaxy formation in a $\Lambda$CDM context
have seen significant improvements over the past decade.  In particular, great
strides forward were made in resolving the so-called angular momentum catastrophy
\citep[the inability to reproduce the Tully-Fisher and rotation speed - angular
momentum relation of observed disks galaxies; ][]{Nav00}, and the overproduction
of stars in both low- and high-mass galaxies.  Cosmological galaxy
formation models still feature variations at the factor of $\sim 2$ level in
for example the peak stellar-to-halo mass ratio reached around
$M_{\rm halo} \sim 10^{12}~{\rm M_{\odot}}$ and possibly more at lower/higher
masses, but they now fall within the range of uncertainties from abundance
matching estimates that traditionally serve as a benchmark.
Today, we face a landscape of theoretical models that can be differentiated by
the physical scales they resolve, the numerical techniques they employ, and the
(astro)physics they implement.  The scales that are resolved dictate which
physical properties can be considered ``imposed'' versus ``emerging'' from
such models \citep[see reviews by][]{Som15,Naab17}.

On the largest scales, semi-analytic models can efficiently imprint
the baryonic growth of galaxies on merger trees extracted from DM-only simulations
with box sizes of $1 - 10$~Gpc (Millennium, Millennium-XXL, Bolshoi, Las Damas).
Effectively resolving individual galaxies at the halo scale, basic structural
properties such as galaxy sizes are then evaluated through analytical recipes
that either assume specific angular momentum conservation \citep{Mo98} or encode
dependencies on both angular momentum and halo concentration \citep{Som18,Jiang19},
and are designed to capture processes such as disk instabilities and mergers.
For the
latter, simple energy conservation arguments are often augmented with calibrations
based on idealized merger simulations to account for the impact of dissipative
processes on the resulting bulges \citep{Cov11}.  While intrinsically unable
to track detailed structural evolution from first principles, such models have
the merit of being computationally cheap (7 CPU hours to execute a single
realization producing over $10^7$ galaxies).  They are therefore the only type
of models for which a full exploration of parameter space and a mapping of its
degeneracies by means of Monte Carlo Markov Chains is feasible
\citep[e.g.,][]{Henriques15}.

Inclusion of hydrodynamics comes at a major computational expense but allows
key processes for structural evolution to be resolved rather than prescribed.
State-of-the-art full cosmological simulations (Illustris, EAGLE, Magneticum,
Horizon-AGN, Illustris TNG, SIMBA) are capable of evolving populations of
$10^{4} - 10^{5}$ galaxies in $\rm 10 - 300~Mpc$ boxes with sufficient resolution
(baryonic particle masses of $\rm \sim 10^{6} - 10^{7}~M_{\odot}$, sub-kpc
gravitational softening lengths) to track their internal structural development
and kinematics.  With a temperature floor of $\rm 10^{4}~K$ and the reliance
on subgrid recipes to infer cold gas fractions, they are complemented by
zoom-in simulations of more than 100 times enhanced mass and spatial resolution,
which are capable of resolving Jeans mass/length scales, giant molecular cloud
formation and a self-consistent modeling of the multi-phase ISM (e.g., FIRE,
Auriga, VELA).  Further down the series of Russian dolls come simulations of
isolated galaxies or ISM slices in an external potential resolved down to parsec
scales (e.g., SILCC).  They are ideally suited to track the multi-phase breakdown
of the ISM including the chemistry of molecular gas formation, the local injection
of energy and momentum by late stages of stellar evolution and its coupling to
the surrounding medium (e.g., effects of peak driving vs.\ supernovae exploding
after stars migrate away from their birthclouds, ISM porosity and the possibility
of feedback energy escaping through the path of least resistance, impact of the
IR opacity on the effectiveness of radiation pressure, ...).

The hydro-solvers employed in generating the above multi-scale simulations range
from grid-based Adaptive Mesh Refinement (AMR) to Smoothed Particle Hydrodynamics
(SPH; with refinements to better capture contact discontinuities and shock fronts,
\citealt{Hopkins15}), and include hybrid moving mesh approaches \citep{Springel10}.
In common between these models, the physics of gravity, hydrodynamics, cooling
and heating, star formation and evolution (SNIa, SNII, AGB), chemical enrichment
(tracking up to 11 individual elements), black hole growth, and stellar and AGN
feedback are now routinely implemented.  Increasingly, also the impact of other
processes, such as magnetic fields, radiation pressure, cosmic rays and even the
formation and destruction of dust are explored, albeit some of them restricted
to the highest resolution simulations only.

Qualitatively, overcoming the hurdles posed by the angular momentum problem
and the observed inefficiency of galaxy formation took the implementation of
strong feedback processes.  How exactly this goal is most realistically achieved
through numerical and/or subgrid recipes remains a matter of intense debate, on
which resolved observations of galaxy structure and kinematics aim to shed light.
Implementations of stellar feedback differ in their injection velocities, mass
loadings and directionality, and whether or not wind particles are temporarily
decoupled from hydrodynamic interactions to prevent numerical overcooling.
Likewise, AGN feedback as a term covers a considerable range of implementations,
starting from the choice of black hole seeding, whether or not boosting factors
are applied to conventional Bondi accretion, directionality and
continuity/stochasticity with which gas particles are being heated or receive
kinetic kicks.  Different choices are made regarding which gas particles this
thermal/kinetic energy is imparted on, and whether (e.g., Illustris TNG) or
not (e.g., EAGLE) different prescriptions are applied in the high vs.\ low
accretion rate regime.  Consequently, predictions on key wind properties are
still in flux, with for example TNG winds being faster but of lower mass loading
than those in its Illustris precursor.  This illustrates the continued need for
empirical guidance.

\looseness=-2
Last but not least, significant work on the interface between
simulations and observations is enabling ever more consistent comparisons.
This starts with the basic question of what it is that constitutes a galaxy's
stellar mass, and related, what it is that observers are measuring.
\citet{Pil18b} illustrate how aperture-based masses (as opposed to total
stellar masses integrated out to the virial radius) can significantly alter
our view on the stellar mass function and SMHM
relation, particularly for the most massive galaxies featuring extended wings,
but even so at the knee of the SMHM~relation.  Bringing the models yet more
into the observational realm, post-processing with advanced radiative transfer
techniques (SKIRT, Sunrise, Powderday) are enabling mock observations accounting
for the effects of light weighting, dust extinction and reprocessing, including
also ionized and molecular gas line emission.  These can aid refined calibrations
of observational diagnostics and SED modeling techniques, and are adopted
in feasibility studies for upcoming observing facilities.

\section{SUMMARY AND OUTLOOK}
\label{outlook.sec}

This article highlighted some of the key insights emerging from increasingly
complete population censuses and increasingly detailed studies of individual
galaxies back to the cosmic noon epoch.
Many global and resolved properties tracing the stars, gas, and kinematics
are well probed down to $\log(M_{\star}/{\rm M_{\odot}})$\,$\sim$\,10 (or below).
Current results draw a consistent broad picture (see Summary Points)
and raise the next questions for future work (see Future Issues
for a selection).
The knowledge gained from these observations has contributed to transform
--- in some aspects, profoundly --- our view of galaxy evolution.  The
emerging picture is encapsulated in the equilibrium growth model summarized
in Section\ \ref{intro_bkgd.sec}, and discussed by \citet{Tac20} in
relation to the evolution of the characteristic timescales of the processes
controlling galaxy growth including cosmic accretion, merging, galactic
gas depletion and star formation, internal dynamics, and gas recycling.

\begin{figure}[!t]
$\begin{array}{ll}
 \includegraphics[scale=0.45,trim={31.0cm -0.5cm 10.0cm 3.0cm},clip=0,angle=0]{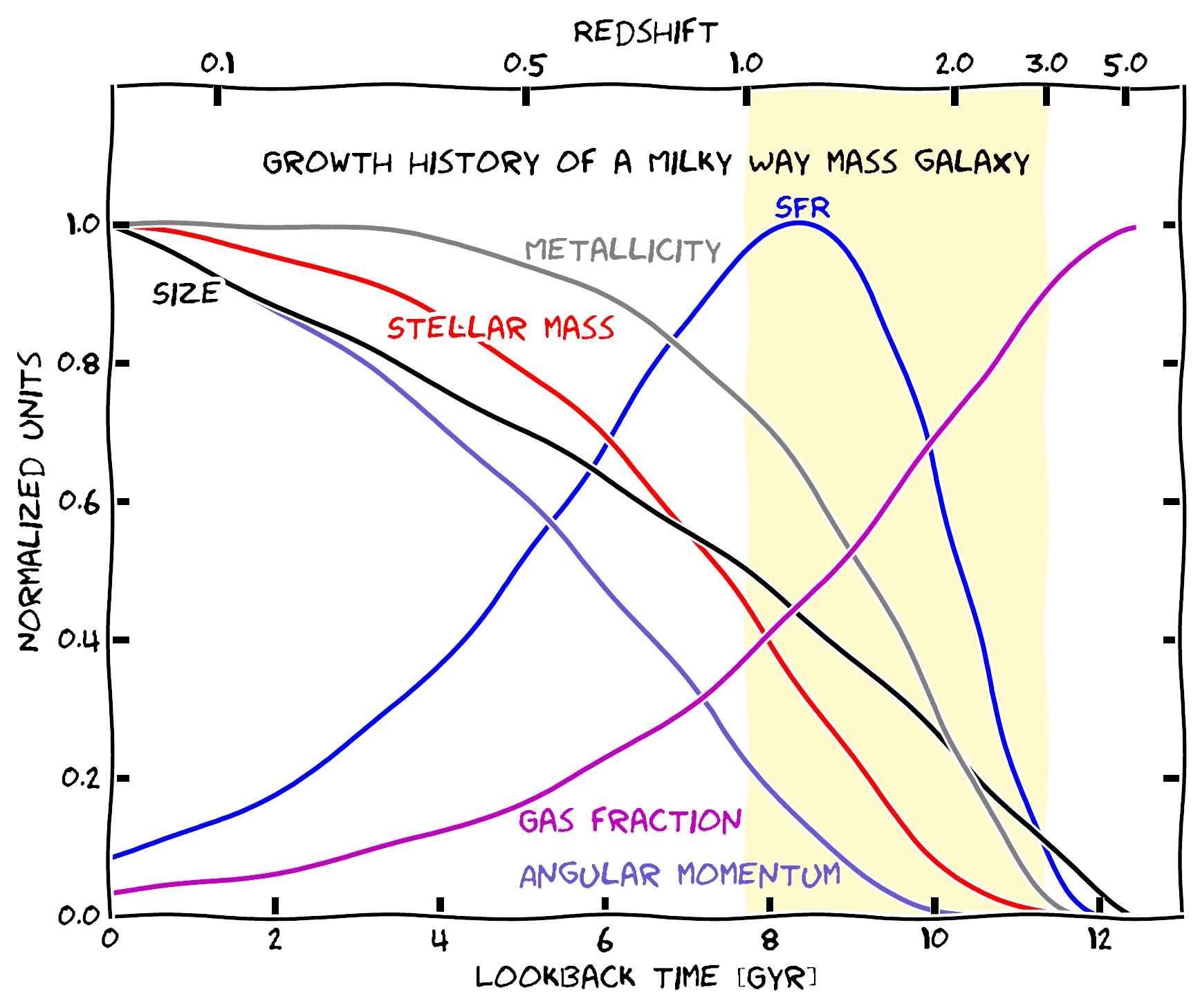}  &
 \includegraphics[scale=0.45,trim={46.5cm -0.5cm 4.0cm 3.0cm},clip=0,angle=0]{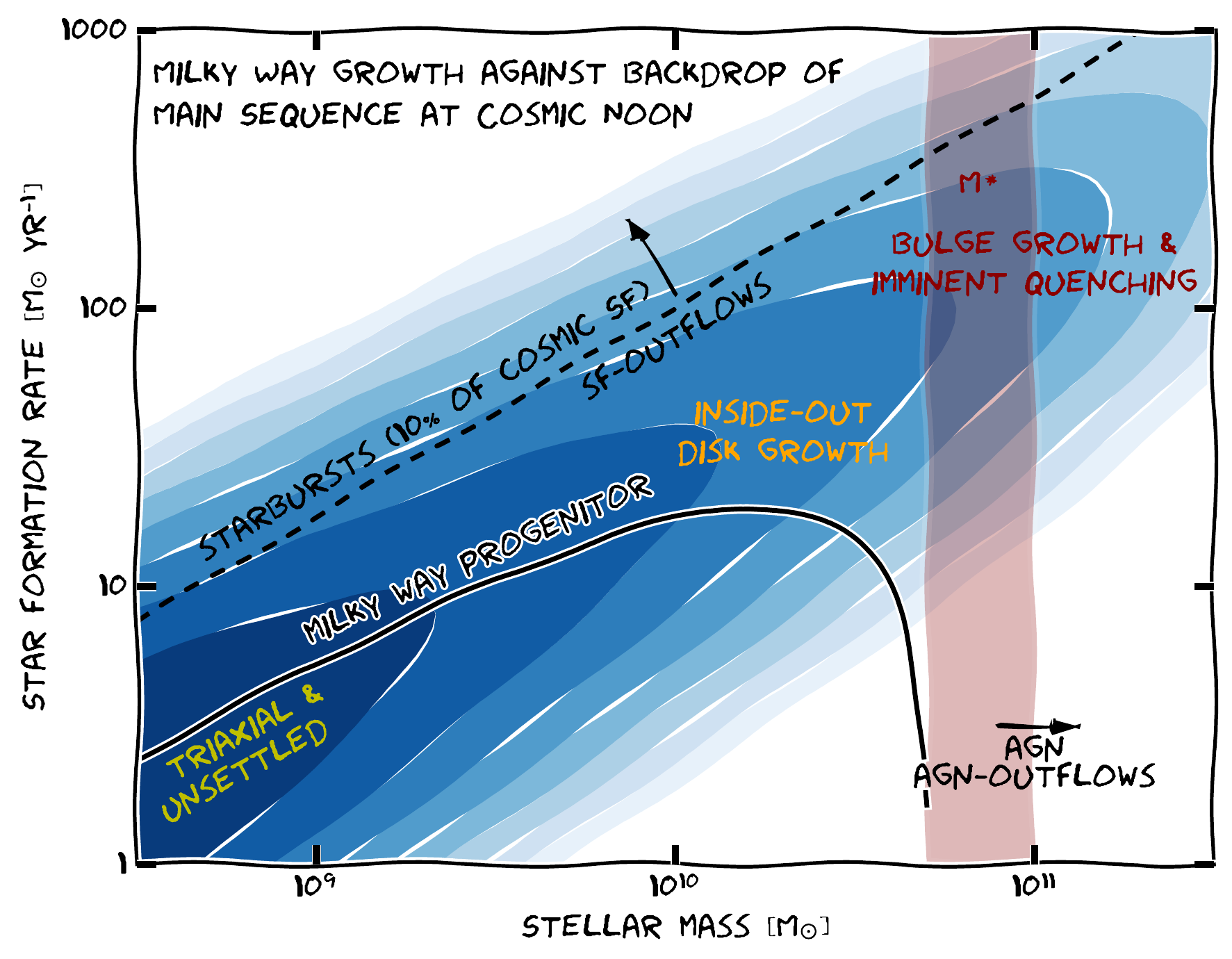}  \\
\end{array}$
\renewcommand\baselinestretch{0.85}
\caption{
\looseness=-2
{\it Left:}
Evolutionary history of a Milky Way-mass progenitor galaxy.
Tracks of different global properties are plotted as follows:
gas mass fraction $f_{\rm gas}$ (magenta), SFR (blue), $M_{\star}$ (red),
gas-phase metallicity (grey), rest-optical $R_{\rm e}$ (black), and stellar
angular momentum $\propto R_{\rm e} M_{\star} v_{\rm rot}$ (purple).
Each curve is normalized to a maximum of unity to highlight the relative
rate of variations between the properties with lookback time.
The stellar mass growth is derived from abundance-matching following
\citet{Hill17},
and the other curves are computed from evolving scaling relations
at the corresponding $M_{\star}(z)$ \citep{Spe14, vdWel14a, Ueb17, Tac18}.
Though simplistic (e.g., the progenitor is assumed
to remain on the MS and other relationships all the time), the plot illustrates
how current empirical censuses and scaling relations allow us to
investigate the
average evolution of individual galaxies.
{\it Right:} The same evolutionary track of a Milky Way-mass progenitor
presented in the $M_{\star} - {\rm SFR}$ diagram, against
the backdrop of the $z \sim 2$ SFG population (blueshades marking logarithmic
steps in number density; based on \citet{Spe14} and \citet{Tom14}).
Markers indicate the structural/dynamical state, mode of star formation,
and where feedback processes become increasingly apparent.
The vertical red bar marks the characteristic mass $M^{\star}$, which has
remained approximately constant since cosmic noon.
}
\label{MWG_MS.fig}
\end{figure}

\begin{figure}[!t]
\centering
\includegraphics[scale=0.48,trim={1.0cm 2.8cm 0.5cm 2.8cm},clip=0,angle=0]
                {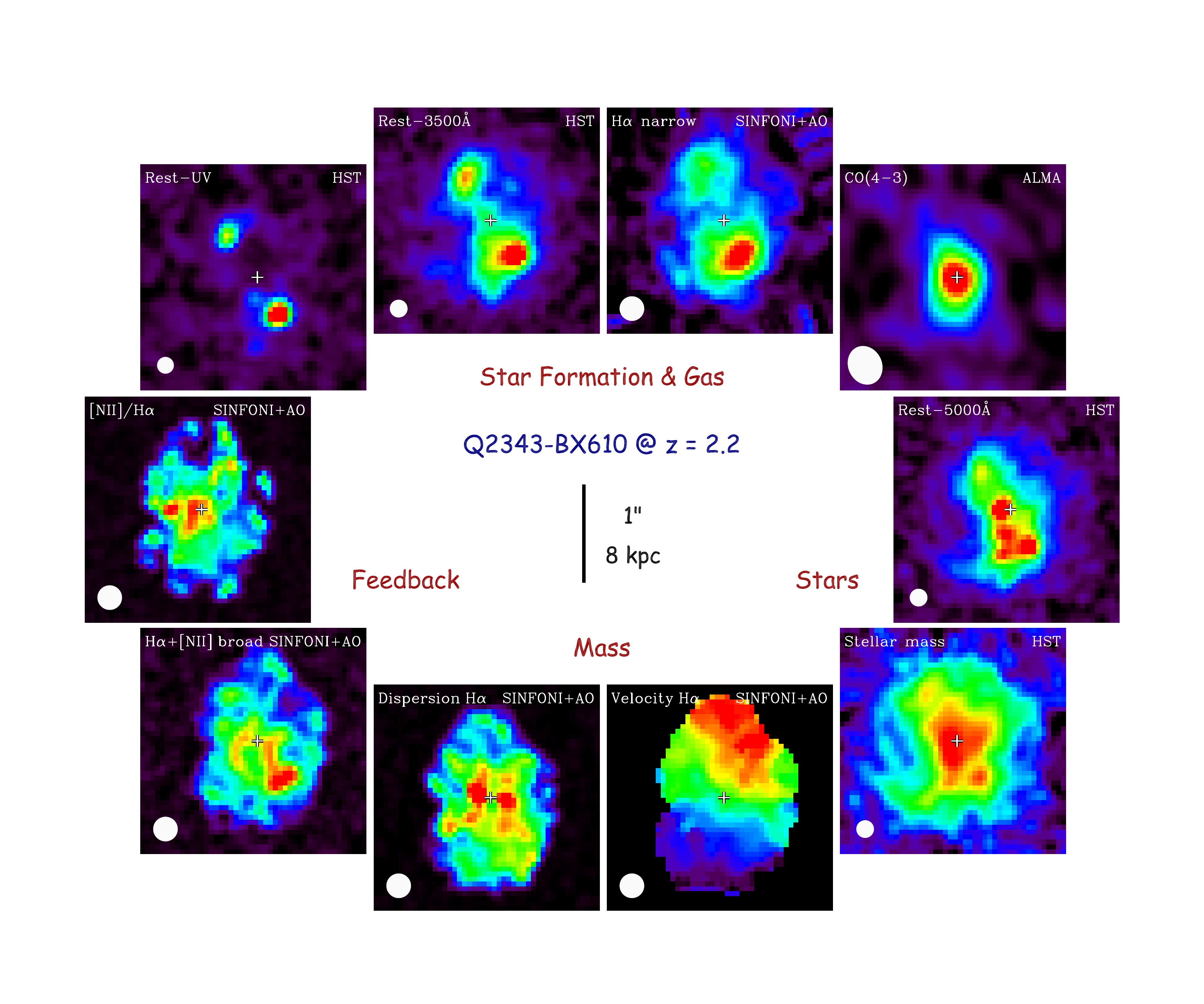}
\renewcommand\baselinestretch{0.85}
\caption{
\looseness=-2
State-of-the-art observations detailing the evolutionary state
and probing the baryon cycle of a $z = 2.2$ massive MS galaxy
($M_{\star}$$\sim$$10^{11}~{\rm M_{\odot}}$).
The maps show, clockwise from the top left, the
rest-frame UV and $U$ band emission dominated by unobscured continuum
light from young massive stars;
H$\alpha$ emission from moderately unobscured HII regions;
$\rm CO(4-3)$ emission revealing the cold molecular gas fueling largely
obscured star formation;
rest-frame $\rm \sim 5000$\,\AA\ light tracing the bulk of stars;
the stellar mass distribution;
H$\alpha$ velocity field and dispersion map tracing gravitational motions;
broad low-amplitude emission in H$\alpha$$+$\nii\ revealing high-velocity
outflowing gas;
\niiha\ ratio sensitive to the excitation and physical conditions
of the nebular gas.
The FWHM resolution is shown by the white ellipse in each panel.
Despite a clumpy appearance in UV/optical stellar light and H$\alpha$,
the kinematics and stellar mass map reveal a massive rotating yet turbulent
disk hosting a dense bulge-like component.  The bulge may still be
growing out of the massive central molecular gas reservoir, which may be
replenished through inward gas streaming along a bar or spiral arms as hinted
at by the inner isovelocity twist, double-peaked central dispersion, and
$\rm \sim 5000$\,\AA\ morphology.
The weak \niiha\ radial gradient in the outer disk could indicate
a shallow metallicity gradient, consistent with efficient metal mixing
within the turbulent gas disk and/or through galactic outflows.
The elevated \niiha\ $\sim 0.7$ at the center
signals the presence of a (low-luminosity) AGN.
Ionized gas is being driven out of the galaxy through star formation feedback
near the location of the brightest UV/optical/H$\alpha$ clump, as well as
through AGN-driven feedback near the nucleus.
Based on data presented by \citet{FS11, FS18, Tacch18},
and obtained from the ALMA archive (program 2013.1.00059.S, PI Aravena).
}
\label{bx610.fig}
\end{figure}

\looseness=-2
The state-of-the-art in our knowledge of the properties of
$z \sim 2$ SFGs is illustrated in Figures~\ref{MWG_MS.fig} and \ref{bx610.fig}.
The censuses and scaling relations allow a depiction of the evolutionary and
dynamical state of SFGs in relation to the MS.
Coupled with the assumption that mass-ranking of galaxies is conserved, this
cross-sectional view of the galaxy population at different epochs can be
translated to tracks representing the average evolution of individual galaxies.
The outcome of such an approach is shown in Figure~\ref{MWG_MS.fig} for
a galaxy reaching the stellar mass of the Milky Way by the present day.
Cosmic noon is the main formation epoch of stars in $z \sim 0$ galaxies
of masses similar and up to $\sim 2\,\times$ higher than the Milky Way
($\rm \log(M_{\star}/M_{\odot}) \sim 10.7 - 11$), which account for as
much as $\sim 25\%$ of the local stellar mass budget.
In turn, resolved mapping is now possible for various tracers of the baryon
cycle from gas and star formation to metal enrichment and feedback, of the
dynamical state, of processes leading to the build-up of galactic
components and their imprint on the distribution of stars.
Such comprehensive sets at the currently best achievable $\rm \sim 1~kpc$
resolution (unlensed) are still limited to small numbers of
$z \sim 2$ SFGs; Figure~\ref{bx610.fig} shows one example.

\begin{summary}[SUMMARY POINTS]
\begin{enumerate}
\item Two key observational aspects have driven major advances in our
      understanding of how galaxies evolved since cosmic noon by providing
      unparalleled comprehensive views of distant galaxies:
      (i) the concentration in ``legacy cosmological fields'' of photometric
          and spectroscopic surveys across the electromagnetic spectrum, and
      (ii) the growing samples with stellar structure, star formation, and
           gas kinematics resolved on subgalactic scales.
      Mass selection is routinely used, allowing more complete
      population-wide censuses of physical processes driving
      galaxy evolution.
\item \looseness=-2
      Scaling relations between galaxy stellar mass, SFR, metallicity, gas content,
      size, structure, and kinematics are in place since at least $z \sim 2.5$,
      indicating that regulatory mechanisms start to act on galaxy growth within
      $\rm 2-3~Gyr$ of the Big Bang.  There is significant evolution in population
      properties: compared to $z \sim 0$, typical SFGs at $z \sim 2$ were forming
      stars and growing their central SMBH $\sim 10\times$ faster from
      $\sim 10\times$ larger cold molecular gas reservoirs.  Disks are prevalent
      but smaller, more turbulent, and thicker that today's spirals.
      Quenching was underway at high masses,
      through mechanisms that appear to largely preserve disky structure.
\item Resolved stellar light, star formation, and kinematics on scales down to
      $\rm \sim 1~kpc$ point to spatial patterns --- more pronounced in higher
      mass SFGs --- from dense and strongly baryon-dominated core regions with
      possibly suppressed star formation to more actively star-forming outskirts.
      Whether these patterns reflect inside-out growth/quenching scenarios, or
      carry the imprint of strong radial gradients in extinction and efficient
      dissipative processes in gas-rich disks is open.  The detection of large
      nuclear concentrations of cold gas and kinematic evidence of radial inflows
      in the most massive galaxies support the latter scenario, in which case
      massive but highly obscured stellar bulges may still be rapidly growing.
\item Outflows traced by warm ionized and neutral high-velocity
      gas act across a wide swath of the galaxy population.
      SF-driven winds dominate below the Schechter mass and are more ubiquitous
      and/or stronger at higher star formation levels, but may largely remain
      bound to the galaxy.
      AGN-driven winds dominate at higher masses, with rapidly rising incidence
      and/or strength with stellar mass and central concentration thereof.
      Improved constraints suggest dense, possibly shocked-compressed ionized
      material in both outflow types, leading to modest 
      sub-unity mass loading factors in the warm ionized phase.
      The high duty-cycle AGN-driven winds are sufficiently fast to
      escape their massive host and heat halo gas, tantalizingly suggesting
      a preventive form of AGN feedback contributes to quenching.
\end{enumerate}
\end{summary}

\begin{issues}[FUTURE ISSUES]
\begin{enumerate}
\item
 What is the origin of scatter in galaxy scaling relations
 (between $M_{\star}$, SFR, size, gas content, metallicity, ...)?
 Is any scatter around the observed relations attributed to short-term
 stochasticity (i.e., the equivalent of ``weather'') or an imprint of a
 long-term differentiation in growth histories among SFGs of the same mass
 at a given epoch?  If the latter, what (halo) property other than mass is
 most appropriate to describe the SFG population as a two-parameter family?
\item
 What is the physics responsible for setting the gas turbulence?
 The redshift evolution of \sigo\ can be understood in the
 framework of marginally stable disks with gas fractions that are dwindling
 with cosmic time.  Yet, at fixed redshift, no clear correlation with galaxy
 properties is emerging that would unambiguously identify the main driver of
 turbulence.  Is this because of limited dynamic range sampled, significant
 contributions from unresolved non-circular motions, other observational
 factors?
 Results from strongly-lensed galaxies indicate elevated dispersions on
 scales down to a few 100~pc, but samples are still small and limited in
 galaxy mass coverage.  Tighter constraints on spatial variations and
 anisotropy (as observed in nearby disks) will be helpful in addressing
 these questions.
\item
 What is the origin of the high baryon fractions and concentrations of SFGs?
 A robust trend of increasing baryon fractions with redshift up to $z \sim 2.5$,
 and a correlation with increasing surface density, are emerging from disk modeling
 of IFU kinematics.  Several lines of empirical evidence, supported by theoretical
 work, point to the important role of efficient transport of material from
 the halo to the disk scale and further inwards to the bulge in the gas-rich
 high $z$ disks.  More direct constraints are needed on gas inflows onto and
 within galaxies, and on the relative importance of radial transport vs.\
 inside-out growth in setting the structure of galactic components and, possibly,
 contributing to star formation quenching.
\item
 Where do massive $z \sim 2$ SFGs form their last stars before they get
 quenched?  Balmer decrement maps for individual galaxies and bolometric
 UV+IR SFR maps accounting for potential gradients in dust temperature will
 be required to address whether half-SFR sizes at the tip of the MS are smaller
 than, equal to, or larger than the half-stellar mass sizes inferred from
 multi-wavelength {\it HST\/} imagery.
\item
 What are the total mass loading and energetics of galactic-scale winds,
 and the breakdown into multi-phase components?  Much of our knowledge
 about wind properties and demographics is based on the warm ionized and
 neutral phase.  A more holistic view on wind properties and their impact
 on galaxies will strongly benefit from the combination of
 multi-phase tracers, still limited to small numbers of more extreme objects
 and very few normal MS SFGs at high $z$.
 A few pilot programs
 suggest that, akin to what is seen in nearby starbursts, the bulk
 of the mass flow may be in the molecular phase, highlighting the importance
 of cold molecular gas kinematics to fully capture their role in
 galaxy evolution and baryon cycling.
\item
 What are the exact mechanisms responsible for the shutdown of star formation
 in massive galaxies?
 The increase in the prevalence of massive bulges, dense cores, and powerful AGN
 and AGN-driven outflows at high galaxy masses, where the specific SFR and cold
 gas mass fractions drop, suggest they likely play a role in galaxy quenching.
 The evidence of an association with quenching, however, remains to date largely
 circumstancial, and further observational constraints are needed to pin down
 the mechanism(s) at play and establish causality.
\item
\looseness=-2
 How do galaxies below $\rm \sim 10^{9-9.5}~M_{\odot}$ fit into
 the emerging picture anchored in the properties of higher mass populations?
 Low-mass galaxies are still poorly explored because of current observational
 limitations.  If an increasing proportion of the low-mass population has
 prolate/triaxial structure, how can we interpret their kinematics until
 we can fully resolve them?  Do scaling relations break down at these masses?
\end{enumerate}
\end{issues}

\begin{figure}[!t]
\centering
\includegraphics[scale=0.63,trim={0.0cm 4.5cm 0.0cm 4.5cm},clip=0,angle=0]
                {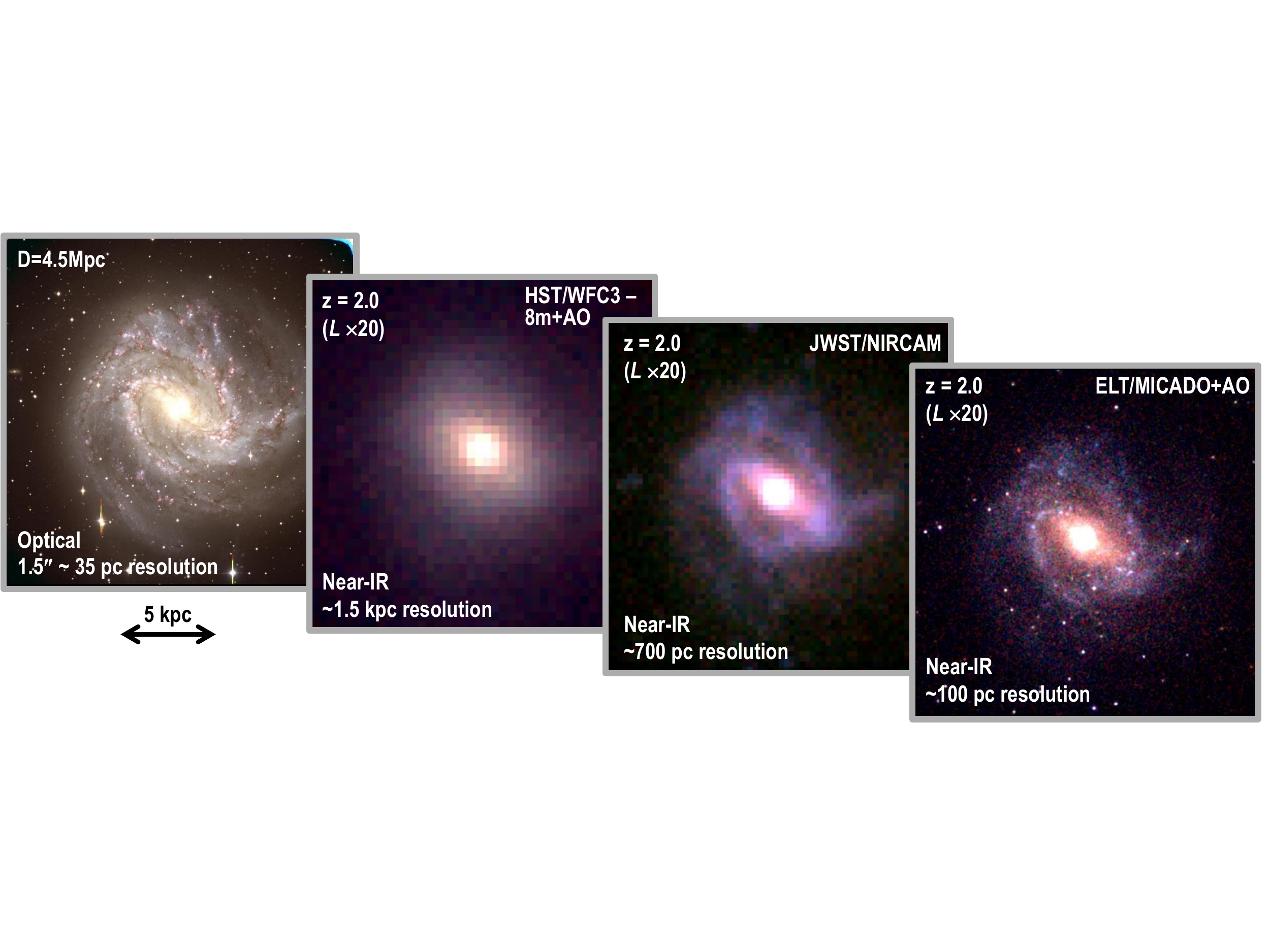}
\renewcommand\baselinestretch{0.85}
\caption{
Illustration of the gain in angular resolution from current to future
facilities.
For this simple illustration, optical imaging of the nearby M83 spiral
galaxy (at a distance of 4.5~Mpc, based on data presented by \citealt{Larsen99})
is redshifted to $z = 2$ and boosted up
in luminosity by a factor of $\sim 20$ (following the MS evolution) but
no other evolution is considered (e.g., in size or gas fraction).
The left panel shows the original color-composite map at a resolution 
corresponding to 35~pc.  Successive panels to the right are simulated
color-composite images for observations with {\it HST\/}
and AO-assisted instruments on 8\,m-class telescopes at a resolution of
$\rm \sim 1.5~kpc$, with the {\it JWST\/}/NIRCAM imager at a resolution of
$\rm \sim 700~pc$, and with the ELT/MICADO first-light instrument reaching
a diffraction-limited resolution of $\rm \sim 100~pc$ (pixel sampling is
adjusted for each instrument).
Simulations with the SimCADO software \citep{Les16} indicate that compact
cluster-like sources with luminosities comparable to those of bright super
star clusters in nearby starburst galaxies can be detected and characterized
with on-source integrations of a few hours.  Such objects at $z \sim 2$ might
be progenitors to today's metal-rich globular cluster population
\citep[e.g.][]{Sha10}.
\vspace{5ex}
}
\label{outlook.fig}
\end{figure}

The outlined questions, among others, frame the observational (and theoretical)
landscape for the next decade, with exciting progress anticipated from
developments on the instrumentation scene. 
NOEMA and ALMA are leveraging our knowledge about the stellar component
and ionized gas with that of the cold molecular gas.
The combination of the multi-IFU KMOS and the new sensitive AO-assisted
ERIS single-IFU at the VLT will expand samples with kinematics, star formation,
and ISM conditions from near-IR observations, and resolve them on sub-galactic
scales down to $\rm \sim 1~kpc$.
{\it JWST\/} at near- and mid-IR wavelengths will open up an unprecedented
window on the earliest stages of galaxy evolution, charting the progenitor
populations of cosmic noon galaxies.
The giant leap in resolution afforded by diffraction-limited instruments
on the next generation of  25-40\,m-class telescopes, such as the first-light
imager and spectrometer MICADO and the IFU HARMONI at the ELT, will be the
next game-changers (Figure\ \ref{outlook.fig}).
With unparalleled sharp views of the galaxy population on the scales of
individual giant molecular cloud and star-forming complexes, the era of
extremely large telescopes
will undoubtedly dramatically boost our knowledge and change our approach
to studying galaxy evolution across all times.

\vspace{45ex}

\section*{DISCLOSURE STATEMENT}
The authors are not aware of any affiliations, memberships, funding, or
financial holdings that might be perceived as affecting the objectivity
of this review.

\vspace{-2ex}

\section*{ACKNOWLEDGMENTS}

We are grateful to our many colleagues and friends for stimulating, critical,
and inspiring discussions throughout the years, which have all contributed to
shape the present work.
We thank the members of the SINS/zC-SINF, \kmostd, PHIBSS, and 3D-HST teams
for their input and involvement in various aspects covered in this article.
We give our special thanks to Ralf Bender, Michele Cirasuolo, Ric Davies,
Sandy Faber, Marijn Franx, Reinhard Genzel, Dieter Lutz, Trevor Mendel,
Sedona Price, Alvio Renzini, Mara Salvato, Alice Shapley, Taro Shimizu,
Linda Tacconi, Hannah \"Ubler, Pieter van Dokkum, Emily Wisnioski for
discussions and comments that were extremely useful while writing
this manuscript, and for input for Figures.
We are also grateful to Mark Swinbank, Marianne Girard,
and Chris Harrison for sharing information in advance of publication.
Star-forming galaxies at cosmic noon is a vast topic that rests on a much
richer body of work than can be included in a single article within space
allocation --- we have strived to provide useful references through which
further work can be found.




\clearpage


\clearpage

\section*{SUPPLEMENTAL TABLES}
\label{suppl_tab.sec}

The Tables below are associated with Figures\ 2 and 3 of the main article,
which feature a selection of extragalactic surveys providing relevant samples
at cosmic noon epochs, either specifically targeting objects or having a
significant number of sources overlapping with the $1 \leq z \leq 3$ interval.
Table\ \ref{suppl_tab_surveys} lists the photometric and spectroscopic surveys,
acronyms or brief description, and the main reference for the source catalogs
used in Figure\ 2.  Table\ \ref{suppl_tab_ifusurveys} focusses on the near-IR
IFU surveys plotted in Figure\ 3, with their acronyms or brief description,
the main IFU instrument and observing mode used, and the reference for the
published galaxy sample properties.

\clearpage

\newenvironment{changemargin}[2]{%
\begin{list}{}{%
\setlength{\leftmargin}{#1}%
\setlength{\rightmargin}{#2}%
}%
\item[]}{\end{list}}


\begin{sidewaystable}[hp]
\vspace{1.0cm}
\begin{changemargin}{-2.5cm}{2.0cm}
\tabcolsep4.0pt
\caption{Photometric and Spectroscopic Surveys}
\label{tab1}
\begin{center}
\begin{tabular}{@{}lll@{}}
\hline\hline
Survey$^{\rm a}$ & Full Name or Description & Reference(s) \\
\hline\hline
\multicolumn{3}{c}{Wide-field Ground-based Multi-band Photometric Imaging Surveys}\\
\hline
KiDS$+$VIKING-450$^{\rm b}$ (KV450) &
  Kilo Degree Survey $+$ VISTA Kilo degree Infrared Galaxy &
  \citet{Wright19} \\
HSC-SSP$^{\rm b}$ (HSC-W/-D/-UD) &
  Hyper Suprime-Cam Subaru Strategic Program (Wide, Deep, UltraDeep) &
  \citet{Aih18} \\
BM/BX/LBG          &
  Imaging campaigns centered on QSO fields &
  \citet{Ste03,Ste04} \\
\hline
\multicolumn{3}{c}{Wide- to Deep-field Ground-/space-based Multi-band
                   Photometric Surveys, including HST Imaging}\\
\hline
COSMOS        &
  Cosmic Evolution Survey (COSMOS2015 catalog) &
  \citet{Sco07,Lai16} \\
CANDELS            &
  Cosmic Assembly Near-infrared Deep Extragalactic Legacy Survey &
  \citet{Gro11,Koe11} \\
MUSYC/ECDFS        &
  Multi-wavelength Survey by Yale-Chile / Extended Chandra Deep Field South &
  \citet{Car10} \\
zFOURGE            &
  FourStar Galaxy Evolution survey &
  \citet{Straat16} \\
AEGIS          &
  All-wavelength Extended Groth strip International Survey, NEWFIRM Medium Band Survey &
  \citet{Davis07,Whi11} \\
HFF                &
  Hubble Frontier Fields Treasury Survey (HFF-DeepSpace catalog) &
  \citet{Lotz17,Shi18} \\
\hline
\multicolumn{3}{c}{HST Grism Spectroscopic Imaging Surveys}\\
\hline
3D-HST/AGHAST          &
  HST WFC3/G141 grism spectroscopic surveys  &
  \citet{Mom16} \\
WISPs              &
  WFC3 Infrared Spectroscopic Parallel surveys &
  \citet{Atek10,Col13} \\
GLASS              &
  Grism Lens-Amplified Survey from Space &
  \citet{Treu15} \\
\hline
\multicolumn{3}{c}{Ground-based Optical and Near-IR Spectroscopic Surveys}\\
\hline
DEEP2/DEEP3        &
  Keck/DEIMOS spectroscopic surveys &
  \citet{Newman13,Coo12} \\
VVDS               &
  VIMOS VLT Deep Survey &
  \citet{LeF13} \\
zCOSMOS            &
  Bright and Deep spectroscopic survey in COSMOS (COSMOS2015 catalog) &
  \citet{Lil09,Lai16} \\
BM/BX/LBG          &
  Optical spectroscopy of rest-UV-selected galaxies &
  \citet{Ste03,Ste04} \\
VUDS$^{\rm b}$     &
  VIMOS Ultra-Deep Survey &
  \citet{LeF15,Tas17} \\
VANDELS$^{\rm b}$   &
  VIMOS survey of the CANDELS CDFS and UDS fields &
  \citet{McL18,Pen18} \\
GOODS-N            &
  Compilation of optical spectroscopic redshifts from multiple programs in GOODS-North &
  \citet{Bar08} \\
CDFS              &
  Compilation of optical spectroscopic redshifts from multiple programs in GOODS/CDF-South &
  Master Catalog v3.0$^{\rm c}$ (excluding VVDS) \\
KBSS-UV/-MOSFIRE &
  Keck Baryonic Structure Survey with LRIS (optical), MOSFIRE (near-IR) &
  \citet{Ste14,Strom17} \\
FastSound             &
  Subaru/FMOS near-IR galaxy redshift survey  &
  \citet{Ton15,Oka16} \\
FMOS-COSMOS           &
  Subaru/FMOS near-IR spectroscopic survey in COSMOS &
  \citet{Kash19} \\
MOSDEF                &
  MOSFIRE Deep Evolution Field survey (near-IR) &
  \citet{Kriek15} \\
zFIRE                 &
  Keck/MOSFIRE near-IR spectroscopic survey in rich environments &
  \citet{Nana16} \\
TKRS2         &
  Team Keck Redshift Survey in near-IR with MOSFIRE &
  \citet{Wirth15} \\
\hline
\end{tabular}
\end{center}
\begin{tabnote}
$^{\rm a}$Label used in Figure~2 is given in parenthesis if different from
the survey name.
$^{\rm b}$On-going observations or final data processing; information included in
Figure~2 is based on interim source catalog releases presented in the reference(s)
in the last column.
$^{\rm c}$The compilation and references are available at
 https://www.eso.org/sci/activities/garching/projects/goods/MasterSpectroscopy.html;
 the VVDS survey is plotted separately in Figure~2 of the main article. \\
\end{tabnote}
\label{suppl_tab_surveys}
\end{changemargin}
\end{sidewaystable}


\clearpage

\begin{sidewaystable}[hp]
\begin{changemargin}{-0.3cm}{2.0cm}
\tabcolsep4.0pt
\caption{Near-IR Integral Field Unit Spectroscopic Surveys}
\label{tab1}
\begin{center}
\begin{tabular}{@{}lllll@{}}
\hline\hline
Survey & Full Name or Description & Instrument and mode$^{\rm a}$ &
  Redshifts & Reference(s) \\
\hline\hline
\kmostd               &
  KMOS IFU survey across cosmic noon &
  KMOS, no-AO  & $0.6 - 2.7$ &
  \citet{Wis15,Wis19} \\
KROSS             &
  KMOS Redshift One Spectroscopic Survey  &
  KMOS, no-AO  & $0.6 - 1.0$ &
  \citet{Sto16,Har17} \\
KGES              &
  KMOS Galaxy Evolution Survey  &
  KMOS, no-AO  & $1.3 - 1.5$ &
  \citet{Gill20} \\
KDS               &
  KMOS Deep Survey  &
  KMOS, no-AO  & $3.1 - 3.8$ &
  \citet{Turner17} \\
KLEVER$^{\rm b}$  &
  KMOS LEnsed galaxies Velocity and Emission line Review &
  KMOS, no-AO  & $1.2 - 2.5$ &
  \citet{Curti20} \\
KLASS$^{\rm c}$ &
  KMOS Lens-Amplified Spectroscopic Survey &
  KMOS, no-AO  & $0.6 - 2.3$ &
  \citet{Mas17,Gir20} \\
KLENS   &
  KMOS LENsing Survey &
  KMOS, no-AO  & $1.2 - 3.6$ &
  \citet{Gir18} \\
KASHz$^{\rm b}$       &
  KMOS AGN Survey at High redshift &
  KMOS, no-AO  & $0.6 - 2.6$ &
  \citet{Har16,Scholtz20} \\
SINS/zC-SINF      &
  Spectroscopic Imaging survey in the Near-IR with SINFONI &
  SINFONI, no-AO and AO & $1.3 - 2.6$ &
  \citet{FS09} \\
                  &
  zCOSMOS-SINFONI project &
  SINFONI, no-AO      & $1.3 - 2.6$ &
  \citet{Man11} \\
SINS/zC-SINF AO   &
  AO follow-up of SINS/zC-SINF sample   &
  SINFONI, AO & $1.4 - 2.5$ &
  \citet{FS18} \\
MASSIV            &
  Mass Assembly Survey with SINFONI in VVDS &
  SINFONI, no-AO and AO & $0.9 - 2.2$ &
  \citet{Cont12,Epi12} \\
AMAZE            &
  Assessing the Mass-Abundance redshift[-Z] Evolution &
  SINFONI, no-AO    & $3.0 - 5.1$ &
  \citet{Tro14} \\
LSD              &
  Lyman-break galaxies Stellar populations and Dynamics &
  SINFONI, AO & $2.4 - 3.4$ &
  \citet{Tro14} \\
Law09             &
  IFU observations of rest-UV-selected galaxies &
  OSIRIS, AO & $2.0 - 3.3$ &
  \citet{Law09,Law12} \\
W09               &
  IFU observations of rest-UV-selected galaxies &
  OSIRIS, AO & $1.5 - 1.7$ &
  \citet{Wri09} \\
IROCKs        &
  Intermediate Redshift OSIRIS Chemo-Kinematic Survey &
  OSIRIS, AO & $0.8 - 1.4$ &
  \citet{Mie16} \\
WiggleZ           &
  IFU follow-up of rest-UV selected galaxies &
  OSIRIS, AO & $1.3 - 1.5$ &
  \citet{Wis11} \\
SHiZELS            &
  SINFONI - High-Z Emission Line Survey &
  SINFONI, OSIRIS, NIFS, AO & $0.8 - 3.3$ &
  \citet{Gill19} \\
Lensed             &
  Collection of strongly-lensed targets & 
  SINFONI, OSIRIS, NIFS, AO & $1.0 - 3.7$ &
  \citet{Jones10,Jones13} \\
                   &  &  &  &
  \citet{Yua11,Yua12} \\
                   &  &  &  &
  \citet{EWuy14} \\
                   &  &  &  &
  \citet{Liv15} \\
                   &  &  &  &
  \citet{Leetho16} \\
\hline
\end{tabular}
\end{center}
\begin{tabnote}
$^{\rm a}$ KMOS: multi-IFU at the VLT; 
           SINFONI: single-IFU at the VLT;
           OSIRIS: single-IFU at Keck;
           NIFS: single-IFU at Gemini.
Seeing-limited and adaptive optics-assisted surveys are distinguished (no-AO, AO).
$^{\rm b}$ Survey on-going or recently completed; band, on-source integration
times, number of objects targeted, redshift range, and galaxy properties of
detected subsets in Figure\ 3 and in fourth column refer to the published samples
in the references given in the last column.
$^{\rm c}$ KLASS also targeted $z>7$ galaxies.
\end{tabnote}
\label{suppl_tab_ifusurveys}
\end{changemargin}
\end{sidewaystable}


\clearpage

\section*{SUPPLEMENTAL TEXT: SPECTRAL AND KINEMATIC MODELING}
\label{suppl_mod.sec}

The past decade has seen important developments in modeling of the spectral
energy distribution (SED) and kinematics data of distant galaxies, to derive
their stellar populations properties such as stellar mass, age, star formation
rate and history as well as their dynamical properties such as circular velocity
and dynamical mass.
Deriving these fundamental properties is essential to place observed galaxies
in the theoretical framework of galaxy evolution through comparisons with
(semi-)analytical models and numerical cosmological simulations.
As spectral and kinematic data sets are growing rapidly in both sample size and
detail of information, increasingly sophisticated approaches are being developed
to improve the efficiency of modeling codes and treat adequately the various
parameter correlations involved.
Here we summarize basic ingredients and methods employed in state-of-the-art
SED and kinematic modeling applied to data of high-redshift galaxies.

\subsection*{Spectral Modeling}
  \label{suppl_mod_SED.sec}

The translation from SEDs to physical quantities describing a galaxy's stellar
mass, star formation rate or history requires the use of stellar population
synthesis (SPS), dust, and ideally photoionization models.  This is the case
for SEDs sampled at any spectral resolution, and we therefore discuss these
techniques indiscriminately of $R$.

The ingredients to SPS models include a stellar spectral library, a set of
isochrones, an IMF, and a star formation history (SFH).  Each of these components
is discussed in depth in the review by \citet{Conroy13}.  Here we highlight a few
succinct aspects of particular relevance to the study of distant galaxies.
The stellar library is to cover a range in stellar metallicities, effective
temperatures and surface gravities appropriate for the stellar population hosted
by the galaxy under consideration.  Since empirical libraries are composed from
spectral observations of stars in the Solar neighborhood, they may lack or cover
too sparsely certain regions of parameter space that potentially could contribute
significantly to the integrated emission of early galaxies.  In order to include
very sub- or super-Solar metallicities, or stars caught during short-lived
evolutionary phases such as Wolf-Rayet (WR) or thermally-pulsating asymptotic
giant branch (TP-AGB) phases, theoretical libraries can be employed instead,
even though these are not without flaws themselves, ranging from the treatment of
convection to the quality and completeness of atomic and molecular line lists
underpinning them.  A hybrid approach has been applied as well, in which
theoretically motivated differential corrections are applied to empirical spectra
to provide a denser and more complete sampling of parameter space, e.g., in
metallicity and elemental abundance \citep{Con12}.  Short-lived evolutionary
phases also pose a challenge when pairing stellar libraries with isochrones to
construct so-called single (i.e., mono-age) stellar populations.  Approaches
alternative to the ``isochrone synthesis'' technique have been explored by,
e.g., \citet{Maraston05} who adopted the fuel consumption theorem in which the
amount of hydrogen and/or helium consumed is taken as integration variable, in
principle allowing luminous, short-lived evolutionary stages such as TP-AGB stars
to be captured more fully.  With substantial contributions to the rest-frame
near-IR they were argued to significantly impact inferred galaxy stellar ages and
masses, particularly at cosmic noon where characteristic stellar population ages
match the phase where TP-AGB stars are most prominent
($\rm \sim 3 \times 10^8 - 2 \times 10^9~yr$).  That said, observational efforts
at intermediate \citep{Kriek10} and higher spectral resolution \citep{Zibetti13}
failed to find strong TP-AGB spectral signatures in those galaxies where they
ought to be most prominent, potentially explained by (self-produced) dust
attenuating the TP-AGB light.  An extensive review on the IMF, and evidence for
potential deviations from the standard \citet{Cha03} IMF, is presented by
\citet{Hopkins18}.  Claims of non-universality of the IMF based on observations
of distant galaxies themselves (e.g., a top-heavy IMF in order to reconcile number
counts of submillimeter galaxies with models \citep{Baugh05},
or to reconcile a census of
the cosmic SFR density and stellar mass assembly history \citep{Wilkins08}) are
not unambiguous in their interpretation \citep[see, e.g.,][]{Safarzadeh17,Lej19b}.
More convincing evidence, of a more bottom-heavy IMF in nearby ellipticals with
high velocity dispersions, was provided based on three orthogonal lines of
inquiry: IR spectroscopy \citep{Con12}, dynamical modeling \citep{Cap13} and
gravitational lensing \citep{Treu10}.  While the peak in SFH of these galaxies
can be traced back to the cosmic noon era, an application of such IMF variations
has yet to find its way into direct look-back studies, with as additional
complication that the respective IMF changes may be confined to the central
regions of these galaxies \citep{Con17}.

Finally, SFGs are not well represented by single stellar populations, and need
to be modeled with extended SFHs.  Here, a common approach has originally been
to parametrize the SFH by an exponentially declining, so-called $\tau$ model,
largely because of its historical roots in SPS modeling of nearby early-type
galaxies to which the technique was applied first.  \citet{Ren09} made the
case that rising SFHs may be more appropriate for SFGs at cosmic noon, and hence
more flexible functional forms (delayed $\tau$ models, log-normal SFHs, or double
power laws) are increasingly being adopted.  Offering yet more freedom,
\citet{Pacifici15} adopt a more extensive and physically motivated library of
SFHs drawn from a semi-analytical model of galaxy formation, and conclude that
a quantification of the normalization, slope and scatter of the stellar mass - SFR
relation can be severely biased if both quantities are inferred from a common,
oversimplified approach.
In the same vein \citet{Lej19a} advocate the use of more flexible non-parametric
(i.e., piecewise constant) star formation histories, and stress the importance
of adopting appropriate priors.

Attenuation by dust, present in copious amounts within massive SFGs at cosmic
noon, has a dimming and reddening effect on the emerging SED.  With the exception
of the potential presence of a bump at 2175\AA, often attributed to PAH molecules,
its wavelength dependence is smooth, but nevertheless leaves a signature that is
highly degenerate with variations in stellar age and/or metallicity.  Whereas the
most common approach is to adopt the \citet{Calz00} reddening law calibrated
locally on a sample of starbursting galaxies, in recent years first strides are
made to map the attenuation curves at high $z$ and their variation as a function
of galaxy type directly \citep[e.g.,][]{Kriek13, Red15}.  As an aid in breaking
age-metallicity-dust degeneracies, SPS modeling codes increasingly are capable
of accounting for far-IR constraints, where available.  Any emission absorbed
at short wavelengths should contribute to dust heating with associated
reprocessed emission at long wavelengths.  Several state-of-the-art SPS modeling
codes such as {\tt MAGPHYS} \citep{daCunha15}, {\tt BEAGLE} \citep{Chevallard16},
{\tt Prospector} \citep{Lej17}, and {\tt BAGPIPES} \citep{Carn18} now incorporate
such energy balance arguments as well as Bayesian inference to explore parameter
space.  If not known spectroscopically, redshifts can be fit for simultaneously
by these codes, enabling a self-consistent assessment of the error budget,
including covariances.

As a third component besides the SPS and dust models, photoionization codes such
as {\tt CLOUDY} \citep{Ferland17} or {\tt MAPPINGS} \citep{Sutherland17} can be
employed to superpose on the stellar emission the anticipated nebular lines.
This is indispensable for full spectral fitting, but contributions from nebular
line emission can also matter (and provided proper modeling even help) at lower
spectral resolutions, especially when medium- or narrow-bands are included
or for galaxies with high specific SFRs
\citep[e.g.,][]{vdWel11}.  \citet{Kew19}
and \citet{Mai19} present comprehensive overviews of the ingredients to
photoionization models and recent advances in their application and calibration
to galaxies across cosmic time.  The need for redshift-appropriate calibrations
was brought to light by the observation of systematic shifts in the characteristic
strong rest-optical line ratios captured in excitation diagrams, revealing the
evolving ISM conditions (see Section 3.6) as well as the changing shapes of the
ionizing radiation field.  Topics of
current debate in this regard entail, from a modeling perspective, the role
attributed to stellar rotation, binary evolution, and stellar mass loss in
determining the amount of ionizing photons and their hardness
\citep[e.g.,][]{Eldridge12}.  SPS codes equipped with grids from photoionization
models generally implement this in a self-consistent manner such that line
intensities are tied to the metallicity and star formation history of the
stellar population, but nevertheless the dimensionality of the problem is
typically increased by the introduction of additional free parameters, such
as the extra attenuation to \hii\ regions.

Overall, it is well established that stellar mass represents the quantity on
which SPS techniques can place the tightest constraints, as its inference
requires an assessment of the mass-to-light (M/L) ratio only, to zeroth-order
blind to the physical conditions responsible for setting this M/L (i.e., the
balance of age, metallicity and dust attenuation).  Whereas systematic
differences arise depending on the assumptions made, code-by-code comparisons
at various levels of control suggest that at least in terms of mass ranking a
high degree of consistency is reached \citep{Mobasher15}.  SFRs can be more
challenging in the presence of large columns of dust, in which case panchromatic
information aids greatly.  Star formation histories represent the most
challenging inference in the case of SFGs.

Looking ahead, a few avenues can be identified for future progress in this area.
First, with the increasing availability (and with the advent of JWST
also wavelength coverage) of spatially resolved information, SPS modeling can
be applied to SEDs extracted on sub-galactic scales.  This has the merit of
allowing to trace the stellar build-up in situ, but in addition can mitigate
outshining effects.  Whereas resolved SED modeling may go at the expense of
wavelength coverage and sampling, galaxy-integrated constraints can be imposed
\citep[see, e.g.,][]{Wuy12}.
Second, in almost all applications to date a uniform
metallicity is adopted for the entire stellar population.  In future work, one
could envision star formation and chemical enrichment histories to be coupled
self-consistently, an approach that several of the aforementioned codes already
allow for in principle.  Exactly what constitutes a self-consistent treatment
is an issue that may not be addressed trivially, as the connection between the
two histories is modulated by gaseous in- and outflows, both of which are
ubiquitous around cosmic noon.  Finally, a full interpretation of galaxy spectra
and emission lines would ideally account not only for full SPS but also for
radiative transfer.  Such full-fledged 3D radiative transfer modeling is to
date restricted to a handful of very nearby galaxies for which very
high-resolution datasets are available \citep[e.g.,][]{DeLooze14}.
Much simplified analytical descriptions of absorption and scattering under
different geometries such as homogeneous mixtures, (clumpy) foreground screens,
and mixtures thereof can be applied via analytical recipes to interpret the
distribution of line strengths and ratios resolved on kiloparsec scales
within 100s of nearby galaxies \citep[e.g.,][]{Li19}.

\subsection*{Kinematic Measurements and Modeling}
  \label{suppl_mod_kin.sec}

To date, kinematics of distant star-forming galaxies come exclusively from
observations of emission lines, mostly H$\alpha$ or other rest-optical nebular
lines, or CO transitions in the submillimeter regime.
The best constraints are obtained from integral field unit (IFU) spectroscopy
or interferometry, providing simultaneously the full three-dimensional (3D) 
data, which is the focus in what follows.
Galaxy-integrated and slit spectra have been used to derive kinematic properties,
and slit spectra were also modeled, following similar approaches as outlined
below adapted for that type of data \citep[e.g.,][]{Wei06,Pri16}.

The data are usually interpreted in the framework of axisymmetric rotating
disks motivated by the observations (Section~4 of the main article) where
physical quantities of interest include for instance the intrinsic peak
rotation velocity \vrot\ and local velocity dispersion \sigo
\footnote{
 For an exponential model, the maximum velocity is reached at a radius
 $R_{\rm max} = 2.2\,R_{\rm d} = 1.3\,R_{\rm e}$, where $R_{\rm d}$ is the disk
 scale length and $R_{\rm e}$ the effective radius enclosing half the light,
 in which case measuring a $v_{2.2}$ at $2.2\,R_{\rm d}$ is the same as
 \vrot.  The choice between \vrot\ and $v_{2.2}$ depends on the goal of the
 analysis.  Deviations from an exponential distribution change the
 $R_{\rm max}/R_{\rm e}$ \citep{Bin08}, such that measuring the maximum
 \vrot\ is ideally done from the velocity curve rather than at fixed radius.
 The quantity \sigo\ refers to the velocity dispersion across the galaxy as
 a measure of ``turbulence,'' which, in the case of a disk and isotropic
 dispersion, is related to its geometrical thickness.
 It is to be distinguished from the total velocity dispersion \sigt\ measured
 from the line width in source-integrated spectra (which includes line broadening
 from galaxy-wide velocity gradients) and from the central velocity dispersion
 commonly employed in the analysis of early-type systems (which would be strongly
 dominated by beam-smearing of the steep inner velocity gradient for a disk).
 To minimize line broadening caused by inner disk velocity gradients,
 \sigo\ is best measured away from the central regions.  
 In data of high-$z$ galaxies, \sigo\ may also include contributions by
 noncircular motions on scales below the resolution element
 ($\rm \la 1 - 5~kpc$ depending on data set).
},
and the total dynamical mass of the system \mdyn.
Various approaches are followed, ranging from simple determinations based
on the observed maximum velocity difference and line widths measured directly
from the data or estimated by adjusting a parametric representation of the
velocity curve and dispersion profile (e.g., computed for an exponential
distribution, or approximated by an arctan function) to full forward modeling
of the data.  The simpler methods use one-dimensional (1D) major-axis profiles
or 2D maps extracted from the data cubes.  The flux, velocity, and dispersion
are usually obtained by fitting the observed line emission with a single
Gaussian, which was shown to be adequate for the typical resolved scales
and S/N levels of high-$z$ galaxy data \citep{FS18,Tiley19}.
Deviations from a single Gaussian may become appreciable in galaxy-integrated
spectra depending, for instance, on the spatial distribution of the tracer
emission line, the local intrinsic gas velocity dispersion, and the possible
presence of strong galactic-scale winds; these effects should be taken into
account in estimating rotation velocities from galaxy-integrated line widths
\citep[e.g.,][]{deBlok14,Wis18}.
Spatial beam smearing, instrumental spectral resolution, and galaxy
inclination $i$ are treated explicitly by rescaling of the observed maximum
velocity and local dispersion through functions or lookup tables based on mock
beam-smeared rotating disk models parametrized in terms of $R_{\rm beam}/R_{\rm e}$
and galaxy properties (such as mass and inclination for \sigo), by subtracting
in quadrature the instrumental broadening from the measured dispersion, and by
dividing the projected velocity by $\sin(i)$ derived from the morphology
\citep[e.g.,][]{Bur16,John18}.
Studies applying forward modeling perform fits of 1D profiles, 2D maps, or 3D
cubes.  The effects of resolution and inclination are treated implicitly by
convolving the intrinsic, inclined 3D model with a kernel representing the
point and line spread functions
\citep[PSF and LSF; e.g.,][]{Cre09,diT15}.

Kinematic modeling codes developed specifically for application to
observations at high redshift have become increasingly sophisticated in recent
years to allow more flexibility in model assumptions, more efficient parameter
space exploration, and better quantification of uncertainties of the best-fit
values accounting for covariances.  Examples include the {\tt DYSMAL} code
\citep[e.g.,][with recent updates described by
 \citealt{Wuy16}, \citealt{Ueb18}, \citealt{Genz20}]{Davies11},
{\tt GalPaK}$^{\rm 3D}$ \citep{Bou15}, and $\rm ^{3D}${\tt BAROLO} \citep{diT15},
all based on axisymmetric models but differing in ingredients and
dimensional space in which fits are performed.
The most recent version of {\tt DYSMAL} allows to fit multiple mass components,
such as a disk and a bulge, with relative mass ratios specified and
parametrized as S\'ersic profiles; the code self-consistently accounts
for finite thickness (turbulent disk, flattened rotatin bulge) based on
\citet{Noor08}.  The baryonic component(s) can be embedded in a dark matter
halo with a choice of profile parametrizations
(e.g., ``NFW,'' \citealt{Nav96}; double power-law; cored \citealt{Bur95} profile).
The line-of-sight velocity distribution is computed from the total
mass model and relative weights can be applied to the light of different
components.  {\tt DYSMAL} is optimized to fit in 1D or 2D although 3D fitting
also is possible.
$\rm GalPaK^{3D}$ is designed to fit simultaneously structural and kinematic
parameters directly in 3D data cubes, assuming a light/mass component
among several choices (e.g., exponential, Gaussian, and de Vaucouleurs profiles),
different parametrizations of the velocity profile (e.g., arctan, inverted
exponential, hyperbolic, or computed from the 3D mass model).
Both {\tt DYSMAL} and $\rm GalPaK^{3D}$ employ Markov chain Monte Carlo (MCMC)
algorithms in a Bayesian framework to derive the best-fit parameters and
uncertainties thereof.
$\rm ^{3D}${\tt BAROLO} fits tilted ring models to 3D data, where each concentric
ring
is parametrized independently and is randomly populated with line emitting clouds
in six dimensions (three in each of spatial and velocity space), from which line
profiles are built and projected into the model cube.  This method can more
naturally account for possible variations in orbits with radius, such as warps.
The code uses a multidimensional downhill simplex solver for the minimization
of non-analytic models, with uncertainties estimated via a Monte Carlo method.

\looseness=-2
In principle, fitting in 3D space offers a number of advantages as it avoids
the necessary loss of information in extracting the projected 2D maps or 1D
profile from both data and model.  In practice, the success of the fits can
be hampered by low S/N and irregular or clumpy light distributions.
For axisymmetric mass distributions, most of the information is encoded along
the line of nodes, such that the parameters can be well determined from 1D fits;
for sufficiently high S/N and well resolved galaxies, 2D maps can constrain
more accurately the inclination.
Especially at high redshift, the morphology in line emission can be quite
different from the underlying mass distribution and cannot be captured by
simple representations, let alone in 3D (which ideally would best account
for projection and light weighting effects); in such cases, fits are best
performed only in velocity and dispersion.
Despite the flexibility afforded by the above models, the observations may
not allow to constrain well all possible parameters but the implementation
of Bayesian analysis and MCMC algorithms have brought a major improvement
over previous modeling by allowing for priors and propagation of uncertainties
including covariances rather than simply fixing values.

The residuals between observed data and best-fit kinematic model can be used
in implementing the kinematic classification scheme discussed in Section~4.3.
An alternative classification method relies on kinemetry, introduced by
\citet{Kraj06} to analyze data of nearby early-type galaxies and
adapted for applications to IFU studies of distant SFGs by \citet{Sha08}.
Kinemetry is a generalization of surface photometry to the higher-order moments
of the line-of-sight distribution, where the degree of (a)symmetry in the velocity
field and dispersion map along best-fit ellipses 
is quantified through harmonic expansions.
The exact values of the parameters will depend on the resolution and S/N regime
of the data, such that the boundaries to distinguish between disks and mergers
need to be appropriately calibrated for the data sets under analysis.





\begin{thebibliography}{00}

\bibitem[Abraham \etal(1996)]{Abraham96}
  Abraham RG, van den Bergh S, Glazebrook K, \etal~1996.
  \apjs, 107:1--17
\bibitem[Abramson \etal(2014)]{Abr14}
  Abramson LE, Kelson DD, Dressler A, \etal~2014.
  \apj, 785:36
\bibitem[Abramson \etal(2015)]{Abr15}
  Abramson LE, Gladders MD, Dressler A, \etal~2015.
  \apjl, 801:L12
\bibitem[Abramson \etal(2016)]{Abr16}
  Abramson LE, Gladders MD, Dressler A, \etal~2016.
  \apj, 832:7
\bibitem[Abramson \& Morishita(2018)]{Abr18}
  Abramson LE, Morishita T. 2018.
  \apj, 858:40
\bibitem[Abruzzo \etal(2018)]{Abruzzo18}
 Abruzzo MW, Narayanan D, Dav\'e R, Thompson R. 2018.
 arXiv1803.02374
\bibitem[Adamo \etal(2013)]{Adamo13}
 Adamo A, \"Ostlin G, Bastian N, \etal~2013.
 \apj, 766:105
\bibitem[Adamo \etal(2017)]{Adamo17}
 Adamo A, Ryon JE, Messa M, \etal~2017.
 \apj, 841:131
\bibitem[Aihara \etal(2018)]{Aih18}
  Aihara H, Armstrong R, Bickerton S, \etal~2018.
  \pasj, 70:S8
\bibitem[Aird \etal(2012)]{Aird12}
  Aird J, Coil AL, Moustakas J, \etal~2012.
  \apj, 746:90
\bibitem[Aird \etal(2018)]{Aird18}
  Aird J, Coil AL, Georgakakis A. 2018.
  \apj, 775:41
\bibitem[Allen \etal(2017)]{Allen17}
  Allen RJ, Kacprzak GG, Glazebrook K, \etal~2017.
  \apj, 834:11
\bibitem[Aumer \etal(2010)]{Aumer10}
  Aumer M, Burkert A, Johansson PH, Genzel R. 2010.
  \apj, 719:1230-43
\bibitem[Barro \etal(2013)]{Bar13}
 Barro G, Faber SM, P\'erez-Gonz\'alez PG, \etal~2013.
 \apj, 765:104
\bibitem[Barro \etal(2014a)]{Bar14a}
 Barro G, Faber SM, P\'erez-Gonz\'alez PG, \etal~2014a.
 \apj, 791:52
\bibitem[Barro \etal(2014b)]{Bar14b}
 Barro G, Trump JR, Koo DC, \etal~2014b.
 \apj, 795:145
\bibitem[Barro \etal(2016)]{Bar16}
 Barro G, Kriek M, P\'erez-Gonz\'alez PG, \etal~2016.
 \apj, 827:32
\bibitem[Barro \etal(2017a)]{Bar17a}
  Barro G, Faber SM, Koo DC, \etal~2017a.
  \apj, 840:47
\bibitem[Barro \etal(2017b)]{Bar17b}
 Barro G, Kriek M, P\'erez-Gonz\'alez PG, \etal~2017b.
 \apjl, 851:L40
\bibitem[Behroozi \etal(2013a)]{Beh13a}
  Behroozi PS, Wechsler RH, Conroy C. 2013a.
  \apjl, 762:L31
\bibitem[Behroozi \etal(2013b)]{Beh13b}
  Behroozi PS, Wechsler RH, Conroy C. 2013b.
  \apj, 770:57
\bibitem[Beifiori \etal(2017)]{Bei17}
  Beifiori A, Mendel JT, Chan JCC, \etal~2017.
  \apj, 846:120
\bibitem[Bell \& de Jong(2001)]{Bell01}
 Bell EF, de Jong RS. 2001.
 \apj, 550:212--29
\bibitem[Belli \etal(2017a)]{Belli17a}
 Belli S, Newman AB, Ellis RS. 2017.
 \apj, 834:18
\bibitem[Belli \etal(2017b)]{Belli17b}
 Belli S, Genzel R, F\"orster Schreiber NM, \etal~2017b.
 \apjl, 841:L6
\bibitem[Bellocchi \etal(2016)]{Bello16}
 Bellocchi E, Arribas S, Colina L. 2016.
 \aap, 591:A85
\bibitem[Bezanson \etal(2019)]{Bez19}
 Bezanson R, Spilker J, Williams CC, \etal~2019.
 \apj, 873:L19
\bibitem[Binney \& Tremaine(2008)]{Bin08}
  Binney J, Tremaine S. 2008.
  Galactic Dynamics (2nd ed.; Princeton, NJ: Princeton Univ. Press)
\bibitem[Bland-Hawthorn \& Gerhard(2016)]{BH16}
  Bland-Hawthorn J, Gerhard O. 2016.
  \araa, 54:529--96
\bibitem[Boada \etal(2015)]{Boada15}
 Boada S, Tilvi V, Papovich C, \etal~2015.
 \apj, 803:104
\bibitem[Bolatto \etal(2015)]{Bolatto15}
 Bolatto AD, Warren SR, Leroy AK, \etal~2015.
 \apj, 809:175
\bibitem[Bolatto \etal(2013)]{Bolatto13}
 Bolatto AD, Wolfire M, Leroy AK. 2013.
 \araa, 51:207--68
\bibitem[Bouch\'e \etal(2010)]{Bou10}
  Bouch\'e N, Dekel A, Genzel R, \etal~2010.
  \apj, 718:1001--18
\bibitem[Bournaud \etal(2007)]{Bournaud07}
 Bournaud F, Elmegreen BG, Elmegreen DM. 2007.
 \apj, 670:237--48
\bibitem[Bournaud \etal(2014)]{Bournaud14}
 Bournaud F, Perret V, Renaud F, \etal~2014.
 \apj, 780:57
\bibitem[Bournaud(2016)]{Bournaud16}
 Bournaud F. 2016.
 in {\it Galactic Bulges\/}, \assl, 418:355--90.
 Eds. E. Laurikainen \etal~(City:Springer)
\bibitem[Bower \etal(2017)]{Bow17}
 Bower RG, Schaye J, Frenk CS, \etal~2017
 \mnras, 465:32--44
\bibitem[Brandt \& Alexander(2015)]{Brandt15}
 Brandt WN, Alexander DM. 2015.
 \aarev, 23:1
\bibitem[Brinchmann \etal(2004)]{Brinch04}
  Brinchmann J, Charlot S, White SDM, \etal~2004.
  \mnras, 351:1151--79
\bibitem[Brusa \etal(2015a)]{Bru15a}
  Brusa M, Bongiorno A, Cresci G, \etal~2015a.
  \mnras, 446:2394--2417
\bibitem[Brusa \etal(2018)]{Bru18}
  Brusa M, Cresci G, Daddi E, \etal~2018.
  \aap, 612:A29
\bibitem[Buck \etal(2017)]{Buck17}
  Buck T, Macci\`o AV, Obreja A, \etal~2017.
  \mnras, 468:3628--49
\bibitem[Bullock \etal(2001)]{Bullock01}
  Bullock JS, Dekel A, Kolatt TS, \etal~2001.
  \apj, 555:240--57
\bibitem[Burkert \etal(2016)]{Bur16}
  Burkert A, F\"orster Schreiber NM, Genzel R, \etal~2016.
  \apj, 826:214
\bibitem[Cano-Di\'az \etal(2012)]{Can12}
  Cano-Di\'az M, Maiolino R, Marconi A, \etal~2012.
  \aap, 537:L8
\bibitem[Caplar \& Tacchella(2019)]{Cap19}
  Caplar N, Tacchella S. 2019.
  \mnras, 487:3845--69
\bibitem[Cappellari \etal(2013)]{Cap13}
 Cappellari M, Scott N, Alatalo K, \etal~2013.
 \mnras, 432:1709--41
\bibitem[Cappellari(2016)]{Cap16}
  Cappellari M. 2016.
  \araa, 54:597--665
\bibitem[Carnall \etal(2019)]{Carn19}
  Carnall AC, Leja J, Johnson BD, \etal~2019
  \apj, 873:44
\bibitem[Carnall \etal(2018)]{Carn18}
  Carnall AC, McLure RJ, Dunlop JS, Dav\'e R. 2018
  \mnras, 480:4379--401
\bibitem[Carniani \etal(2016)]{Car16}
  Carniani S, Marconi A, Maiolino R, \etal~2016
  \aap, 591:A28
\bibitem[Carollo \etal(2013)]{Carollo13}
 Carollo CM, Bschorr TJ, Renzini A, \etal~2013.
 \apj, 773:112
\bibitem[Cava \etal(2018)]{Cava18}
 Cava A, Schaerer D, Richard J, \etal~2018.
 Nature Astronomy, 2:76--82
\bibitem[Ceverino \etal(2010)]{Cev10}
 Ceverino D, Dekel A, Bournaud F. 2010.
 \mnras, 404:2151--69
\bibitem[Ceverino \etal(2015)]{Cev15}  
 Ceverino D, Dekel A, Tweed D, Primack J. 2015.
 \mnras, 447:3291--310
\bibitem[Chabrier(2003)]{Cha03}
 Chabrier G. 2003
 \pasp, 115, 763--95
\bibitem[Chandar \etal(2014)]{Chandar14}
 Chandar R, Whitmore, BC, Calzetti D, O'Connell R. 2014.
 \apj, 787:17
\bibitem[Chang \etal(2013)]{Chang13}
 Chang YY, van der Wel A, Rix HW, \etal~2013.
 \apj, 762:83
\bibitem[Chen \etal(2020)]{Chen20}
  Chen Z, Faber SM, Koo DC, \etal~2020.
  \apj, 897:102
\bibitem[Cheung \etal(2012)]{Cheung12}
  Cheung E, Faber SM, Koo DC, \etal~2012.
  \apj, 760:131
\bibitem[Cibinel \etal(2015)]{Cib15}
  Cibinel A, Le Floc'h E, Perret V, \etal~2015.
  \apj, 805:181
\bibitem[Ciesla \etal(2017)]{Cie17}
  Ciesla L, Elbaz D, Fensch J. 2017.
  \aap, 608:A41
\bibitem[Clauwens \etal(2017)]{Clauwens17}
 Clauwens B, Hill A, Franx M, Schaye J. 2017.
 \mnras, 469:58--62
\bibitem[Combes(2018)]{Com18}
  Combes, F. 2018
  \aarev, 26:5
\bibitem[Condon(1992)]{Condon92}
  Condon JJ. 1992.
  \araa, 30:575--611
\bibitem[Conroy(2013)]{Conroy13}
  Conroy C. 2013.
  \araa, 51:393--455
\bibitem[Conselice(2003)]{Con03}
 Conselice CJ. 2003.
 \apjs, 147:1--28
\bibitem[Conselice(2014)]{Con14}
  Conselice C. 2014.
  \araa, 52:291--337
\bibitem[Courteau \& Dutton(2015)]{Cou15}
  Courteau S, Dutton AA. 2015.
  \apj, 801:L20
\bibitem[Covington \etal(2011)]{Cov11}
  Covington MD, Primack JR, Porter LA, \etal~2011.
  \mnras, 415:3135--52
\bibitem[Cresci \etal(2009)]{Cre09}
  Cresci C, Hicks EKS, Genzel R, \etal~2009.
  \apj, 697:115--32
\bibitem[Cresci \etal(2015)]{Cre15}
  Cresci C, Mainieri V, Brusa M, \etal~2015.
  \apj, 799:82
\bibitem[Daddi \etal(2007)]{Dad07}
  Daddi E, Dickinson M, Morrison G, \etal~2007.
  \apj, 670:156
\bibitem[Danovich \etal(2015)]{Dan15}
  Danovich M, Dekel A, Hahn O, Ceverino D, Primack J. 2015.
  \mnras, 449:2087--111
\bibitem[Dav\'e \etal(2017)]{Dav17}         
  Dav\'e R, Rafieferantsoa MH, Thompson RJ, Hopkins PF. 2017.
  \mnras, 467:115--32
\bibitem[Davies \etal(2019)]{Dav19}
  Davies RL, F\"orster Schreiber NM, \"Uebler H, \etal~2019
  \apj, 873, 122
\bibitem[Dekel \& Silk(1986)]{Dek86}
  Dekel A, Silk J. 1986.
  \apj, 303:39--55
\bibitem[Dekel \etal(2009a)]{Dek09a}
  Dekel A, Sari R, Ceverino D. 2009a.
  \apj, 703:785--801
\bibitem[Dekel \& Burkert(2014))]{Dek14}
  Dekel A, Burkert A. 2014.
  \mnras, 438:1870--79
\bibitem[Delhaize \etal(2017)]{Delh17}
  Delhaize J, Smol\v{c}i\'c V, Delvecchio I, \etal~2017.
  \aap, 602:A4
\bibitem[Dessauges-Zavadsky \etal(2017)]{Dess17}
  Dessauges-Zavadsky M, Schaerer D, Cava A, Mayer L, Tamburello V. 2017.
  \apj, 836:22
\bibitem[Dessauges-Zavadsky \& Adamo(2018)]{Dess18}
  Dessauges-Zavadsky M, Adamo A. 2018.
  \mnras, 479:118--22
\bibitem[Diemer \etal(2017)]{Die17}
  Diemer B, Sparre M, Abramson LE, Torrey P. 2017.
  \apj, 839:26
\bibitem[Donley \etal(2018)]{Don18}
  Donley JL, Kartaltepe J, Kocevski D, \etal~2018.
  \apj, 853:63
\bibitem[Dutton \etal(2007)]{Dut07}
  Dutton AA, van den Bosch FC, Dekel A, Courteau S. 2007.
  \apj, 654:27--52
\bibitem[Dutton \& van den Bosch(2009)]{Dut09}
  Dutton AA, van den Bosch FC. 2009.
  \mnras, 396:141--64
\bibitem[Dutton \& Macci\`o(2014)]{Dut14}
  Dutton AA, Macci\`o AV. 2014.
  \mnras, 441:3359--74
\bibitem[Eales \etal(2014)]{Eal14}
  Eales S, de Vis P, Smith MW, \etal~2017.
  \mnras, 465:3125--33
\bibitem[Elbaz \etal(2007)]{Elbaz07}
  Elbaz D, Daddi E, Le Borgne D, \etal~2007.
  \aap, 468:33
\bibitem[Elbaz \etal(2011)]{Elbaz11}
 Elbaz D, Dickinson M, Hwang HS, \etal~2011.
 \aap, 533:119
\bibitem[Elmegreen(2009)]{Elm09}
  Elmegreen BG. 2009.
  In IAU Symp. 254, The Galaxy Disk in Cosmological Context,
  ed. J. Anderson, J. Bland-Hawthorn, \& B. Nordstr\"om
  (Cambridge: Cambridge Univ. Press), 289
\bibitem[Elmegreen \etal(2008)]{Elm08}    
 Elmegreen BG, Bournaud F, Elmegreen DM. 2008.
 \apj, 688:67--77
\bibitem[Elmegreen \& Elmegreen(2017)]{Elm17}   
  Elmegreen DB, Elmegreen BG.  2017.
  \apjl, 851, L44
\bibitem[Elmegreen \& Elmegreen(2005)]{Elm05}   
  Elmegreen BG, Elmegreen DM. 2005.
  \apj, 627:632--46
\bibitem[\'Epinat \etal(2010)]{Epi10}
  \'Epinat B, Amram P, Balkowski C, Marcelin M. 2010.
  \mnras, 401:2113--47
\bibitem[\'Epinat \etal(2012)]{Epi12}
  \'Epinat B, Tasca L, Amram P, \etal~2012.
  \aap, 539:A92
\bibitem[Erb(2008)]{Erb08}
  Erb DK. 2008.
  \apj, 674:151--6
\bibitem[Erb \etal(2012)]{Erb12}
  Erb DK, Quider AM, Henry AL, Martin CL. 2012.
  \apj, 759:26
\bibitem[Fabian(2012)]{Fab12}
  Fabian AC. 2012
  \araa, 50:455--89
\bibitem[Fall \& Romanowsky(2013)]{Fall13}
  Fall SM, Romanowsky AJ. 2013.
  \apj, 769:L26
\bibitem[Fang \etal(2018)]{Fang18}
  Fang JJ, Faber SM, Koo DC, \etal~2018.
  \apj, 858:100
\bibitem[Ferguson \etal(2004)]{Ferg04}
  Ferguson HC, Dickinson M, Giavalisco M, \etal~2004.
  \apjl, 600:L107--10
\bibitem[F\"orster Schreiber \etal(2009)]{FS09}
  F\"orster Schreiber NM, Genzel R, Bouch\'e N, \etal~2009.
  \apj, 706:1364--428
\bibitem[F\"orster Schreiber \etal(2011)]{FS11}
  F\"orster Schreiber NM, Shapley AE, Genzel R, \etal~2011.
  \apj, 739:45
\bibitem[F\"orster Schreiber \etal(2014)]{FS14}
  F\"orster Schreiber NM, Genzel R, Newman SF, \etal~2014.
  \apj, 787:38
\bibitem[F\"orster Schreiber \etal(2018)]{FS18}
  F\"orster Schreiber NM, Renzini A, Mancini C, \etal~2018.
  \apjs, 238:21
\bibitem[F\"orster Schreiber \etal(2019)]{FS19}
  F\"orster Schreiber NM, \"Ubler H, Davies RL, \etal~2019.
  \apj, 875:21
\bibitem[Franx \etal(1991)]{Fra91}
  Franx M, Illingworth G, de Zeeuw T. 1991.
  \apj, 383:112--34
\bibitem[Franx \etal(2008)]{Fra08}
  Franx M, van Dokkum PG, F\"{o}rster Schreiber NM, \etal~2008.
  \apj, 688:770--88
\bibitem[Freeman(1970)]{Free70}
  Freeman KC. 1970.
  \apj, 160:811--30
\bibitem[Freeman \etal(2019)]{Free19}
  Freeman WR, Siana B, Kriek M, \etal~2019.
  \apj, 873:102
\bibitem[Gatto \etal(2015)]{Gatto15}
  Gatto A, Walch S, Mac Low M-M, \etal~2015
  \mnras, 449:1057--75
\bibitem[Genel \etal(2008)]{Genel08}
  Genel S, Genzel R, Bouch\'e N, \etal~2008.
  \apj, 688:789--93
\bibitem[Genel \etal(2015)]{Genel15}
  Genel S, Fall SM, Hernquist L, \etal~2015.
  \apj, 804:L40
\bibitem[Genzel \etal(2006)]{Genz06}
  Genzel R, Tacconi LJ, Eisenhauer F, \etal~2006.
  \nat, 442:786--9
\bibitem[Genzel \etal(2008)]{Genz08}
  Genzel R, Burkert A, Bouch\'e N, \etal~2008.
  \apj, 687:59--77
\bibitem[Genzel \etal(2011)]{Genz11}
  Genzel R, Newman S, Jones T, \etal~2011.
  \apj, 733:101
\bibitem[Genzel \etal(2012)]{Genz12}
  Genzel R, Tacconi LJ, Combes F, \etal~2012.
  \apj, 746:69
\bibitem[Genzel \etal(2014b)]{Genz14b}
  Genzel R, F\"orster Schreiber NM, Rosario D, \etal~2014b.
  \apj, 796:7
\bibitem[Genzel \etal(2015)]{Genz15}
  Genzel R, Tacconi LJ, Lutz D, \etal~2015.
  \apj, 800:20
\bibitem[Genzel \etal(2017)]{Genz17}
  Genzel R, F\"orster Schreiber NM, \"Ubler H. \etal~2017.
  \nat, 543:397--401
\bibitem[Genzel \etal(2020)]{Genz20}
  Genzel R, Price SH, \"Ubler H, \etal~2020
  \apj, in press (arXiv:2006.03046)
\bibitem[Giavalisco \etal(1996)]{Gia96}
   Giavalisco M, Steidel CC, Macchetto FD. 1996.
   \apj, 470:189--94
\bibitem[Gillman \etal(2019)]{Gill19}
   Gillman S, Swinbank AM, Tiley AL, \etal~2019.
   \mnras, 486:175--94
\bibitem[Girard \etal(2018a)]{Gir18a}
   Girard M, Dessauges-Zavadsky M, Schaerer D, \etal~2018a.
   \aap, 613:A72
\bibitem[Gladders \etal(2013)]{Gla13}
  Gladders MD, Oemler A, Dressler A, \etal~2013.
  \apj, 770:64
\bibitem[Glazebrook(2013)]{Glaze13}
  Glazebrook K 2013.
  \pasa, 30:56
\bibitem[Goldbaum \etal(2015)]{Gold15}
  Goldbaum NJ, Krumholz MR, Forbes JC. 2015.
  \apj, 814:131
\bibitem[Goldbaum \etal(2016)]{Gold16}
  Goldbaum NJ, Krumholz MR, Forbes JC. 2015.
  \apj, 827:28
\bibitem[Graham \& Driver(2005)]{Graham05}
  Graham AW, Driver SP. 2005.
  \pasa, 22:118--27
\bibitem[Griffiths \etal(1994)]{Griffiths94}
  Griffiths RE, Casertano S, Ratnatunga KU, \etal~1994.
  \apj, 435:19--22
\bibitem[Grogin \etal(2011)]{Gro11}
  Grogin NA, Kocevski DD, Faber SM, \etal~2011.
  \apjs, 197:35
\bibitem[Guo \etal(2012)]{Guo12}
 Guo Y, Giavalisco M, Ferguson HC, Cassata P, Koekemoer AM. 2012.
 \apj, 757:120
\bibitem[Guo \etal(2015)]{Guo15}
 Guo Y, Ferguson HC, Bell EF, \etal~2015.
 \apj, 800:39
\bibitem[Guo \etal(2018)]{Guo18}
 Guo Y, Rafelski M, Bell EF, \etal~2018.
 \apj, 853:108
\bibitem[Harrison \etal(2016)]{Har16}
  Harrison CM, Alexander DM, Mullaney JR, \etal~2016.
  \mnras, 456:1195--220
\bibitem[Harrison \etal(2017)]{Har17}
  Harrison CM, Johnson HL, Swinbank AM, \etal~2017.
  \mnras, 467:1965--83
\bibitem[Harrison \etal(2012)]{Har12}
  Harrison CM, Alexander DM, Swinbank AM, \etal~2012.
  \mnras, 426:1073--96
\bibitem[Hayward \etal(2018)]{Hayward18}
  Hayward CC, Chapman SC, Steidel CC, \etal~2018.
  \mnras, 476:2278--87
\bibitem[Heckman \& Best(2014)]{Hec14}
  Heckman TM, Best PN. 2014.
  \araa, 52:589--660
\bibitem[Heckman \& Thompson(2017)]{Hec17}
  Heckman TM, Thompson TA. 2017.
  Galactic Winds and the Role Played by Massive Stars.
  In: Alsabti A., Murdin P. (eds) Handbook of Supernovae.
  Springer, Cham.  pp. 2431--54
\bibitem[Henriques \etal(2015)]{Henriques15}
 Henriques BMB, White SDM, Thomas PA, \etal~2015.
 \mnras, 451:2663--80
\bibitem[Herrera-Camus \etal(2019)]{Her19}
  Herrera-Camus R, Tacconi L, Genzel R, \etal~2010.
  \apj, 871:37
\bibitem[Hickox \etal(2014)]{Hic14}
  Hickox RC, Mullaney JR, Alexander DM, \etal~2014.
  \apj, 782:9
\bibitem[Hill \etal(2017)]{Hill17}
  Hill AR, Muzzin A, Franx M, Marchesini D. 2017.
  \apjl, 849:L26
\bibitem[Hocking \etal(2018)]{Hocking18}
  Hocking A, Geach JE, Sun Y, Davey N. 2018.
  \mnras, 473:1108--29
\bibitem[Hopkins \etal(2006)]{Hopkins06}
  Hopkins PF, Hernquist L, Cox TJ, \etal~2006.
  \apjs, 163:1--49
\bibitem[Hopkins \etal(2012)]{Hopkins12}
  Hopkins PF, Quataert E, Murray N.  2012.
  \mnras, 421:3522--37
\bibitem[Hopkins(2015)]{Hopkins15}
 Hopkins PF. 2015.
 \mnras, 450:53--110
\bibitem[Huang \etal(2013)]{Huang13}
  Huang JS, Faber SM, Willmer CNA, \etal~2013.
  \apj, 766:21
\bibitem[Huang \etal(2017)]{Huang17}
  Huang K-H, Fall SM, Ferguson HC, \etal~2017.
  \apj, 838:6
\bibitem[Huertas-Company \etal(2015)]{Huert15}
  Huertas-Company M, Gravet R, Cabrera-Vives G, \etal~2015.
  \apjs, 221:8 
\bibitem[Huertas-Company \etal(2016)]{Huert16}
  Huertas-Company M, Bernardi M, P\'erez-Gonz\'alez PG, \etal~2016.
  \mnras, 462:4495--516
\bibitem[Hung \etal(2019)]{Hung19}
  Hung C.-L., Hayward CC, Yuan T, \etal~2019.
  \mnras, 482:5125--37
\bibitem[Ilbert \etal(2009)]{Ilb09}
  Ilbert O, Capak P, Salvato M, \etal~2009.
  \apj, 690:1236--49
\bibitem[Ilbert \etal(2013)]{Ilb13}
  Ilbert O, McCracken HJ, Le F\`evre O, \etal~2013.
  \aap, 556:A55
\bibitem[Illingworth \etal(2013)]{Illing13}
  Illingworth GD, Magee D, Oesch PA, \etal~2013.
  \apjs, 209:6
\bibitem[Ivison \etal(2007)]{Ivi07}
  Ivison RJ, Magnelli B, Ibar E, \etal~2010.
  \aap, 518:L31
\bibitem[Jiang \etal(2019)]{Jiang19}
  Jiang F, Dekel A, Kneller, O, \etal~2019.
  \mnras, 488:4801--15
\bibitem[Johnson \etal(2018)]{John18}
  Johnson , \etal~2015.
  \mnras, 453:2540--57
\bibitem[Johnston \etal(2015)]{Joh15}
  Johnston R, Vaccari M, Jarvis M, \etal~2015.
  \mnras, 453:2540--57
\bibitem[Jones \etal(2010a)]{Jones10a}
  Jones TA, Swinbank AM, Ellis RS, Richard J, Stark DP. 2010a.
  \mnras, 404:1247--62
\bibitem[Kaasinen \etal(2017)]{Kaa17}
  Kaasinen M, Bian F, Groves B, Kewley LJ, Gupta A. 2017.
  \mnras, 465:3220--34
\bibitem[Kakkad \etal(2016)]{Kak16}
  Kakkad D, Mainieri V, Padovani P, \etal~2016.
  \aap, 592:A148
\bibitem[Kartaltepe \etal(2015)]{Kart15}
  Kartaltepe JS, Mozena M, Kocevski D, \etal~2015.
  \apjs, 221:11
\bibitem[Kashino \etal(2019)]{Kash19}
  Kashino D, Silverman JD, Sanders D, \etal~2019.
  \apjs, 241:10
\bibitem[Kassin \etal(2007)]{Kassin07}
  Kassin SA, Weiner BJ, Faber SM, \etal~2007.
  \apj, 660:L35--38
\bibitem[Kassin \etal(2012)]{Kassin12}
  Kassin SA, Weiner BJ, Faber SM, \etal~2012.
  \apj, 758:106
\bibitem[Kelson(2014)]{Kel14}
  Kelson DD. 2014.
  \arxiv1406.5191
\bibitem[Kelson \etal(2016)]{Kel16}
  Kelson DD, Benson AJ, Abramson LE. 2016.
  \arxiv1610.06566
\bibitem[Kewley \etal(2019)]{Kew19}
  Kewley LJ, Nicholls DC, Sutherland RS. 2019.
  \araa, 57:511--70
\bibitem[Kocevski \etal(2012)]{Koc12}
  Kocevski DD, Faber SM, Mozena M, \etal~2012.
  \apj, 744:148
\bibitem[Kocevski \etal(2017)]{Koc17}
  Kocevski DD, Barro G, Faber SM, \etal~2017.
  \apj, 846:112
\bibitem[Koekemoer \etal(2011)]{Koe11}
  Koekemoer AM, Faber SM, Ferguson HC, \etal~2011.
  \apjs, 197:36
\bibitem[Kornei \etal(2012)]{Kor12}
  Kornei KA, Shapley AE, Martin CL, \etal~2012.
  \apj, 758:135
\bibitem[Kravtsov(2013)]{Kra13}
  Kravtsov AV. 2013.
  \apjl, 764:L31
\bibitem[Krumholz \etal(2018)]{Kru18}
  Krumholz MR, Blakesley B, Forbes JC, Crocker RM. 2018.
  \mnras, 477:2716--40
\bibitem[Laigle \etal(2016)]{Lai16}
  Laigle C, McCracken HJ, Ilbert O, \etal~2016.
  \apjs, 224:24
\bibitem[Lang \etal(2014)]{Lang14}
  Lang P, Wuyts S, Somerville RS, \etal~2014.
  \apj, 788:11
\bibitem[Lang \etal(2017)]{Lang17}
  Lang P, F\"orster Schreiber NM, Genzel R, \etal~2017.
  \apj, 840:92
\bibitem[Larsen \& Richtler(1999)]{Larsen99}
  Larsen SS, Richtler T. 1999.
  \aap, 345:59--72
\bibitem[Law \etal(2009)]{Law09}
  Law DR, Steidel CC, Erb DK, \etal~2009
  \apj, 697:2057--82
\bibitem[Law \etal(2012a)]{Law12a}
  Law DR, Shapley AE, Steidel CC, \etal~2012a
  \nat, 487:338--40
\bibitem[Law \etal(2012b)]{Law12b}             
  Law DR, Steidel CC, Shapley AE, \etal~2012b
  \apj, 745:85
\bibitem[Law \etal(2012c)]{Law12c}             
  Law DR, Steidel CC, Shapley AE, \etal~2012c
  \apj, 759:29
\bibitem[Leitner(2012)]{Lei12}
  Leitner SN. 2012.
  \apj, 745:149
\bibitem[Leja \etal(2015)]{Lej15}
  Leja J, van Dokkum PG, Franx M, Whitaker KE. 2015.
  \apj, 798:115
\bibitem[Leja \etal(2019a)]{Lej19a}
 Leja J, Carnall AC, Johnson BD, Conroy C, Speagle JS. 2019a.
 \apj, 876:3
\bibitem[Leja \etal(2019b)]{Lej19b}
 Leja J, Johnson BD, Conroy C, \etal~2019b.
 \apj, 877:140
\bibitem[Leschinski \etal(2016)]{Les16}
  Leschinski K, Czoske O, K\"ohler R, \etal~2016.
  Proc. SPIE, 9911, 991124
\bibitem[Leung \etal(2019)]{Leu19}
  Leung GCK, Coil AL, Aird J, \etal~2019.
  \apj, 886:11
\bibitem[Lilly \etal(2013)]{Lil13}
  Lilly SJ, Carollo CM, Pipino A, Renzini A, Peng Y. 2013.
  \apj, 772:119
\bibitem[Lilly \& Carollo(2016)]{Lil16}
  Lilly SJ, Carollo CM. 2016.
  \apj, 833:1
\bibitem[Lindroos \etal(2018)]{Lindroos18}
  Lindroos L, Knudsen KK, Stanley, F, \etal~2018.
  \mnras, 476:3544--54
\bibitem[Liu \etal(2016)]{Liu16}
  Liu FS, Jiang D, Guo Y, \etal~2016.
  \apj, 822:L25
\bibitem[Liu \etal(2017)]{Liu17}
  Liu FS, Jiang D, Faber SM, \etal~2017.
  \apjl, 844:L2 
\bibitem[Liu \etal(2018)]{Liu18}
  Liu FS, Jia M, Yesuf HM, \etal~2018.
  \apj, 860:60
\bibitem[Livermore \etal(2015)]{Liv15}
  Livermore RC, Jones TA, Richard J, \etal~2015.
  \mnras, 450:1812--35
\bibitem[L\'opez-Sanjuan \etal(2013)]{Lop13}
  L\'opez-Sanjuan C, Le F\`evre O, Tasca LAM, \etal~2013.
  \aap, 553, A78
\bibitem[Lotz \etal(2004)]{Lotz04}
  Lotz JM, Primack J, Madau P. 2004.
  \aj, 128:163--82
\bibitem[Lotz \etal(2017)]{Lotz17}
  Lotz JM, Koekemoer A, Coe D, \etal~2017.
  \aj, 837:97
\bibitem[Lovell \etal(2018)]{Lov18}
  Lovell MR, Pillepich A, Genel S, \etal~2018.
  \mnras, 481:1950--75
\bibitem[Luo \etal(2017)]{Luo17}
  Luo B, Brandt WN, Xue YQ, \etal~2017.
  \apjs, 228:2
\bibitem[Lutz(2014)]{Lut14}
  Lutz D. 2014.
  \araa, 52:373--414
\bibitem[Madau \& Dickinson(2014)]{Mad14}
  Madau P, Dickinson M. 2014.
  \araa, 52:415--86
\bibitem[Magnelli \etal(2014)]{Magn14}
  Magnelli B, Lutz D, Saintonge A, \etal~2014.
  \aap, 561:86
\bibitem[Magnelli \etal(2015)]{Magn15}
  Magnelli B, Ivison RJ, Lutz D, \etal~2015.
  \aap, 573:A45
\bibitem[Maiolino \& Mannucci(2019)]{Mai19}
  Maiolino R, Mannucci F. 2019.
  \aap Review, 27:3
\bibitem[Mandelker \etal(2014)]{Mandelker14}
  Mandelker N, Dekel A, Ceverino D, \etal~2014.
  \mnras, 443:3675--702
\bibitem[Mandelker \etal(2017)]{Mandelker17}
  Mandelker N, Dekel A, Ceverino D, \etal~2017.
  \mnras, 464:635--65
\bibitem[Mannucci \etal(2010)]{Man10}
  Mannucci F, Cresci G, Maiolino R, Marconi A, Gnerucci A. 2010.
  \mnras, 408:2115--27
\bibitem[Marchesini \etal(2012)]{March12}
  Marchesini D, Stefanon M, Brammer GB, Whitaker KE, 2012.
  \apj, 748:126
\bibitem[Martinsson \etal(2013)]{Mart13}
  Martinsson TPK, Verheijen MAW, Westfall KB, \etal~2013.
  \aap, 557:A131
\bibitem[Martizzi \etal(2012)]{Mart12}
  Martizzi D, Teyssier R, Moore B, Wentz T. 2012.
  \mnras, 422:3081--91
\bibitem[Mason \etal(2017)]{Mas17}
  Mason CA, Treu T, Fontana A, \etal~2017.
  \apj, 838:14
\bibitem[Masters \etal(2016)]{Masters16}
  Masters D, Faisst A, Capak P. 2016.
  \apj, 828:18
\bibitem[Matthee \& Schaye(2019)]{Mat19}
  Matthee J, Schaye J. 2019.
  \mnras 484:915--32
\bibitem[McGaugh(2012)]{McG12}
  McGaugh S. 2012.
  \aj, 143:40
\bibitem[Mendel \etal(2015)]{Mendel15}
  Mendel JT, Saglia RP, Bender R, \etal~2015.
  \apjl, 804:L4
\bibitem[Mendel \etal(2020)]{Mendel20}
  Mendel JT, Beifiori A, Saglia RP, \etal~2020.
  \apj, 899, 87
\bibitem[Micha\l owski \etal(2014)]{Mich14}
  Micha\l owski MJ, Hayward CC, Dunlop JS, \etal~2014.
  \aap, 571:75
\bibitem[Mieda \etal(2016)]{Mie16}
  Mieda E, Wright SA, Larkin JE, \etal~2016.
  \apj, 831:78
\bibitem[Mihos \& Hernquist(1994)]{Mihos94}
  Mihos JC, Hernquist L. 1994.
  \apj, 431:9--12
\bibitem[Miller \etal(2012)]{Miller12}
  Miller SH, Ellis RS, Sullivan M, \etal~(2012)
  \apj, 753:74
\bibitem[Mo \etal(1998)]{Mo98}
  Mo HJ, Mao S, White SDM. 1998.
  \mnras, 295:319--36
\bibitem[Momcheva \etal(2016)]{Mom16}
  Momcheva IG, Brammer GB, van Dokkum PG, \etal~2016.
  \apjs, 225:27
\bibitem[Morrison \etal(2010)]{Morr10}
  Morrison GE, Owen FN, Dickinson M, Ivison RJ, Ibar E. 2010.
  \apjs, 188:178--86
\bibitem[Moster \etal(2013)]{Mos13}
  Moster BP, Naab T, White SDM. 2013
  \mnras, 428:3121--38
\bibitem[Mowla \etal(2019a)]{Mowla19a}
  Mowla L, van der Wel A, van Dokkum P, Miller TB. 2019a.
  \apj, 872:13
\bibitem[Mowla \etal(2019b)]{Mowla19b}
  Mowla L, van Dokkum P, Brammer GB. 2019b.
  \apj, 880:57
\bibitem[Mullaney \etal(2012a)]{Mull12a}
  Mullaney JR, Pannella M, Daddi E, \etal~2012a.
  \mnras, 419:95--115
\bibitem[Mullaney \etal(2012b)]{Mull12b}
  Mullaney JR, Daddi E, B\'ethermin M, \etal~2012b.
  \apj, 753:L30
\bibitem[Muratov \etal(2015)]{Mur15}
  Muratov AL, Kere\v{s} D, Faucher-Gigu\`ere C.-A. \etal~2015.
  \mnras, 454:2691--713
\bibitem[Murray \etal(2011)]{Mur11}
  Murray N, M\'enard B, Thompson TA.  2011.
  \apj, 735:66
\bibitem[Murray \etal(2005)]{Mur05}
  Murray N, Quataert E, Thompson TA.  2005.
  \apj, 618:569--85
\bibitem[Naab \& Ostriker(2017)]{Naab17}
  Naab T, Ostriker JP. 2017.
  \araa, 55:59--109
\bibitem[Navarro \& Steinmetz(2000)]{Nav00}
 Navarro JF, Steinmetz M. 2000.
 \apj, 538:477--88
\bibitem[Nelson \etal(2018)]{Nels18}
  Nelson D, Pillepich A, Springel V, \etal~2018.
  \mnras, 475:624--47
\bibitem[Nelson \etal(2013)]{Nel13}
  Nelson EJ, van Dokkum PG, Momcheva I, \etal~2013.
  \apjl, 763:L16
\bibitem[Nelson \etal(2014)]{Nel14}
  Nelson E, van Dokkum P, Franx M, \etal~2014.
  \nat, 513:394--97
\bibitem[Nelson \etal(2016a)]{Nel16a}
  Nelson EJ, van Dokkum PG, Momcheva, IG, \etal~2016a.
  \apjl, 817:L9
\bibitem[Nelson \etal(2016b)]{Nel16b}
  Nelson EJ, van Dokkum PG, F\"orster Schreiber NM, \etal~2016b.
  \apj, 828:27
\bibitem[Nesvadba \etal(2008)]{Nes08}
  Nesvadba\,N,\,Lehnert\,M,\,DeBreuck\,C,\,Gilbert\,A,\,vanBreugel\,W.\,2008.\aap,\,491:407-424
\bibitem[Newman \etal(2015)]{Newman15}
  Newman AB, Belli S, Ellis RS. 2015.
  \apjl, 813:L7
\bibitem[Newman \etal(2018b)]{Newman18b}
  Newman AB, Belli S, Ellis RS, Patel SG. 2018b.
  \apjl, 862:126
\bibitem[Newman \etal(2012a)]{New12a}
  Newman SF, Genzel R, F\"orster Schreiber NM, \etal~2012a.
  \apj, 761:43
\bibitem[Newman \etal(2013)]{New13}
  Newman SF, Genzel R, F\"orster Schreiber NM, \etal~2013.
  \apj, 767:104
\bibitem[Newman \etal(2012b)]{New12b}
  Newman SF, Shapiro Griffin K, Genzel R, \etal~2012b.
  \apj, 752:111
\bibitem[Noeske \etal(2007)]{Noe07}
  Noeske KG, Faber SM, Weiner BJ, \etal~2007.
  \apjl, 660:L43--6
\bibitem[Noordermeer(2008)]{Noor08}
  Noordermeer E. 2008.
  \mnras, 385:1359--64
\bibitem[Nordon \etal(2012)]{Nordon12}
  Nordon R, Lutz D, Genzel R, \etal~2012.
  \apj, 745:182
\bibitem[Nordon \etal(2013)]{Nordon13}
  Nordon R, Lutz D, Saintonge A, \etal~2013.
  \apj, 762:125
\bibitem[Obreschkow \& Glazebrook(2014)]{Obr14}
  Obreschkow D, Glazebrook K. 2014.
  \apj, 784:26
\bibitem[Oklop\v{c}i\'{c} \etal(2017)]{Okl17}
  Oklop\v{c}i\'{c} A, Hopkins PF, Feldmann R, \etal~2017.
  \mnras, 465:952--69
\bibitem[Oppenheimer \& Dav\'e(2006)]{Opp06}
  Oppenheimer BD, Dav\'e R. 2006.
  \mnras, 373:1265-92
\bibitem[Padovani \etal(2017)]{Pad17}
  Padovani P, Alexander DM, Assef RJ, \etal~2017.
  \aarev, 25:2
\bibitem[Parsa \etal(2016)]{Parsa16}
  Parsa S, Dunlop JS, McLure RJ, Mortlock A, 2016.
  \mnras, 456:3194--211
\bibitem[Patel \etal(2013a)]{Patel13a}
  Patel SG, van Dokkum PG, Franx M, \etal~2013a.
  \apj, 766:15
\bibitem[Patel \etal(2013b)]{Patel13b}
  Patel SG, Fumagalli M, Franx M, \etal~2013b.
  \apj, 778:115
\bibitem[Pearson \etal(2018)]{Pearson18}
  Pearson WJ, Wang L, Hurley PD, \etal~2018.
  \aap, 615:146
\bibitem[Peng \etal(2010)]{Pen10}
  Peng Y, Lilly SJ, Kova\v{c} K, \etal~2010.
  \apj, 721:193--221
\bibitem[P\'eroux \& Howk(2020)]{Per20}
  P\'eroux C, Howk JC. 2020.
  \araa, this volume
\bibitem[Peth \etal(2016)]{Peth16}
  Peth MA, Lotz JM, Freeman PE, \etal~2016.
  \mnras, 458:963-987
\bibitem[Pettini \& Pagel(2004)]{Pet04}
  Pettini M, Pagel BEJ. 2004.
  \mnras, 348:L59--63
\bibitem[Pillepich \etal(2018a)]{Pil18a}
  Pillepich A, Springel V, Nelson D, \etal~2018a
  \mnras, 473:4077--106
\bibitem[Pillepich \etal(2018b)]{Pil18b}
  Pillepich A, Nelson D, Hernquist L, \etal~2018b.
  \mnras, 475:648--75
\bibitem[Pillepich \etal(2019)]{Pil19}
  Pillepich A, Nelson D, Springel V, \etal~2019.
  \mnras, 490:3196--233
\bibitem[Popping \etal(2017)]{Popping17}
  Popping G, Decarli R, Man AWS, \etal~2017.
  \aap, 602:11
\bibitem[Price \etal(2020)]{Pri20}
  Price SH, Kriek M, Barro G, \etal~2020
  \apj, 894:91
\bibitem[Puglisi \etal(2017)]{Pug17}
  Puglisi A, Daddi E, Renzini A, \etal~2017.
  \apj, 838:18
\bibitem[Rangel \etal(2014)]{Ran14}
  Rangel C, Nandra K, Barro G, \etal~2014.
  \mnras, 440:
\bibitem[Reddy \etal(2015)]{Red15}
  Reddy NA, Kriek M, Shapley AE, \etal~2015.
  \apj, 806, 259
\bibitem[Reddy \etal(2018)]{Red18}
  Reddy NA, Oesch PA, Bouwens RJ, \etal~2018.
  \apj, 853:56
\bibitem[Renzini(2009)]{Ren09}
  Renzini A. 2009.
  \mnras, 398:58--62
\bibitem[Ribeiro \etal(2016)]{Ribeiro16}
  Ribeiro B, Le F\`{e}vre O, Tasca LAM, \etal~2016.
  \aap, 593:A22
\bibitem[Ribeiro \etal(2017)]{Ribeiro17}
  Ribeiro B, Le F\`{e}vre O, Cassata P, \etal~2017.
  \aap, 608:A16
\bibitem[Rigby \etal(2017)]{Rigby17}
  Rigby JR, Johnson TL, Sharon K, \etal~2017.
  \apj, 843:79
\bibitem[Rodighiero \etal(2011)]{Rod11}
  Rodighiero G, Daddi E, Baronchelli I, \etal~2011.
  \apjl, 739:L40
\bibitem[Rodighiero \etal(2014)]{Rod14}
  Rodighiero G, Renzini A, Daddi E, \etal~2014.
  \mnras, 443:19--30
\bibitem[Rodrigues \etal(2018)]{Rod18}
  Rodrigues M, Puech M, Flores H, Hammer F, Pirzkal N. 2018.
  \mnras, 475:5133--43
\bibitem[Rujopakarn \etal(2016)]{Rujo16}
  Rujopakarn W, Dunlop JS, Rieke GH, \etal~2016.
  \apj, 833:12
\bibitem[Salmon \etal(2016)]{Sal16}
  Salmon B, Papovich C, Long J, \etal~2016.
  \apj, 827:20
\bibitem[Salvato \etal(2019)]{Salv19}
  Salvato M, Ilbert O, Hoyle B. 2019.
  {\it Nature Astronomy\/}, 3:212--22
\bibitem[Sanders \etal(2016a)]{San16a}
  Sanders RL, Shapley AE, Kriek M, \etal~2016a.
  \apj, 816:23
\bibitem[Sanders \etal(2017)]{San17}
  Sanders RL, Shapley AE, Zhang K, Yan R. 2017.
  \apj, 850:136
\bibitem[Sanders \etal(2018)]{San18}
  Sanders RL, Shapley AE, Kriek M, \etal~2018.
  \apj, 858:99
\bibitem[Santini \etal(2012)]{Sant12}
  Santini P, Rosario DJ, Shao L, \etal~2012.
  \aap, 540:A109
\bibitem[Sargent \etal(2012)]{Sargent12}
  Sargent MT, B\'ethermin M, Daddi E, Elbaz D. 2012.
  \apj, 747:31
\bibitem[Schechter(1976)]{Schechter76}
  Schechter P. 1976.
  \apj, 203:297--306
\bibitem[Schreiber \etal(2015)]{Schreiber15}
  Schreiber C, Pannella M, Elbaz D, \etal~2015.
  \aap, 575:74
\bibitem[Scoville \etal(2007)]{Sco07}
  Scoville N, Abraham RG, Aussel H, \etal~2007.
  \apjs, 172:38--45
\bibitem[Scoville \etal(2017)]{Sco17}
  Scoville N, Lee N, Vanden Bout P, \etal~2017.
  \apj, 837:150
\bibitem[Seon \& Draine(2016)]{Seon16}
  Seon K-I, Draine BT. 2016.
  \apj, 833:201
\bibitem[Shapiro \etal(2008)]{Sha08}
   Shapiro KL, Genzel R, F\"orster Schreiber, \etal~2008.
   \apj, 682:231--51
\bibitem[Shapiro \etal(2010)]{Sha10}
   Shapiro KL, Genzel R, F\"orster Schreiber NM. 2010.
   \mnras, 403:L36--40
\bibitem[Shapley(2011)]{Sha11}
  Shapley AE. 2011.
  \araa, 49:525--80
\bibitem[Shapley \etal(2003)]{Sha03}
  Shapley AE, Steidel CC, Pettini M, Adelberger KL.  2003.
  \apj, 588:65--89
\bibitem[Shapley \etal(2015)]{Sha15}
  Shapley AE, Reddy NA, Kriek M, \etal~2015.
  \apj, 801:88
\bibitem[Shapley \etal(2019)]{Sha19}
  Shapley AE, Sanders RL, Shao P, \etal~2019.
  \apj, 881:L35
\bibitem[Shivaei \etal(2015)]{Shi15}
  Shivaei I, Reddy NA, Shapley AE, \etal~2015.
  \apj, 815:98
\bibitem[Silverman \etal(2009)]{Silv09}
  Silverman JD, Lamareille F, Maier C, \etal~2009.
  \apj, 696:396--410
\bibitem[Silverman \etal(2015)]{Silv15}
  Silverman JD, Daddi E, Rodighiero G, \etal~2015.
  \apj, 812:L23
\bibitem[Simmons \etal(2017)]{Simmons17}
  Simmons BD, Lintott C, Willett KW, \etal~2017.
  \mnras, 464:4420--47
\bibitem[Simons \etal(2017)]{Sim17}
  Simons RC, Kassin SA, Weiner BJ, \etal~2017.
  \apj, 843:46
\bibitem[Simons \etal(2019)]{Sim19}
  Simons RC, Kassin SA, Snyder GF, \etal~2019.
  \apj, 874:59
\bibitem[Skelton \etal(2014)]{Ske14}
  Skelton RE, Whitaker KE, Momcheva IG, \etal~2014.
  \apjs, 214:24
\bibitem[Smail \etal(1997)]{Smail97}
  Smail I, Ivision RJ, Blain AW. 1997.
  \apj, 490:5--8
\bibitem[Smol\v{c}i\'c \etal(2017)]{Smol17}
  Smol\v{c}i\'c V, Novak M, Bondi M, \etal~2017.
  \aap, 602:A1
\bibitem[Snyder \etal(2015)]{Snyder15}
  Snyder GF, Lotz JM, Moody C, \etal~2015.
  \mnras, 451:4290--310
\bibitem[Snyder \etal(2017)]{Snyder17}
  Snyder GF, Lotz JM, Rodriguez-Gomez V, \etal~2017.
  \mnras, 468:207--16
\bibitem[Somerville \& Dav\'e(2015)]{Som15}
  Somerville RS, Dav\'e R. 2015.
  \araa, 53:51--113
\bibitem[Somerville \etal(2008)]{Som08}
  Somerville RS, Barden M, Rix H-W., \etal~2008.
  \apj, 672:776--86
\bibitem[Somerville \etal(2018)]{Som18}
  Somerville RS, Behroozi P, Pandya V, \etal~2018.
  \mnras, 473:2714--36
\bibitem[Soto \etal(2017)]{Soto17}
  Soto E, de Mello DF, Rafelski M, \etal~2017.
  \apj, 834:6
\bibitem[Speagle \etal(2014)]{Spe14}
  Speagle JS, Steinhardt CL, Capak PL, Silverman JD. 2014.
  \apjs, 214:15
\bibitem[Spilker \etal(2015)]{Spilker15}
  Spilker JS, Aravena M, Marrone DP, \etal~2015.
  \apj, 811:124
\bibitem[Spilker \etal(2016)]{Spilker16}
  Spilker JS, Bezanson R, Marrone DP, \etal~2016.
  \apj, 832:19
\bibitem[Springel(2010)]{Springel10}
 Springel V. 2010.
 \mnras, 401:791--851
\bibitem[Stach \etal(2018)]{Stach18}
  Stach SM, Smail I, Swinbank AM, \etal~2018.
  \apj, 860:161
\bibitem[Steidel \etal(2016)]{Ste16}
  Steidel CC, Strom AL, Pettini M, \etal~2016.
  \apj, 826:159
\bibitem[Stott \etal(2016)]{Sto16}
  Stott JP, Swinbank AM, Johnson HL, \etal~2016.
  \mnras, 457:1888--904
\bibitem[Strickland \etal(2004)]{Str04}
  Strickland DK, Heckman TM, Colbert EJM, Hoopes CG, Weaver KA. 2004.
  \apj, 606:829--52
\bibitem[Strom \etal(2018)]{Strom18}
  Strom AL, Steidel CC, Rudie GC \etal~2018.
  \apj, 836:164
\bibitem[Suess \etal(2019)]{Suess19}
  Suess KA, Kriek M, Price SH, Barro G. 2019.
  \apj, 877:103
\bibitem[Swinbank \etal(2017)]{Swi17}
  Swinbank AM, Harrison CM, Trayford J, \etal~2017.
  \mnras, 467:3140--59
\bibitem[Swinbank \etal(2019)]{Swi19}
  Swinbank AM, Harrison CM, Tiley AL, \etal~2019.
  \mnras, 487:381--93
\bibitem[Szomoru \etal(2013)]{Szomoru13}
  Szomoru D, Franx M, van Dokkum PG, \etal~2013.
  \apj, 763:73
\bibitem[Tacchella \etal(2015)]{Tacch15}
 Tacchella S, Carollo CM, Renzini A, \etal~2015.
 \sci, 348:314--7
\bibitem[Tacchella \etal(2016)]{Tacch16}
 Tacchella S, Dekel A, Carollo CM, \etal~2016.
 \mnras, 458:242--63
\bibitem[Tacchella \etal(2018)]{Tacch18}
 Tacchella S, Carollo CM, F\"orster Schreiber NM, \etal~2018.
 \apj, 859:56
\bibitem[Tacconi \etal(2006)]{Tac06}
 Tacconi LJ, Neri R, Chapman, SC, \etal~2006.
 \apj, 640:228--40
\bibitem[Tacconi \etal(2008)]{Tac08}
 Tacconi LJ, Genzel R, Smail I, \etal~2008.
 \apj, 680:246--62
\bibitem[Tacconi \etal(2013)]{Tac13}
 Tacconi LJ, Neri R, Genzel R, \etal~2013.
 \apj, 768:74
\bibitem[Tacconi \etal(2018)]{Tac18}
  Tacconi LJ, Genzel R, Saintonge A, \etal~2018.
  \apj, 853:179
\bibitem[Tacconi \etal(2020)]{Tac20}
  Tacconi LJ, Genzel R, Sternberg A. 2020.
  \araa, this volume
\bibitem[Tadaki \etal(2017a)]{Tad17a}
  Tadaki K, Genzel R, Kodama T, \etal~2017a.
  \apj, 834:135
\bibitem[Tadaki \etal(2017b)]{Tad17b}
  Tadaki K, Kodama T, Nelson EJ, \etal~2017b.
  \apjl, 841:L25
\bibitem[Tadhunter(2016)]{Tadh16}
  Tadhunter C. 2016.
  \aarev, 24:10
\bibitem[Talia \etal(2017)]{Tal17}
  Talia M, Brusa M, Cimatti A, \etal~2017
  \mnras, 471:4527--40
\bibitem[Teklu \etal(2018)]{Tek18}
  Teklu AF, Remus R-S, Dolag K, \etal~2018.
  \apj, 854:L28
\bibitem[Thomas \etal(2005)]{Thomas05}
  Thomas D, Maraston C, Bender R, Mendes de Oliveira C. 2005.
  \apj, 621:673-94
\bibitem[Thompson \etal(2015)]{Thompson15}
  Thompson R, Dav\'e R, Huang S, Katz N. 2015.
  arXiv1508.01851
\bibitem[Tiley \etal(2019a)]{Tiley19a}
  Tiley AL, Bureau M, Cortese L, \etal~2019a.
  \mnras, 482:2166--88
\bibitem[Tiley \etal(2019b)]{Tiley19b}
  Tiley AL, Swinbank AM, Harrison CM, \etal~2019b.
  \mnras, 485:934--60
\bibitem[Toft \etal(2014)]{Toft14}
  Toft S, Smol\v{c}i\'{c} V, Magnelli B., \etal~2014.
  \apj, 782:68
\bibitem[Toft \etal(2017)]{Toft17}
  Toft S, Zabl J, Richard J, \etal~2017.
  \nat, 546:510--3
\bibitem[Tomczak \etal(2014)]{Tom14}
  Tomczak AR, Quadri RF, Tran K-VH, \etal~2014
  \apj, 783, 85
\bibitem[Tomczak \etal(2016)]{Tom16}
  Tomczak AR, Quadri RF, Tran K-VH, \etal~2016
  \apj, 817, 118
\bibitem[Torrey \etal(2017)]{Torrey17}
  Torrey P, Wellons S, Ma CP, Hopkins PF, Vogelsberger M. 2017.
  \mnras, 467:4872--85
\bibitem[Turner \etal(2017)]{Turner17}
  Turner OJ, Cirasuolo M, Harrison CM, \etal~2017
  \mnras, 471, 1280--320
\bibitem[\"Ubler \etal(2014)]{Ueb14}
  \"Ubler H, Naab T, Oser L, \etal~2014.
  \mnras, 443:2092--111
\bibitem[\"Ubler \etal(2017)]{Ueb17}
  \"Ubler H, F\"orster Schreiber NM, Genzel R, \etal~2017.
  \apj, 842:121
\bibitem[\"Ubler \etal(2018)]{Ueb18}
  \"Ubler H, Genzel R, Tacconi LJ, \etal~2018.
  \apj, 854:L24
\bibitem[\"Ubler \etal(2019)]{Ueb19}
  \"Ubler H, Genzel R, Wisnioski E, \etal~2019.
  \apj, 880:48
\bibitem[van der Kruit \& Allen(1978)]{vdK78}
  van der Kruit PC, Allen RJ. 1978.
  \araa, 16:103--39
\bibitem[van der Wel \etal(2011)]{vdWel11}
 van der Wel A, Rix, HW, Wuyts S, \etal~2011.
 \apj, 730:38
\bibitem[van der Wel \etal(2014a)]{vdWel14a} 
  van der Wel A., Franx M, van Dokkum PG, \etal~2014a.
  \apj, 788:28
\bibitem[van der Wel \etal(2014b)]{vdWel14b} 
  van der Wel A., Chang Y, Bell EF, \etal~2014b.
  \apjl, 792:L6
\bibitem[van de Voort(2016)]{vdVoort16}
 van de Voort F. 2016.
 \mnras, 462:778--93
\bibitem[van Dokkum \etal(2010)]{vD10} 
 van Dokkum PG, Whitaker KE, Brammer G, \etal~2010.
 \apj, 709:1018--41
\bibitem[van Dokkum \etal(2013)]{vD13} 
  van Dokkum PG, Leja J, Nelson EJ, \etal~2013.
  \apjl, 771:L35
\bibitem[van Dokkum \etal(2015)]{vD15}
 van Dokkum PG, Nelson EJ, Franx M, \etal~2015.
 \apj, 813:23
\bibitem[Veilleux \etal(2005)]{Vei05}
 Veilleux S, Cecil G, Bland-Hawthorn J. 2005.
 \araa, 43:769--826
\bibitem[Wang \etal(2017)]{Wang17}
 Wang W, Faber SM, Liu FS, \etal~2017.
 \mnras, 469:4063--82
\bibitem[Weiner \etal(2009)]{Wei09}
  Weiner BJ, Coil AL, Prochaska JX, \etal~2009.
  \apj, 692:187--211
\bibitem[Weiner \etal(2006a)]{Wei06a}
  Weiner BJ, Willmer CNA, Faber SM, \etal~2006a.
  \apj, 653:1027--48
\bibitem[Wellons \etal(2015)]{Wellons15}
  Wellons S, Torrey P, Ma CP, \etal~2015.
  \mnras, 449:361--72
\bibitem[Wellons \& Torrey(2017)]{Wellons17}
  Wellons S, Torrey P. 2017.
  \mnras, 467:3887--97
\bibitem[Wellons \etal(2020)]{Wellons20}
  Wellons S, Faucher-Gigu\'ere C-A, Angl\'es-Alc\`azar\,D, \etal~2020.
  \mnras, 497, 4051--65
\bibitem[Whitaker \etal(2011)]{Whi11}
  Whitaker KE, Labb\'e I, van Dokkum PG, \etal~2011.
  \apj, 735:86
\bibitem[Whitaker \etal(2012)]{Whi12}
  Whitaker KE, van Dokkum PG, Brammer G, Franx M. 2012.
  \apj, 754:29
\bibitem[Whitaker \etal(2014)]{Whi14}
  Whitaker KE, Franx M, Leja J, \etal~2014.
  \apj, 795:104
\bibitem[Williams \etal(2009)]{Williams09}
  Williams RJ, Quadri RF, Franx M, van Dokkum PG, Labb\'{e} I. 2009.
  \apj, 691:1879--95
\bibitem[Wilman \etal(2020)]{Wilman20}
  Wilman DJ, Fossati M, Mendel JT, \etal~2020.
  \apj, 892:1
\bibitem[Windhorst \etal(1995)]{Windhorst95}
 Windhorst RA, Fomalont EB, Kellermann KI, \etal~1995.
 \nat, 375:471--4
\bibitem[Wisnioski \etal(2015)]{Wis15}
  Wisnioski E, F\"orster Schreiber NM, Wuyts S, \etal~2015.
  \apj, 799:209
\bibitem[Wisnioski \etal(2018)]{Wis18}
 Wisnioski E, Mendel JT, F\"orster Schreiber NM, \etal~2018.
 \apj, 855:97
\bibitem[Wisnioski \etal(2019)]{Wis19}
  Wisnioski E, F\"orster Schreiber NM, Fossati M, \etal~2019.
  \apj, 886:124
\bibitem[Wright \etal(2019)]{Wright19}
 Wright AH, Hildebrandt H, Kuijken K, \etal~2019.
 \aap, 632:A34
\bibitem[Wuyts \etal(2014)]{EWuy14}
 Wuyts E, Kurk JD, F\"orster Schreiber NM, \etal~2014.
 \apj, 789:L40
\bibitem[Wuyts \etal(2010)]{Wuy10}
 Wuyts S, Cox TJ, Hayward CC, \etal~2010.
 \apj, 722:1666--84
\bibitem[Wuyts \etal(2011a)]{Wuy11a}
  Wuyts S, F\"orster Schreiber NM, Lutz D, \etal~2011a.
  \apj, 738:106
\bibitem[Wuyts \etal(2011b)]{Wuy11b}
  Wuyts S, F\"orster Schreiber NM, van der Wel A, \etal~2011b.
  \apj, 742:96
\bibitem[Wuyts \etal(2012)]{Wuy12}
 Wuyts S, F\"orster Schreiber NM, Genzel R, \etal~2012.
 \apj, 753:114
\bibitem[Wuyts \etal(2013)]{Wuy13}
 Wuyts S, F\"orster Schreiber NM, Nelson EJ, \etal~2013.
 \apj, 779:135
\bibitem[Wuyts \etal(2016b)]{Wuy16b}
 Wuyts S, F\"orster Schreiber NM, Wisnioski E, \etal~2016b.
 \apj, 831:149
\bibitem[Zahid \etal(2014)]{Zahid14}
 Zahid HJ, Dima GI, Kudritzki R-P, \etal~2014.
 \apj, 791:130
\bibitem[Zanella \etal(2015)]{Zan15}
  Zanella A, Daddi E, Le Floc'h E, \etal~2015.
  \nat, 521:54--6
\bibitem[Zavala \etal(2016)]{Zav16}
  Zavala J, Frenk CS, Bower R, \etal~2016.
  \mnras, 460:4466--82
\bibitem[Zhang \etal(2019)]{Zhang19}
  Zhang H, Primack JR, Faber SM, \etal~2019.
  \mnras, 484:5170--91
\bibitem[Zolotov \etal(2015)]{Zol15}
  Zolotov A, Dekel A, Mandelker N, \etal~2015.
  \mnras, 450:2327--53

\end{thebibliography}

\begin{thebibliography}{00}

\bibitem[Aihara \etal(2018)]{Aih18}
  Aihara H, Armstrong R, Bickerton S, \etal~2018.
  \pasj, 70:S8
\bibitem[Atek \etal(2010)]{Atek10}
  Atek H, Malkan M, McCarthy P, \etal~2010.
  \apj, 723:104-15
\bibitem[Barger \etal(2008)]{Bar08}
 Barger AJ, Cowie LL, Wang W-H. 2008.
 \apj, 689:687--708
\bibitem[Baugh \etal(2005)]{Baugh05}
 Baugh CM, Lacey CG, Frenk CS, \etal~2005.
 \mnras, 356:1191--1200
\bibitem[Binney \& Tremaine(2008)]{Bin08}
  Binney J, Tremaine S. 2008.
  Galactic Dynamics (2nd ed.; Princeton, NJ: Princeton Univ. Press)
\bibitem[Bouch\'e \etal(2015)]{Bou15}
  Bouch\'e N, Carfantan H, Schroetter I, Michel-Dansac L, Contini T. 2015.
  \aj, 150:92
\bibitem[Burkert(1995)]{Bur95}
  Burkert A. 1995.
  \apj, 447:L25--8
\bibitem[Burkert \etal(2016)]{Bur16}
  Burkert A, F\"orster Schreiber NM, Genzel R, \etal~2016.
  \apj, 826:214
\bibitem[Calzetti \etal(2000)]{Calz00}
  Calzetti D, Armus L, Bohlin RC, \etal~2000.
  \apj, 533:682--95
\bibitem[Cappellari \etal(2013)]{Cap13}
 Cappellari M, McDermid RM, Alatalo K, \etal~2013.
 \mnras, 432:1862--93
\bibitem[Cardamone \etal(2010)]{Car10}
  Cardamone CN, van Dokkum PG, Urry CM, \etal~2010.
  \apjs, 189:270--85
\bibitem[Carnall \etal(2018)]{Carn18}
  Carnall AC, McLure RJ, Dunlop JS, Dav\'e R. 2018
  \mnras, 480:4379--401
\bibitem[Chabrier(2003)]{Cha03}
 Chabrier G. 2003
 \pasp, 115, 763--95
\bibitem[Chevallard \& Charlot(2016)]{Chevallard16}
  Chevallard J, Charlot S. 2016.
  \mnras, 462:1415--43
\bibitem[Colbert \etal(2013)]{Col13}
 Colbert JW, Teplitz H, Atek H, \etal~2013.
 \apj, 779:34
\bibitem[Conroy \& van Dokkum(2012)]{Con12}
 Conroy C, van Dokkum PG. 2012.
 \apj, 760:71--87
\bibitem[Conroy(2013)]{Conroy13}
  Conroy C. 2013.
  \araa, 51:393--455
\bibitem[Conroy \etal(2017)]{Con17}
 Conroy C, van Dokkum PG, Villaume A. 2017.
 \apj, 837:166
\bibitem[Contini \etal(2012)]{Cont12}
  Contini T, Garilli B, Le F\`evre O, \etal~2012.
  \aap, 539:A91
\bibitem[Cooper \etal(2012)]{Coo12}
  Cooper MC, Griffith RL, Newman JA, \etal~2012.
  \mnras, 419:3018--27
\bibitem[Cresci \etal(2009)]{Cre09}
  Cresci C, Hicks EKS, Genzel R, \etal~2009.
  \apj, 697:115--32
\bibitem[Curti \etal(2020)]{Curti20}
  Curti M, Maiolino R, Cirasuolo M, \etal~2020.
  \mnras, 492:821--42
\bibitem[da Cunha \etal(2015)]{daCunha15}
  da Cunha E, Walter F, Smail IR, \etal~2015.
  \apj, 806:110
\bibitem[Davies \etal(2011)]{Davies11}
  Davies RI, F\"orster Schreiber NM, Cresci G, \etal~2011.
  \apj, 741:69
\bibitem[Davies \etal(2019)]{Dav19}
  Davies RL, F\"orster Schreiber NM, \"Uebler H, \etal~2019
  \apj, 873, 122
\bibitem[Davis \etal(2007)]{Davis07}
  Davis M, Guhathakurta P, Konidaris NP, \etal~2007.
  \apj, 660:L1--6
\bibitem[de Blok \& Walter(2014)]{deBlok14}
  de Blok WJG, Walter F. 2014.
  \aj, 147:96
\bibitem[De Looze \etal(2014)]{DeLooze14}
  De Looze I, Fritz J, Baes M, \etal~2014.
  \aap, 571:69
\bibitem[di Teodoro \& Fraternali(2015)]{diT15}
  Di Teodoro EM, Fraternali F. 2015.
  \mnras, 451:3021--33
\bibitem[Eldridge \& Stanway(2012)]{Eldridge12}
  Eldridge JJ, Stanway ER. 2012.
  \mnras, 419:479--89
\bibitem[\'Epinat \etal(2012)]{Epi12}
  \'Epinat B, Tasca L, Amram P, \etal~2012.
  \aap, 539:A92
\bibitem[Ferland \etal(2017)]{Ferland17}
  Ferland GJ, Chatzikos M, Guzm\'an F, \etal~2017.
  \RMxAA, 53:385
\bibitem[F\"orster Schreiber \etal(2009)]{FS09}
  F\"orster Schreiber NM, Genzel R, Bouch\'e N, \etal~2009.
  \apj, 706:1364--428
\bibitem[F\"orster Schreiber \etal(2018)]{FS18}
  F\"orster Schreiber NM, Renzini A, Mancini C, \etal~2018.
  \apjs, 238:21
\bibitem[Genzel \etal(2020)]{Genz20}
  Genzel R, Price SH, \"Ubler H, \etal~2020.
  \apj, submitted (arXiv:2006.03046)
\bibitem[Gillman \etal(2019)]{Gill19}
   Gillman S, Swinbank AM, Tiley AL, \etal~2019.
   \mnras, 486:175--94
\bibitem[Gillman \etal(2020)]{Gill20}
   Gillman S, Tiley AL, Swinbank AM, \etal~2020.
   \mnras, 492:1492--512
\bibitem[Girard \etal(2018)]{Gir18}
   Girard M, Dessauges-Zavadsky M, Schaerer D, \etal~2018.
   \aap, 613:A72
\bibitem[Girard \etal(2020)]{Gir20}
   Girard M, \etal~2020.
   \mnras, submitted
\bibitem[Grogin \etal(2011)]{Gro11}
  Grogin NA, Kocevski DD, Faber SM, \etal~2011.
  \apjs, 197:35
\bibitem[Harrison \etal(2016)]{Har16}
  Harrison CM, Alexander DM, Mullaney JR, \etal~2016.
  \mnras, 456:1195--220
\bibitem[Harrison \etal(2017)]{Har17}
  Harrison CM, Johnson HL, Swinbank AM, \etal~2017.
  \mnras, 467:1965--83
\bibitem[Hopkins \etal(2018)]{Hopkins18}
 Hopkins AM. 2018.
 \pasa, 35:39
\bibitem[Johnson \etal(2018)]{John18}
  Johnson AL, Harrison CM, Swinbank AM, \etal~2018.
  \mnras, 474:5076--104
\bibitem[Jones \etal(2010)]{Jones10}
  Jones TA, Swinbank AM, Ellis RS, Richard J, Stark DP. 2010.
  \mnras, 404:1247--62
\bibitem[Jones \etal(2013)]{Jones13}
  Jones TA, Ellis RS, Richard J, Jullo E. 2013.
  \apj, 765:48
\bibitem[Kashino \etal(2019)]{Kash19}
  Kashino D, Silverman JD, Sanders D, \etal~2019.
  \apjs, 241:10
\bibitem[Kewley \etal(2019)]{Kew19}
  Kewley LJ, Nicholls DC, Sutherland RS. 2019.
  \araa, 57:511--70
\bibitem[Koekemoer \etal(2011)]{Koe11}
  Koekemoer AM, Faber SM, Ferguson HC, \etal~2011.
  \apjs, 197:36
\bibitem[Krajnovi\'c \etal(2006)]{Kraj06}
  Krajnovi\'c D, Capellari M, de Zeeuw PT, Copin Y. 2006.
  \mnras, 366:787--802
\bibitem[Kriek \etal(2010)]{Kriek10}
  Kriek M, Labb\'e I, Conroy C, \etal~2010.
  \apj, 722:64
\bibitem[Kriek \& Conroy(2013)]{Kriek13}
  Kriek M, Conroy C. 2013.
  \apj, 775:16
\bibitem[Kriek \etal(2015)]{Kriek15}
  Kriek M, Shapley AE, Reddy NA, \etal~2015.
  \apjs, 218:15
\bibitem[Laigle \etal(2016)]{Lai16}
  Laigle C, McCracken HJ, Ilbert O, \etal~2016.
  \apjs, 224:24
\bibitem[Law \etal(2009)]{Law09}
  Law DR, Steidel CC, Erb DK, \etal~2009
  \apj, 697:2057--82
\bibitem[Law \etal(2012)]{Law12}
  Law DR, Shapley AE, Steidel CC, \etal~2012
  \nat, 487:338--40
\bibitem[Leethochawalit \etal(2016)]{Leetho16}
  Leethochawalit N, Jones TA, Ellis RS, \etal~2016.
  \apj, 820:84
\bibitem[Le F\`evre \etal(2013)]{LeF13}
  Le F\`evre O, Cassata P, Cucciati O, \etal~2013.
  \aap, 559:A14
\bibitem[Le F\`evre \etal(2015)]{LeF15}
  Le F\`evre O, Tasca LAM, Cassata P, \etal~2015.
  \aap, 576:A79
\bibitem[Leja \etal(2017)]{Lej17}
  Leja J, Johnson BD, Conroy C, van Dokkum PG, Byler N. 2017.
  \apj, 837:170
\bibitem[Leja \etal(2019a)]{Lej19a}
 Leja J, Carnall AC, Johnson BD, \etal~2019a.
 \apj, 876:3
\bibitem[Leja \etal(2019b)]{Lej19b}
 Leja J, Johnson BD, Conroy C, \etal~2019b.
 \apj, 877:140
\bibitem[Li \etal(2019)]{Li19}
  Li H, Wuyts S, Hao L, \etal~2019.
  \apj, 872:63
\bibitem[Lilly \etal(2009)]{Lil09}
  Lilly SJ, Le Brun V, Maier C, \etal~2009.
  \apjs, 184:218--29
\bibitem[Livermore \etal(2015)]{Liv15}
  Livermore RC, Jones TA, Richard J, \etal~2015.
  \mnras, 450:1812--35
\bibitem[Lotz \etal(2017)]{Lotz17}
  Lotz JM, Koekemoer A, Coe D, \etal~2017.
  \aj, 837:97
\bibitem[Maiolino \& Mannucci(2019)]{Mai19}
  Maiolino R, Mannucci F. 2019.
  \aap Review, 27:3
\bibitem[Mancini \etal(2011)]{Man11}
  Mancini C, F\"orster Schreiber NM, Renzini A, \etal~2011.
  \apj, 743:86
\bibitem[Maraston(2005)]{Maraston05}
  Maraston C. 2005.
  \mnras, 362:799--828
\bibitem[Mason \etal(2017)]{Mas17}
  Mason CA, Treu T, Fontana A, \etal~2017.
  \apj, 838:14
\bibitem[McLure \etal(2018)]{McL18}
  McLure RJ, Pentericci L, Cimatti A, \etal~2018.
  \mnras, 479:25--42
\bibitem[Mieda \etal(2016)]{Mie16}
  Mieda E, Wright SA, Larkin JE, \etal~2016.
  \apj, 831:78
\bibitem[Mobasher \etal(2015)]{Mobasher15}
  Mobasher B, Dahlen T, Ferguson HC, \etal~2015.
  \apj, 808:101
\bibitem[Momcheva \etal(2016)]{Mom16}
  Momcheva IG, Brammer GB, van Dokkum PG, \etal~2016.
  \apjs, 225:27
\bibitem[Nanayakkara \etal(2016)]{Nana16}
  Nanayakkara T, Glazebrook K, Kacprzak GG, \etal~2016.
  \apj, 828:21
\bibitem[Navarro \etal(1996)]{Nav96}
  Navarro JF, Frenk CS, White SDM. 1996.
  \apj, 462:563--75
\bibitem[Noordermeer(2008)]{Noor08}
  Noordermeer E. 2008.
  \mnras, 385:1359--64
\bibitem[Newman \etal(2013)]{Newman13}
  Newman JA, Cooper MC, Davis M, \etal~2013.
  \apjs, 208:5
\bibitem[Okada \etal(2016)]{Oka16}
  Okada H, Totani T, Tonegawa M, \etal~2016.
  \pasj, 68:47
\bibitem[Pacifici \etal(2015)]{Pacifici15}
  Pacifici C, da Cunha E, Charlot S, \etal~2015.
  \mnras, 447:786--805
\bibitem[Pentericci \etal(2018)]{Pen18}
  Pentericci L, McLure RJ, Garilli B, \etal~2018.
  \aap, 616:A174
\bibitem[Price \etal(2016)]{Pri16}
  Price SH, Kriek M, Shapley AE, \etal~2016
  \apj, 819:80
\bibitem[Reddy \etal(2015)]{Red15}
  Reddy NA, Kriek M, Shapley AE, \etal~2015.
  \apj, 806, 259
\bibitem[Renzini(2009)]{Ren09}
  Renzini A. 2009.
  \mnras, 398:58--62
\bibitem[Safarzadeh \etal(2017)]{Safarzadeh17}
 Safarzadeh M, Lu Y, Hayward CC. 2017.
 \mnras, 472:2462--67
\bibitem[Scholtz \etal(2020)]{Scholtz20}
  Scholtz J, Harrison CM, Rosario DJ, \etal~2020.
  \mnras, 492:3194--3216
\bibitem[Scoville \etal(2007a)]{Sco07}
  Scoville NZ, Aussel H, Brusa M, \etal~2007.
  \apjs, 172:1--8
\bibitem[Shapiro \etal(2008)]{Sha08}
   Shapiro KL, Genzel R, F\"orster Schreiber, \etal~2008.
   \apj, 682:231--51
\bibitem[Shipley \etal(2018)]{Shi18}
  Shipley HV, Lange-Vagle D, Marchesini D, \etal~2018.
  \apjs, 235:14
\bibitem[Steidel \etal(2003)]{Ste03}
  Steidel CC, Adelberger KL, Shapley AE, \etal~2003.
  \apj, 592:728--54
\bibitem[Steidel \etal(2004)]{Ste04}
  Steidel CC, Shapley AE, Pettini M, \etal~2004.
  \apj, 604:534--50
\bibitem[Steidel \etal(2014)]{Ste14}
  Steidel CC, Rudie GC, Strom AL, \etal~2014.
  \apj, 795:165
\bibitem[Stott \etal(2016)]{Sto16}
  Stott JP, Swinbank AM, Johnson HL, \etal~2016.
  \mnras, 457:1888--904
\bibitem[Straatman \etal(2016)]{Straat16}
  Straatman CMS, Spitler LR, Quadri RF, \etal~2016.
  \apj, 830:51
\bibitem[Strom \etal(2017)]{Strom17}
  Strom AL, Steidel CC, Rudie GC \etal~2017.
  \apj, 836:164
\bibitem[Sutherland \& Dopita(2017)]{Sutherland17}
  Sutherland RS, Dopita MA. 2017.
  \apjs, 229:34
\bibitem[Tasca \etal(2017)]{Tas17}
  Tasca LAM, Le F\`evre O, Ribeiro B, \etal~2017.
  \aap, 600:A110
\bibitem[Tiley \etal(2019)]{Tiley19}
  Tiley AL, Swinbank AM, Harrison CM, \etal~2019.
  \mnras, 485:934--60
\bibitem[Tonegawa \etal(2015)]{Ton15}
  Tonegawa M, Totani T, Okada H, \etal~2015
  \pasj, 67:81
\bibitem[Treu \etal(2010)]{Treu10}
  Treu T, Auger MW, Koopmans LVE, \etal~2010.
  \apj, 709:1195
\bibitem[Treu \etal(2015)]{Treu15}
  Treu T, Schmidt KB, Brammer GB, \etal~2015.
  \apj, 812:114
\bibitem[Troncoso \etal(2014)]{Tro14}
  Troncoso P, Maiolino R, Sommariva V, \etal~2014.
  \aap, 563:A58
\bibitem[Turner \etal(2017)]{Turner17}
  Turner OJ, Cirasuolo M, Harrison CM, \etal~2017
  \mnras, 471, 1280--320
\bibitem[\"Ubler \etal(2018)]{Ueb18}
  \"Ubler H, Genzel R, Tacconi LJ, \etal~2018.
  \apj, 854:L24
\bibitem[van der Wel \etal(2011)]{vdWel11}
 van der Wel A, Straughn AN, Rix, HW, \etal~2011.
 \apj, 742:111
\bibitem[Weiner \etal(2006)]{Wei06}
  Weiner BJ, Willmer CNA, Faber SM, \etal~2006.
  \apj, 653:1027--48
\bibitem[Whitaker \etal(2011)]{Whi11}
  Whitaker KE, Labb\'e I, van Dokkum PG, \etal~2011.
  \apj, 735:86
\bibitem[Wilkins \etal(2008)]{Wilkins08}
 Wilkins SM, Hopkins AM, Trentham N, Tojeiro R. 2008.
 \mnras, 391:363--68
\bibitem[Wirth \etal(2015)]{Wirth15}
 Wirth GD, Trump JR, Barro G, \etal~2015.
 \aj, 150:153
\bibitem[Wisnioski \etal(2011)]{Wis11}
  Wisnioski E, Glazebrook K, Blake C, \etal~2011.
  \mnras, 417:2061--23
\bibitem[Wisnioski \etal(2015)]{Wis15}
  Wisnioski E, F\"orster Schreiber NM, Wuyts S, \etal~2015.
  \apj, 799:209
\bibitem[Wisnioski \etal(2018)]{Wis18}
 Wisnioski E, Mendel JT, F\"orster Schreiber NM, \etal~2018.
 \apj, 855:97
\bibitem[Wisnioski \etal(2019)]{Wis19}
  Wisnioski E, F\"orster Schreiber NM, Fossati M, \etal~2019.
  \apj, 886:124
\bibitem[Wright \etal(2019)]{Wright19}
 Wright AH, Hildebrandt H, Kuijken K, \etal~2019.
 \aap, 632:A34
\bibitem[Wright \etal(2009)]{Wri09}
 Wright SA, Larkin JE, Law DR, \etal~2009.
 \apj, 699:421--440
\bibitem[Wuyts \etal(2014)]{EWuy14}
 Wuyts E, Rigby JR, Gladders MD, Sharon K. 2014.
 \apj, 781:61
\bibitem[Wuyts \etal(2012)]{Wuy12}
 Wuyts S, F\"orster Schreiber NM, Genzel R, \etal~2012.
 \apj, 753:114
\bibitem[Wuyts \etal(2016)]{Wuy16}
 Wuyts S, F\"orster Schreiber NM, Wisnioski E, \etal~2016.
 \apj, 831:149
\bibitem[Yuan \etal(2011)]{Yua11}
  Yuan T.-T., Kewley LJ, Swinbank AM, Richard J, Livermore RC. 2011.
  \apj, 732:L14
\bibitem[Yuan \etal(2012)]{Yua12}
  Yuan T.-T., Kewley LJ, Swinbank AM, Richard J. 2012.
  \apj, 759:66
\bibitem[Zibetti \etal(2013)]{Zibetti13}
 Zibetti S, Gallazzi A, Charlot S, Pierini D, Pasquali A. 2013.
 \mnras, 428:1479--97

\end{thebibliography}
\end{document}